\documentclass[11pt,a4paper]{article}
\pdfoutput=1
\usepackage{jcappub}
\usepackage{ifthen}
\usepackage{epsfig}
\usepackage{booktabs} 
\usepackage{natbib}
\usepackage{bbold}

\newcommand{\beqra}{\begin{flalign}}
\newcommand{\eeqra}{\end{flalign}}
\newcommand{\beq}{\begin{equation}}
\newcommand{\eeq}{\end{equation}}
\usepackage{ccicons} 
\usepackage{listings}
\usepackage{color}   
\usepackage{braket}  
\usepackage{hyperref}
\usepackage{dsfont}  
\usepackage{leftidx} 
\usepackage{bm}
\usepackage{placeins}
\usepackage{float}
\allowdisplaybreaks

\title{WIMP capture and annihilation in the Earth in effective theories}
\author[a]{Riccardo Catena}
\affiliation[a]{Chalmers University of Technology, Department of Physics, SE-412 96 G\"oteborg, Sweden}
\emailAdd{catena@chalmers.se}

\abstract{I calculate the rate of WIMP capture and annihilation in the Earth in the non-relativistic effective theory of dark matter-nucleon interactions.~Neglecting operator interference, I consider all Galilean invariant interaction operators that can arise from the exchange of a heavy particle of spin less than or equal to one when WIMPs have spin 0, 1/2 or 1.~I compute position and shape of the expected resonances in the mass - capture rate plane and show that Iron is not the most important element in the capture process for many currently ignored interaction operators.~I compare these predictions with the recent results of an Earth WIMP analysis of IceCube in the 86-string configuration and set limits on all isoscalar and isovector coupling constants of the effective theory of dark matter-nucleon interactions.~For certain interaction operators and for a dark matter particle mass of about 50 GeV, I find that these limits are stronger than those I have previously derived in an analysis of the solar WIMP search performed at IceCube in the 79-string configuration.}

\keywords{dark matter theory, dark matter experiments} 

\begin{document}
\maketitle

\section{Introduction}
\label{sec:intro}

Understanding the nature of dark matter is an increasingly important research question in Astroparticle Physics~\cite{Bertone:2010at}.~The search for a first unambiguous non-gravitational signal of dark matter is currently pursued through a variety of complementary approaches~\cite{Bertone:2004pz}.~In the standard paradigm of Weakly Interacting Massive Particles (WIMPs) as a dark matter candidate, WIMPs can be detected via scattering by nuclei in underground laboratories (direct detection), through their annihilation or decay in space (indirect detection), or through WIMP production at particle accelerators such as the Large Hadron Collider (LHC)~\cite{Jungman:1995df,Bergstrom:2000pn,Catena:2013pka}.~At the interface of WIMP direct and indirect detection is the search for energetic neutrinos from the annihilation of WIMPs captured in the Sun or Earth via scattering by nuclei~\cite{Silk:1985ax}, or self-interactions~\cite{Zentner:2009is,Catena:2016ckl}.

Crossing the Sun or Earth, WIMPs might lose energy via local interactions, and scatter from gravitationally unbound to gravitationally bound orbits.~In this scenario, WIMPs are expected to accumulate at the Sun's or Earth's centre through subsequent scattering events.~The accumulation of WIMPs at the centre of a celestial body leads to an increase in the local WIMP density.~As a result, WIMPs eventually annihilate at a potentially observable rate, producing Standard Model particles, and in particular neutrinos.~Neutrino observatories such as IceCube, Super-Kamiokande, ANTARES, BAKSAN, and Baikal are currently testing this hypothesis~\cite{Aartsen:2016exj,Choi:2015ara,Adrian-Martinez:2016gti,Boliev:2013ai,Avrorin:2014swy}.~In this study, I primarily focus on the capture and annihilation of WIMPs in the Earth.~For a recent solar WIMP analysis of neutrino telescopes in effective theories see~\cite{Catena:2015iea,Blumenthal:2014cwa,Liang:2013dsa,Guo:2013ypa}.

The first pioneering studies of WIMP capture and annihilation in the Earth by Freese \cite{Freese:1985qw} and others~\cite{Krauss:1985aaa,Gaisser:1986ha,Gould:1987ir} assumed the Earth to be in free space.~Corrections due to the Sun's gravitational field~\cite{Gould:1987ww}, WIMP diffusion in the solar system~\cite{Gould:1991rc}, solar depletion~\cite{Gould:1999je,Lundberg:2004dn}, and WIMP weak scattering in the Sun~\cite{Sivertsson:2012qj} have subsequently been studied in detail.~It has been found that the free space approximation implies a relative error on the capture rate of at most 35\%, and for specific WIMP masses only.~For WIMP masses close to the mass of an element in the Earth, the relative error on the capture rate tends to zero~\cite{Sivertsson:2012qj}.~So far, the expected neutrino flux from WIMP annihilation in the Earth has been computed for the standard spin-independent dark-matter nucleon interaction only.~Here, I assume the Earth to be in free space and extend previous calculations to virtually arbitrary WIMP-nucleon interactions.  

In this work, I compute the rate of WIMP capture and annihilation in the Earth in the non-relativistic effective theory of dark matter-nucleon interactions, formulated in~\cite{Chang:2009yt,Fan:2010gt,Fitzpatrick:2012ix,Fitzpatrick:2012ib} and developed in~\cite{Fornengo:2011sz,Menendez:2012tm,Cirigliano:2012pq,Anand:2013yka,DelNobile:2013sia,Klos:2013rwa,Peter:2013aha,Hill:2013hoa,Catena:2014uqa,Catena:2014hla,Catena:2014epa,Gluscevic:2014vga,Panci:2014gga,Vietze:2014vsa,Barello:2014uda,Catena:2015uua,Schneck:2015eqa,Dent:2015zpa,Catena:2015vpa,Kavanagh:2015jma,D'Eramo:2016atc,Catena:2016hoj,Kahlhoefer:2016eds}.~Neglecting operator interference, the theory includes all Galilean invariant dark matter-nucleon interaction operators that can arise from the exchange of a heavy particle of spin less than or equal to one for WIMPs of spin 0, 1/2 and 1.~For spin 1 WIMPs, two interactions not considered here can arise if operator interference is not negligible~\cite{Dent:2015zpa}.~I compute WIMP capture and annihilation rates considering eleven elements in the Earth's mantle and core, and using nuclear response functions obtained in~\cite{Catena:2015uha} for $^{16}$O, $^{23}$Na, $^{24}$Mg, $^{27}$Al, $^{28}$Si, $^{32}$S, $^{40}$Ca, $^{56}$Fe, and $^{58}$Ni and in this work for $^{31}$P and $^{52}$Cr.~I compare my calculations with the 90\% CL upper limits on the WIMP annihilation rate found in a recent WIMP analysis of IceCube in the 86-string configuration~\cite{Aartsen:2016fep}.~Through this comparison, I set limits on the isoscalar and isovector coupling constants of the non-relativistic effective theory of dark matter-nucleon interactions.~For certain interaction operators and for a dark matter particle mass of about 50 GeV, these limits are stronger than those I have previously found~\cite{Catena:2015iea} in an analysis of the solar WIMP search performed at IceCube in the 79-string configuration~\cite{Aartsen:2012kia}.

The paper is organised as follows.~In Sec.~\ref{sec:earth} I introduce the theoretical framework used to calculate the rate of WIMP capture and annihilation in the Earth.~I perform this calculation in Sec.~\ref{sec:results}, where using data from IceCube in the 79 and 86-string configuration~\cite{Aartsen:2012kia,Aartsen:2016fep}, Super-Kamiokande~\cite{Choi:2015ara} and LUX~\cite{Akerib:2013tjd}, I set 90\% CL upper limits on the coupling constants of the effective theory of Sec.~\ref{sec:earth}.~I conclude in Sec.~\ref{sec:conclusions}.~Appendix~\ref{sec:appDM} contains key equations, while in Appendix~\ref{sec:nuc} I describe the nuclear shell model calculation through which I derive the $^{31}$P and $^{52}$Cr nuclear response functions.~Finally, I collect in Appendix~\ref{app:figures} figures for capture rates and exclusion limits relative to interaction operators which for brevity are not discussed in the body of the paper.

\section{WIMP capture and annihilation in the Earth}
\label{sec:earth}

Galactic WIMPs are expected to interact in the Earth's core and mantle while crossing the planet.~By scattering to gravitationally bound orbits, WIMPs can thus be captured by the Earth.~In this scenario, WIMPs accumulate at the Earth's centre, where they eventually annihilate producing an observable neutrino flux through the decay of their annihilation products.~Neutrino observatories such as IceCube, Super-Kamiokande, ANTARES, BAKSAN and Baikal are currently searching for this signal~\cite{Aartsen:2016exj,Choi:2015ara,Adrian-Martinez:2016gti,Boliev:2013ai,Avrorin:2014swy}.~Below I introduce the equations that govern the scattering, capture and annihilation of WIMPs in the Earth.~Analogous expressions apply to solar WIMPs.

\subsection{Scattering}
\label{sec:scattering}
First, I review the theoretical framework used to calculate the cross-section for WIMP scattering by nuclei in the Earth.~I perform this calculation in the non-relativistic effective theory of dark matter-nucleon interactions~\cite{Chang:2009yt,Fan:2010gt,Fitzpatrick:2012ix,Fitzpatrick:2012ib}.~The theory is applicable in the limit of small momentum transfer, when the energy transferred in the scattering is small compared to the mass of the particle that mediates the interaction.~The theory predicts that the most general Hamiltonian density for dark matter-nucleon interactions is a linear combination of eighteen quantum mechanical operators~\cite{Dent:2015zpa}.~The eighteen operators can be expressed in terms of four building blocks:~the momentum transfer operator ${\bf{\hat{q}}}$, the transverse relative velocity operator ${\bf{\hat{v}}}^{\perp}$, and the dark matter particle and nucleon spin operators, ${\bf \hat{S}}_\chi$ and ${\bf \hat{S}}_\chi$, respectively.~Neglecting operator interference, and terms proportional to $|{\bf{\hat{v}}}^{\perp}|^2$, only fourteen independent operators remain in the Hamiltonian density~\cite{Catena:2015uua}.~I list the operators considered in this investigation in Tab.~\ref{tab:operators}.

\begin{table}[t]
    \centering
    \begin{tabular}{ll}
    \toprule
        $\hat{\mathcal{O}}_1 = \mathbb{1}_{\chi N}$ & $\hat{\mathcal{O}}_9 = i{\bf{\hat{S}}}_\chi\cdot\left({\bf{\hat{S}}}_N\times\frac{{\bf{\hat{q}}}}{m_N}\right)$  \\
        $\hat{\mathcal{O}}_3 = i{\bf{\hat{S}}}_N\cdot\left(\frac{{\bf{\hat{q}}}}{m_N}\times{\bf{\hat{v}}}^{\perp}\right)$ \hspace{2 cm} &   $\hat{\mathcal{O}}_{10} = i{\bf{\hat{S}}}_N\cdot\frac{{\bf{\hat{q}}}}{m_N}$   \\
        $\hat{\mathcal{O}}_4 = {\bf{\hat{S}}}_{\chi}\cdot {\bf{\hat{S}}}_{N}$ &   $\hat{\mathcal{O}}_{11} = i{\bf{\hat{S}}}_\chi\cdot\frac{{\bf{\hat{q}}}}{m_N}$   \\                                                                             
        $\hat{\mathcal{O}}_5 = i{\bf{\hat{S}}}_\chi\cdot\left(\frac{{\bf{\hat{q}}}}{m_N}\times{\bf{\hat{v}}}^{\perp}\right)$ &  $\hat{\mathcal{O}}_{12} = {\bf{\hat{S}}}_{\chi}\cdot \left({\bf{\hat{S}}}_{N} \times{\bf{\hat{v}}}^{\perp} \right)$ \\                                                                                                                 
        $\hat{\mathcal{O}}_6 = \left({\bf{\hat{S}}}_\chi\cdot\frac{{\bf{\hat{q}}}}{m_N}\right) \left({\bf{\hat{S}}}_N\cdot\frac{\hat{{\bf{q}}}}{m_N}\right)$ &  $\hat{\mathcal{O}}_{13} =i \left({\bf{\hat{S}}}_{\chi}\cdot {\bf{\hat{v}}}^{\perp}\right)\left({\bf{\hat{S}}}_{N}\cdot \frac{{\bf{\hat{q}}}}{m_N}\right)$ \\   
        $\hat{\mathcal{O}}_7 = {\bf{\hat{S}}}_{N}\cdot {\bf{\hat{v}}}^{\perp}$ &  $\hat{\mathcal{O}}_{14} = i\left({\bf{\hat{S}}}_{\chi}\cdot \frac{{\bf{\hat{q}}}}{m_N}\right)\left({\bf{\hat{S}}}_{N}\cdot {\bf{\hat{v}}}^{\perp}\right)$  \\
        $\hat{\mathcal{O}}_8 = {\bf{\hat{S}}}_{\chi}\cdot {\bf{\hat{v}}}^{\perp}$  & $\hat{\mathcal{O}}_{15} = -\left({\bf{\hat{S}}}_{\chi}\cdot \frac{{\bf{\hat{q}}}}{m_N}\right)\left[ \left({\bf{\hat{S}}}_{N}\times {\bf{\hat{v}}}^{\perp} \right) \cdot \frac{{\bf{\hat{q}}}}{m_N}\right] $ \\                                                                               
    \bottomrule
    \end{tabular}
    \caption{Interaction operators appearing in Eq.~(\ref{eq:H_chiT}).~For simplicity, I omit the nucleon index $i$ in the expressions above.~In the equations, $m_N$ is the nucleon mass and all interaction operators have the same mass dimension.} 
    \label{tab:operators}
\end{table}

In this study I focus on the following Hamiltonian density for non-relativistic dark matter-nucleus interactions
\begin{equation}
\hat{\mathcal{H}}_{\chi T}= \sum_{i=1}^{A}  \sum_{\tau=0,1} \sum_{j} c_j^{\tau}\hat{\mathcal{O}}_{j}^{(i)} \, t^{\tau}_{(i)} \,,
\label{eq:H_chiT}
\end{equation}
which is valid in the limit of one-body dark matter-nucleon interactions only.~Corrections induced by two-body currents in the WIMP scattering by nuclei are discussed in~\cite{Toivanen:2008zz,Cirigliano:2012pq,Menendez:2012tm,Klos:2013rwa,Hoferichter:2015ipa}.~In Eq.~(\ref{eq:H_chiT}), $A$ is the mass number of the target nucleus, labelled here by $T$, and the operators $t^0_{(i)}=\mathbb{1}_{2\times 2}$ and $t^1_{(i)}=\tau_3$, where $\tau_3$ is the third Pauli matrix, are defined in the isospin space of the $i$-th nucleon.~I denote the isoscalar and isovector coupling constants by $c_j^0$ and $c_j^1$, respectively.~They are linearly related to the coupling constants for protons and neutrons: 
\begin{align}
c^{p}_j=&(c^{0}_j+c^{1}_j)/2 \nonumber\\
c^{n}_j=&(c^{0}_j-c^{1}_j)/2\,,
\end{align}
and have dimension [mass]$^{-2}$.~Neglecting operator interference, Eqs.~(\ref{eq:H_chiT}) includes all Galilean invariant operators that can arise from the exchange of a heavy particle of spin $\le 1$ for WIMPs of spin $\le$1~\cite{Dent:2015zpa}.

\noindent The differential cross-section for WIMP-nucleus scattering can be calculated from the Hamiltonian density in Eq.~(\ref{eq:H_chiT}):
\begin{align}
\frac{{\rm d}\sigma_{\chi T}(q^2,w^2)}{{\rm d}q^2} =\frac{1}{(2J+1) w^2}\sum_{\tau,\tau'} &\bigg[ \sum_{k=M,\Sigma',\Sigma''} R^{\tau\tau'}_k\left(v_T^{\perp 2}, {q^2 \over m_N^2} \right) W_k^{\tau\tau'}(q^2) \nonumber\\
&+{q^{2} \over m_N^2} \sum_{k=\Phi'', \Phi'' M, \tilde{\Phi}', \Delta, \Delta \Sigma'} R^{\tau\tau'}_k\left(v_T^{\perp 2}, {q^2 \over m_N^2}\right) W_k^{\tau\tau'}(q^2) \bigg] \,, \nonumber\\
\label{eq:dsigma} 
\end{align}
where $J$ is the target nucleus spin, $w$ is the WIMP-nucleus relative velocity, and $v_T^{\perp 2}=w^2-q^2/(4\mu_T^2)$, where $q$ is the momentum transfer and $\mu_T$ is the WIMP-nucleus reduced mass.~The eight dark matter response functions $R^{\tau\tau'}_k$ depend on the coupling constants $c_j^\tau$, on $q^2/m_N^2$, where $m_N$ is the nucleon mass, and on $v_T^{\perp 2}$.~They were found in~\cite{Fitzpatrick:2012ix,Anand:2013yka} and are listed in Appendix~\ref{sec:appDM}.

The eight nuclear response functions $W_k^{\tau\tau'}$ in Eq.~(\ref{eq:dsigma}) are defined in Appendix~\ref{sec:nuc}.~They are expressed in terms of reduced matrix elements of nuclear charges and currents, and must be computed numerically.~Within this study I perform the shell model calculation described in Appendix~\ref{sec:nuc} to derive all $W_k^{\tau\tau'}$ relevant for Phosphorus and Chromium, as they were not known previously.~I use the nuclear response functions obtained in~\cite{Catena:2015uha} for the remaining elements in the Earth.

\subsection{Capture}
\label{sec:capture}
The rate of scattering from a velocity $w$ to a velocity less than the local escape velocity $v(r)$ at a distance $r$ from Earth's centre is~\cite{Gould:1987ir}:
\begin{equation}
\Omega_{v}^{-}(w)= \sum_T n_T w\,\Theta\left( \frac{\mu_T}{\mu^2_{+,T}} - \frac{u^2}{w^2} \right)\int_{E u^2/w^2}^{E \mu_T/\mu_{+,T}^2} {\rm d}E_r\,\frac{{\rm d}\sigma_{\chi T}\left(E_r,w^2\right)}{{\rm d}E_r}\,,
\label{eq:omega}
\end{equation}
where $m_\chi$ is the WIMP mass, $E=m_\chi w^2/2$, and $w=\sqrt{u^2+v(r)^2}$, $u$ being the WIMP velocity at infinity.~The sum in Eq.~(\ref{eq:omega}) extends over the most abundant elements in the Earth, with densities at $r$ denoted here by $n_T(r)$ and mass $m_T$.~The dimensionless parameters $\mu_T$ and $\mu_{\pm,T}$ in Eq.~(\ref{eq:omega}) are defined as follows:~$\mu_T\equiv m_\chi/m_T$ and $\mu_{\pm,T}\equiv (\mu_T\pm1)/2$.~Finally, the differential cross-section ${\rm d}\sigma_{\chi T}/{\rm d}E_r$ is computed from Eq.~(\ref{eq:dsigma}) and the identity $q^2=2 m_T E_r$, whereas the energy integration in Eq.~(\ref{eq:omega}) is performed over all kinematically allowed recoil energies $E_r$.

The differential capture rate per unit volume is then obtained from Eq.~(\ref{eq:omega}) through a velocity integral~\cite{Gould:1987ir}:
\begin{equation}\label{dCdV}
    \frac{{\rm d} C}{{\rm d}  V} = \int_0^\infty {\rm d}u \,\frac{f(u)}{u}\,w\Omega_v^{-}(w) \,,
\end{equation}
where $f(u)$ is the WIMP speed distribution at infinity boosted in the Earth's rest frame.~In all numerical applications, I assume a Maxwell-Boltzmann speed distribution truncated at the escape velocity 533~km~s$^{-1}$, a Local Standard of Rest velocity of 220~km~s$^{-1}$, and a local dark matter density of 0.4~GeV~cm$^{-3}$~\cite{Catena:2009mf,Catena:2011kv,Bozorgnia:2013pua}.~The total rate of WIMP capture by the Earth is finally given by
\begin{equation}
\label{C}
    C =\int_0^{R_\oplus} {\rm d} r\,4\pi r^2 \frac{{\rm d} C}{{\rm d} V} \,,
\end{equation}
where spherical symmetry is assumed in the volume integral, and $R_\oplus$ is the radius of the Earth.

For WIMPs gravitationally bound to the Earth, I assume a thermal radial profile given by
\begin{equation}
\label{eq:thermal}
\epsilon(r) \propto \text{exp}\left[-\frac{m_\chi\phi(r)}{T_c}\right],    
\end{equation}
where $\phi(r)$ is the total gravitational potential at $r$ and $T_c\simeq 5 \times 10^3$~K is the Earth core temperature.~Following~\cite{Gondolo:2004sc}, we model the gravitational potential $\phi(r)$ from the Earth's mass profile and the mass fractions of the most abundant elements in the Earth~\cite{raddens}, namely:~$^{16}$O, $^{23}$Na, $^{24}$Mg, $^{27}$Al, $^{28}$Si, $^{31}$P, $^{32}$S, $^{40}$Ca, $^{52}$Cr, $^{56}$Fe, and $^{58}$Ni.~Regarding the thermalisation assumption in Eq.~(\ref{eq:thermal}), this is expected to be valid for most of the parameter values considered in this study.~However, only detailed numerical calculations can determine the actual distribution and thermalisation time of WIMPs in the Earth.~So far, this calculation has been performed for general dark matter-nucleon interactions only in the case of WIMPs trapped in the Sun~\cite{Vincent:2013lua}.

\subsection{Annihilation}
The average number of WIMP annihilations per unit time in the Earth's core and mantle, $\Gamma_a$, is given by
\begin{equation}
\Gamma_a=\frac{1}{2}\int d^3{\bf x} \,\epsilon^2({\bf x}) \,\langle \sigma_{\rm ann} v_{\rm rel} \rangle \,,
\label{eq:Gamma}
\end{equation}
where $\langle \sigma_{\rm ann} v_{\rm rel} \rangle\simeq3\times10^{-26}$~cm$^3$~s$^{-1}$ is the thermal average of the WIMP annihilation cross-section $\sigma_{\rm ann}$ times relative velocity $v_{\rm rel}$, and ${\bf x}$ is the three-dimensional WIMP position vector.~From Eq.~(\ref{eq:Gamma}), the probability of WIMP pair annihilation per unit time, $C_a$, can be written as:~$C_a=2\Gamma_a/N_\chi^2$, where $N_\chi(t)$ is the time dependent number of WIMPs trapped in the Earth at the time $t$.~The definition of $\Gamma_a$ then leads to the following relation between $\langle \sigma_{\rm ann} v_{\rm rel} \rangle$ and $C_a$
\begin{equation}
C_a = \langle \sigma_{\rm ann} v_{\rm rel} \rangle \frac{V_2}{V_1^2}\,,
\label{eq:CA}
\end{equation}
where $V_1$ and $V_2$ are given by
\begin{equation}
\label{eq:vol}
V_1 = \int d^3{\bf x} \,\frac{\epsilon({\bf x})}{\epsilon_0} \,; \qquad \qquad V_2 = \int d^3{\bf x} \,\frac{\epsilon^2({\bf x})}{\epsilon_0^2}\,,
\end{equation}
$\epsilon_0$ is the WIMP density at he Earth's centre, and $V_j=2.3 \times 10^{25} [j m_\chi/(10~{\rm GeV})]^{-3/2}$~cm$^{3}$, with $j=1,2$~\cite{Gondolo:2004sc}. 

The number of WIMPs trapped in the Earth, $N_\chi$, is found by solving the following differential equation
\begin{equation}
\dot{N}_\chi = C - C_a N_\chi^2 \,,
\label{eq:N}
\end{equation}
the solution of which is given by
\begin{equation}
N_\chi(t) = \sqrt{\frac{C}{C_a}}\tanh\left(\sqrt{C C_a}t\right) \,.
\end{equation}
The above expression implies
\begin{equation}
\label{eq:annrate}
\Gamma_a=\frac{C}{2}\tanh^2\left(\sqrt{C C_a}t\right) \,.
\end{equation}
$\Gamma_a$ has to be evaluated at $t_\oplus=4.5\times 10^9$ years, which is the present age of the Earth.~Since $\sqrt{C C_a}t_\oplus$ turns out to be smaller than one in all numerical applications, the approximation $\tanh^2\left(\sqrt{C C_a}t\right) \simeq 1$, often valid in the case of WIMP capture by the Sun, cannot be made in the Earth WIMP analysis.~Equations analogous to those reviewed above apply to WIMP annihilation in the Sun.

The differential neutrino flux from WIMP annihilation in the Earth (or Sun) depends linearly on $\Gamma_a$~\cite{Jungman:1995df}:
\begin{equation}
    \frac{{\rm d} \Phi_\nu}{{\rm d} E_\nu} =
    \frac{\Gamma_a}{4\pi D^2}\sum_f B^f_\chi \frac{{\rm d} N^f_\nu}{{\rm d} E_\nu}\,.
\label{eq:nuflux}
\end{equation}
In Eq.~(\ref{eq:nuflux}), $D$ is the detector's distance to the Earth's (or Sun's) centre,  $B^f_\chi$ is the branching ratio for WIMP pair annihilation into the final state $f$, ${\rm d} N^f_\nu / {\rm d} E_\nu$ is the neutrino energy spectrum at detector from the decay of Standard Model particles in the final state $f$, and $E_\nu$ is the neutrino energy.

Neutrino telescopes search for an upward muon flux induced by charged-current neutrino interactions in ice or water.~The expected WIMP-induced differential muon flux at detector is given by
\begin{equation}
\frac{{\rm d}\Phi_{\mu}}{{\rm d} E_\mu} = N_T \int_{E_\mu^{\rm th}}^{\infty} {\rm d} E_\nu
\int_0^{\infty} {\rm d} \lambda \int_{E_\mu}^{E_\nu} {\rm d} E_\mu^{\prime} \,\mathcal{P}(E_\mu,E_\mu^{\prime};\lambda)\, \frac{{\rm d}\sigma_{{\rm CC} }(E_\nu,E_\mu^{\prime})}{{\rm d} E_{\mu}^{\prime}}  \,\frac{{\rm d \Phi_\nu}}{{\rm d} E_\nu} \,,
\label{eq:Phi}
\end{equation}
where $E_\mu^{\rm th}$ is the detector energy threshold, $\lambda$ is the muon range, $\mathcal{P}(E_\mu,E_\mu^{\prime};\lambda)$ is the probability for a muon of initial energy $E_{\mu}^{\prime}$ to be detected with a final energy $E_\mu$ after traveling a distance $\lambda$ inside the detector, ${\rm d}\sigma_{{\rm CC}}/{\rm d} E_\mu^{\prime}$ is the weak differential cross-section for production of a muon of energy $E_\mu^{\prime}$, and $N_T$ is the number of nucleons per cubic centimetre.~Here I use Eq.~(\ref{eq:nuflux}) and data from a recent Super-Kamiokande solar WIMP search~\cite{Choi:2015ara} to derive limits on the coupling constants in Eq.~(\ref{eq:H_chiT}).~I evaluate Eq.~(\ref{eq:nuflux}) using neutrino yields generated by {\sffamily WimpSim}~\cite{Blennow:2007tw}, and tabulated in {\sffamily darksusy}~\cite{Gondolo:2004sc}.

\section{Results}
\label{sec:results}
In this section I calculate the rate of WIMP capture in the Earth, Eq.~(\ref{C}), for all operators in Tab.~\ref{tab:operators}.~I perform this calculation assuming Earth composition and WIMP speed distribution introduced in Sec.~\ref{sec:capture}.~The nuclear response functions needed for this calculation are computed in~\cite{Catena:2015uha} for $^{16}$O, $^{23}$Na, $^{24}$Mg, $^{27}$Al, $^{28}$Si, $^{32}$S, $^{40}$Ca, $^{56}$Fe, and $^{58}$Ni and in this paper for $^{31}$P and $^{52}$Cr.~I consider one operator at the time, neglecting operator interference patterns.~Interference effects were extensively discussed in the context of dark matter direct detection in~\cite{Catena:2015uua}.~I then use the capture rates derived here in the effective theory of dark matter-nucleon interactions to evaluate the rate of WIMP annihilation in the Earth, Eq.~(\ref{eq:annrate}).~Comparing this prediction with the results recently obtained in a WIMP analysis of IceCube in the 86-string configuration~\cite{Aartsen:2016fep}, I derive upper limits on the isoscalar and isovector coupling constants of the effective theory in Sec.~\ref{sec:scattering} as a function of the dark matter particle mass.

\begin{figure}[t]
\begin{center}
\begin{minipage}[t]{0.49\linewidth}
\centering
\includegraphics[width=\textwidth]{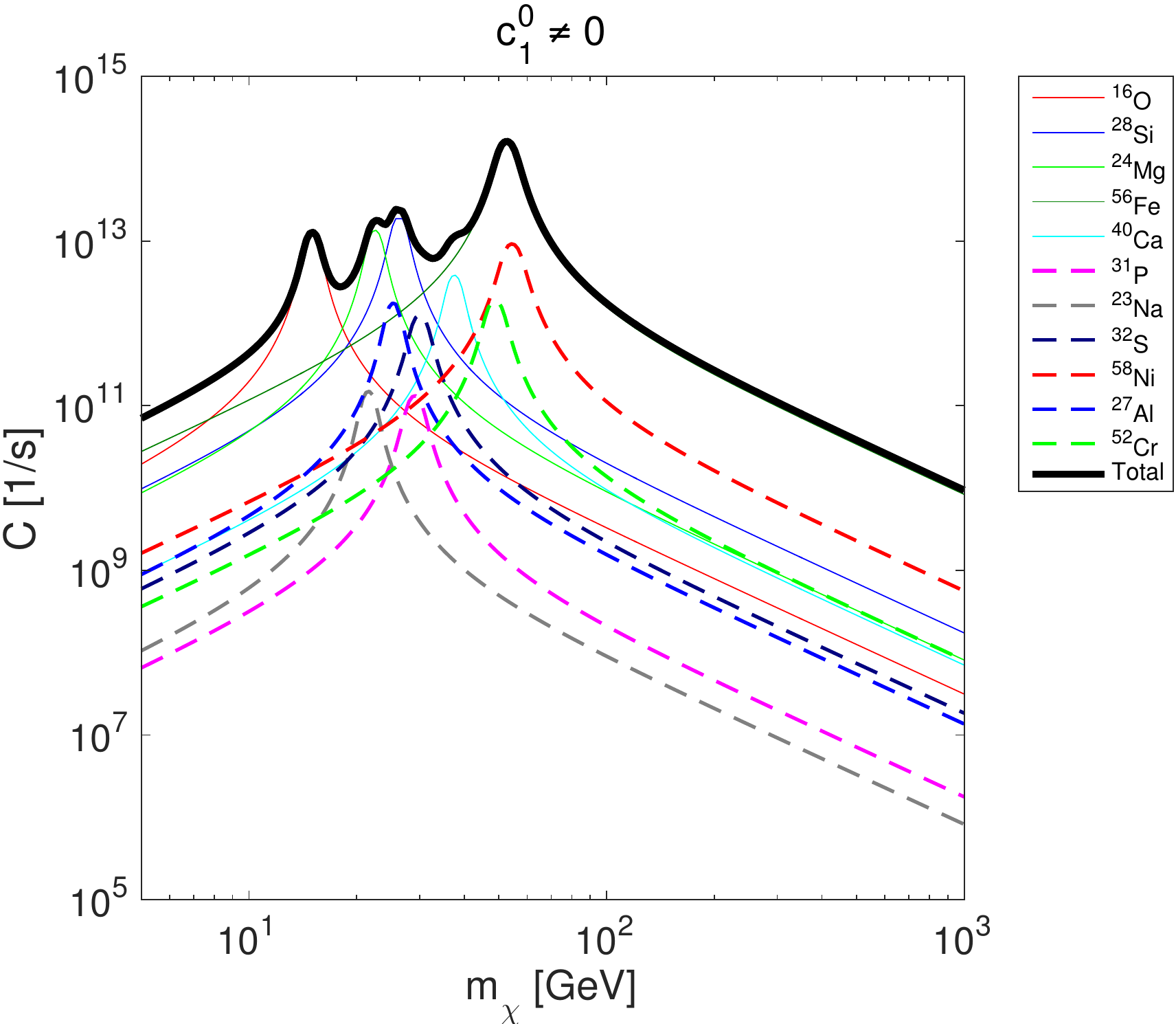}
\end{minipage}
\begin{minipage}[t]{0.49\linewidth}
\centering
\includegraphics[width=\textwidth]{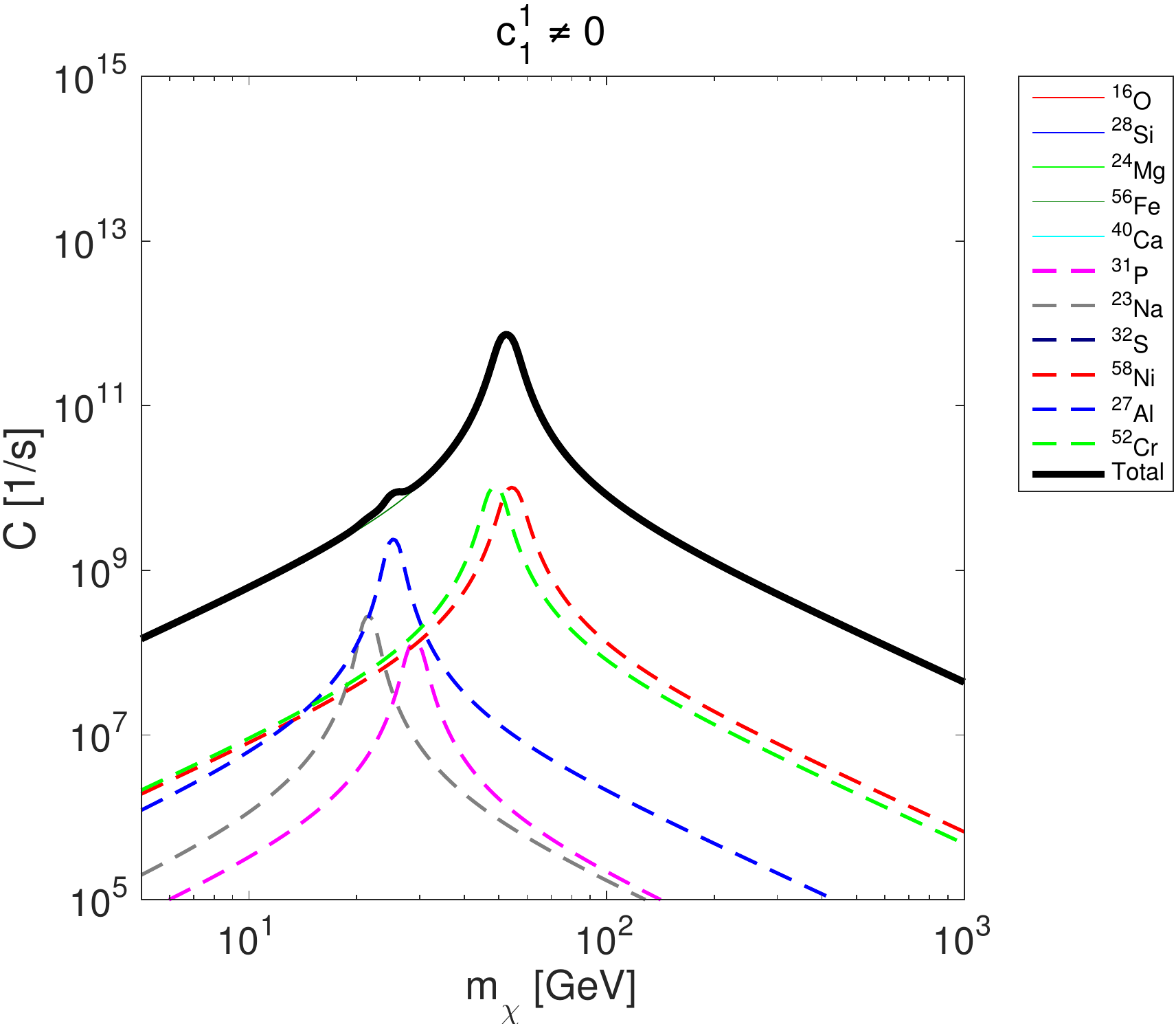}
\end{minipage}
\begin{minipage}[t]{0.49\linewidth}
\centering
\includegraphics[width=\textwidth]{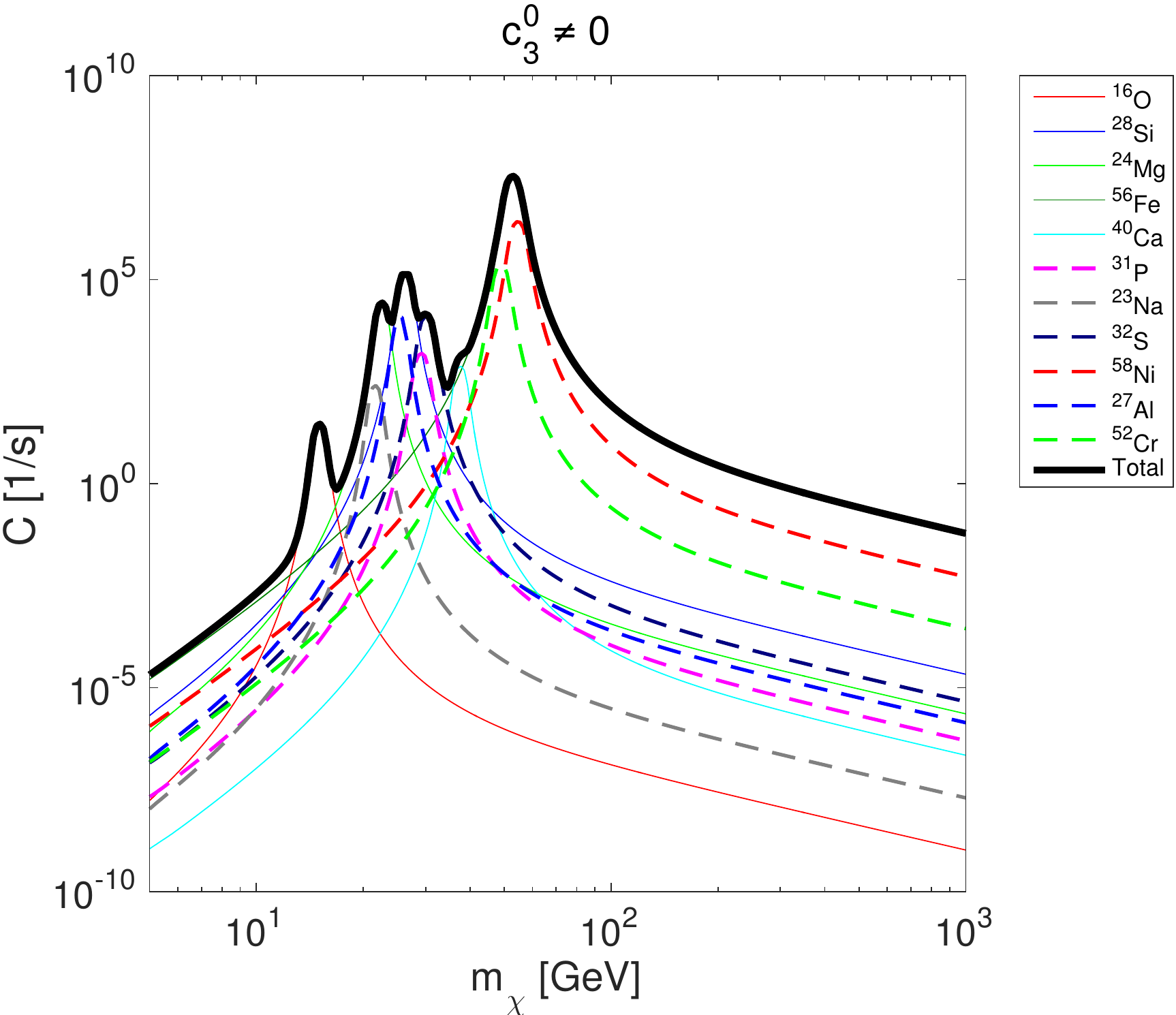}
\end{minipage}
\begin{minipage}[t]{0.49\linewidth}
\centering
\includegraphics[width=\textwidth]{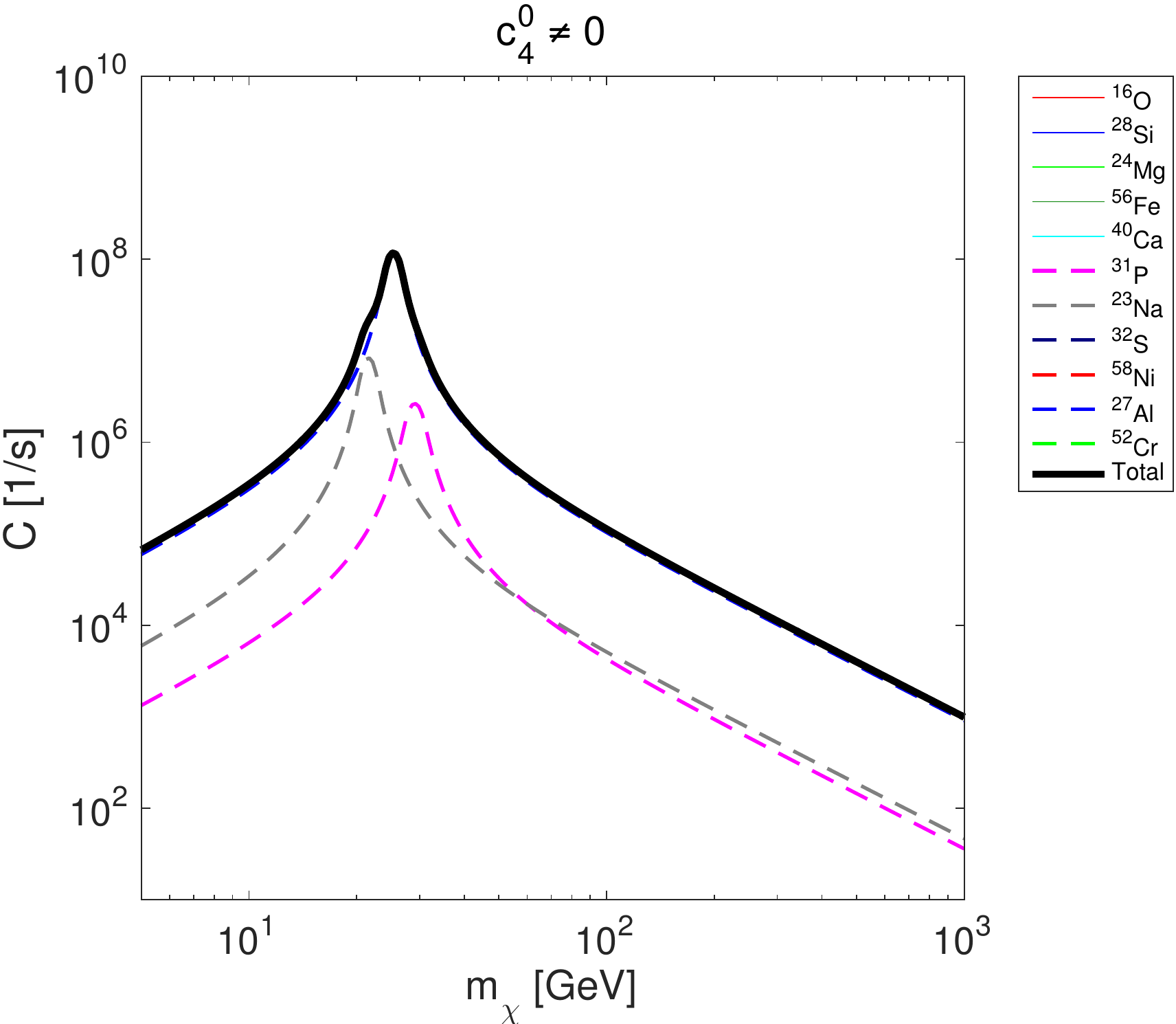}
\end{minipage}
\end{center}
\caption{Rate of WIMP capture by the Earth as a function of the dark matter particle mass $m_{\chi}$ for the interaction operators $\hat{\mathcal{O}}_1 = \mathbb{1}_{\chi N}$ (isoscalar and isovector component), $\hat{\mathcal{O}}_3 = i{\bf{\hat{S}}}_N\cdot({\bf{\hat{q}}}/m_N\times{\bf{\hat{v}}}^{\perp})$ (isoscalar only) and $\hat{\mathcal{O}}_4 = {\bf{\hat{S}}}_{\chi}\cdot {\bf{\hat{S}}}_{N}$ (isoscalar only).~Black solid lines correspond to the total capture rate, whereas coloured lines refer to the contribution to the total capture rate from single elements in the Earth.}  
\label{fig:C4op1}
\end{figure}
\begin{figure}[t]
\begin{center}
\begin{minipage}[t]{0.49\linewidth}
\centering
\includegraphics[width=\textwidth]{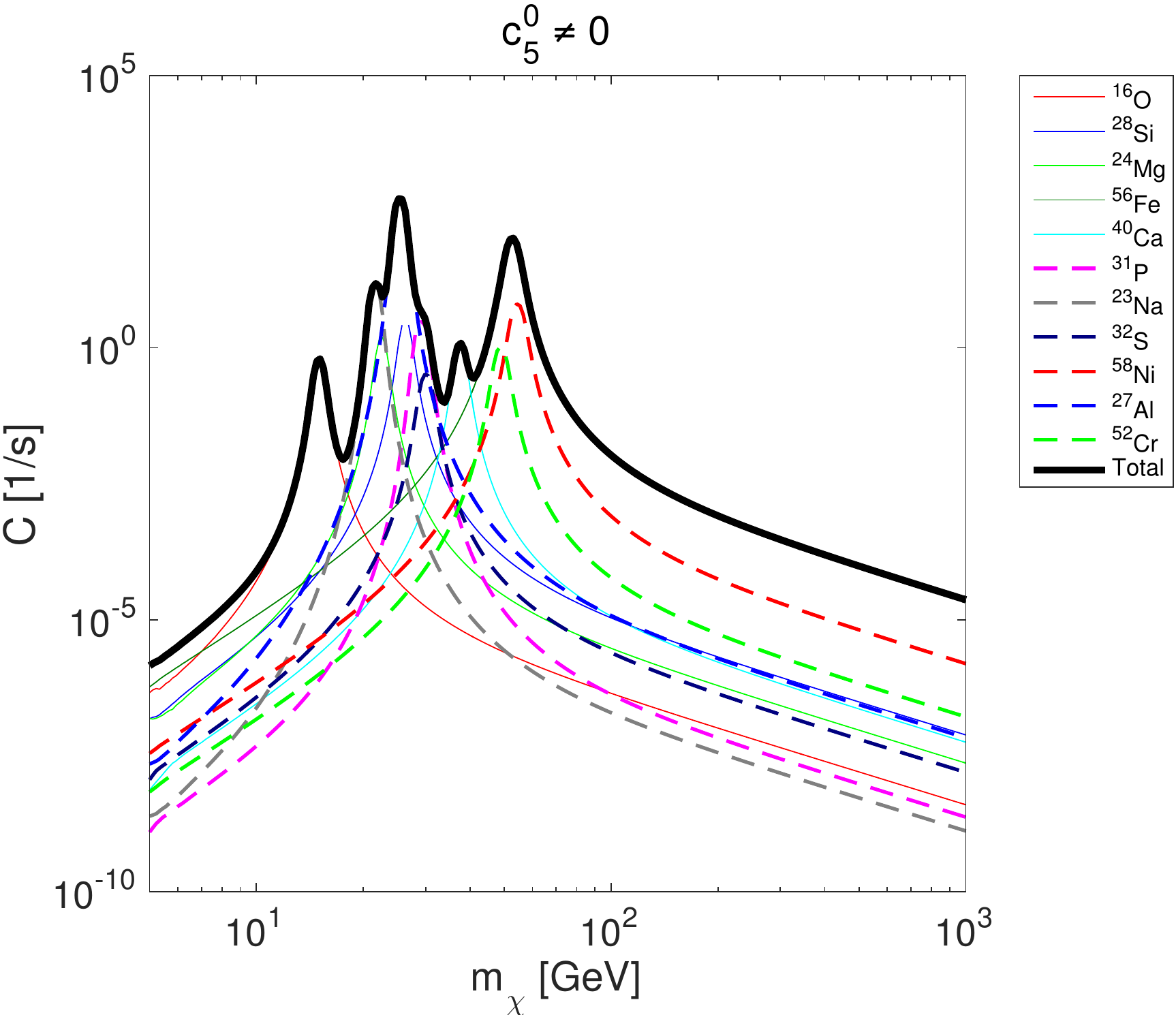}
\end{minipage}
\begin{minipage}[t]{0.49\linewidth}
\centering
\includegraphics[width=\textwidth]{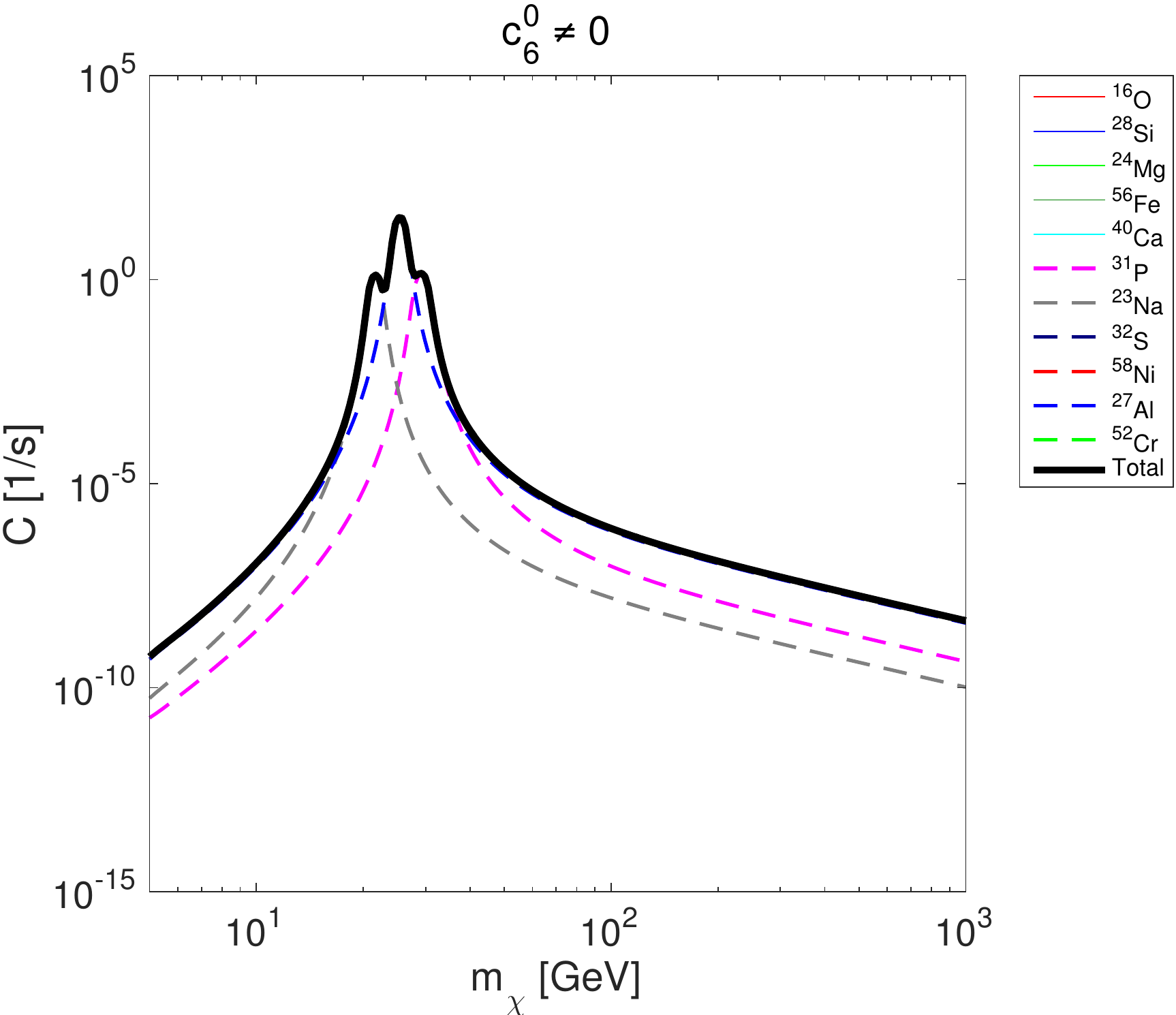}
\end{minipage}
\begin{minipage}[t]{0.49\linewidth}
\centering
\includegraphics[width=\textwidth]{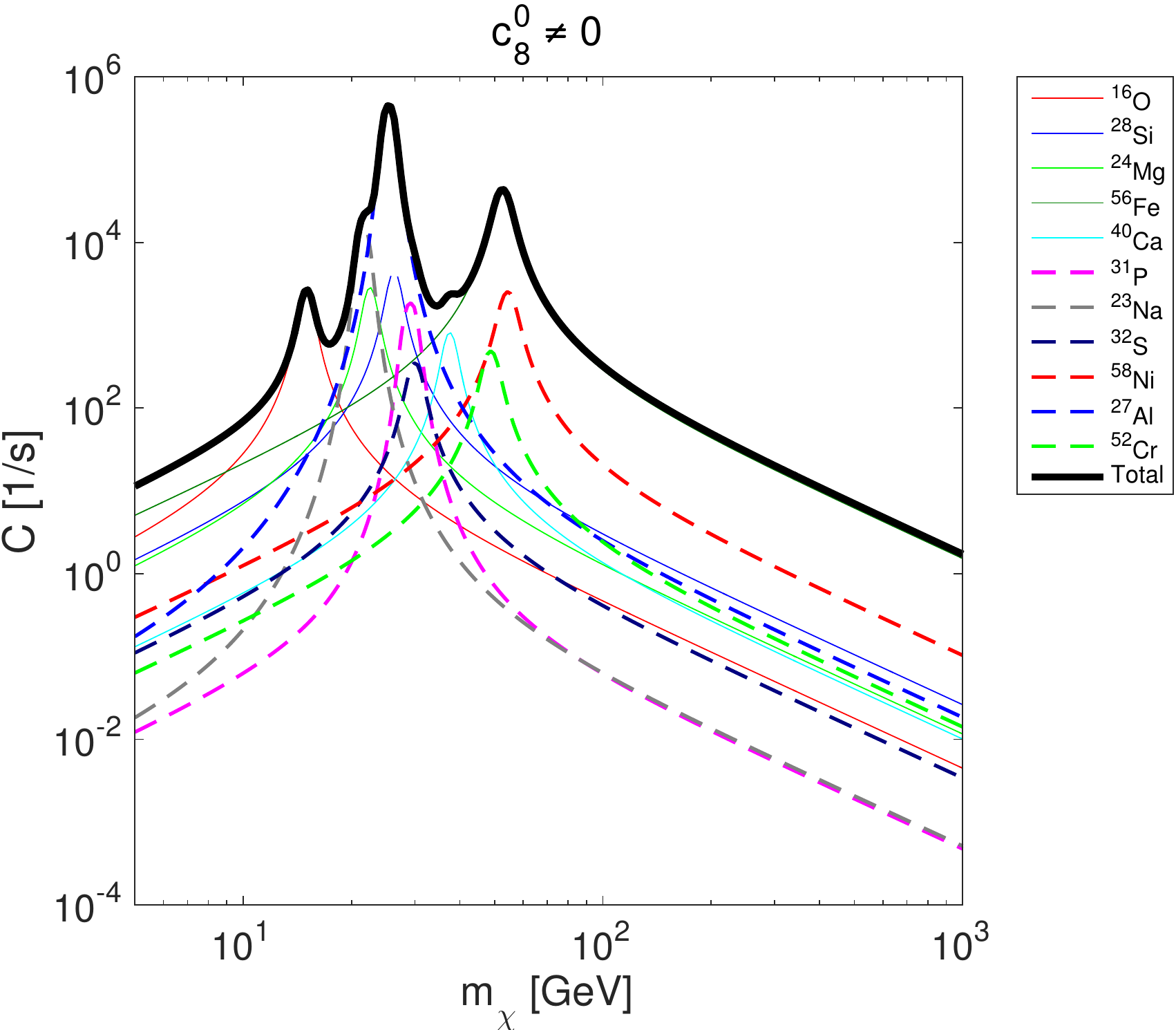}
\end{minipage}
\begin{minipage}[t]{0.49\linewidth}
\centering
\includegraphics[width=\textwidth]{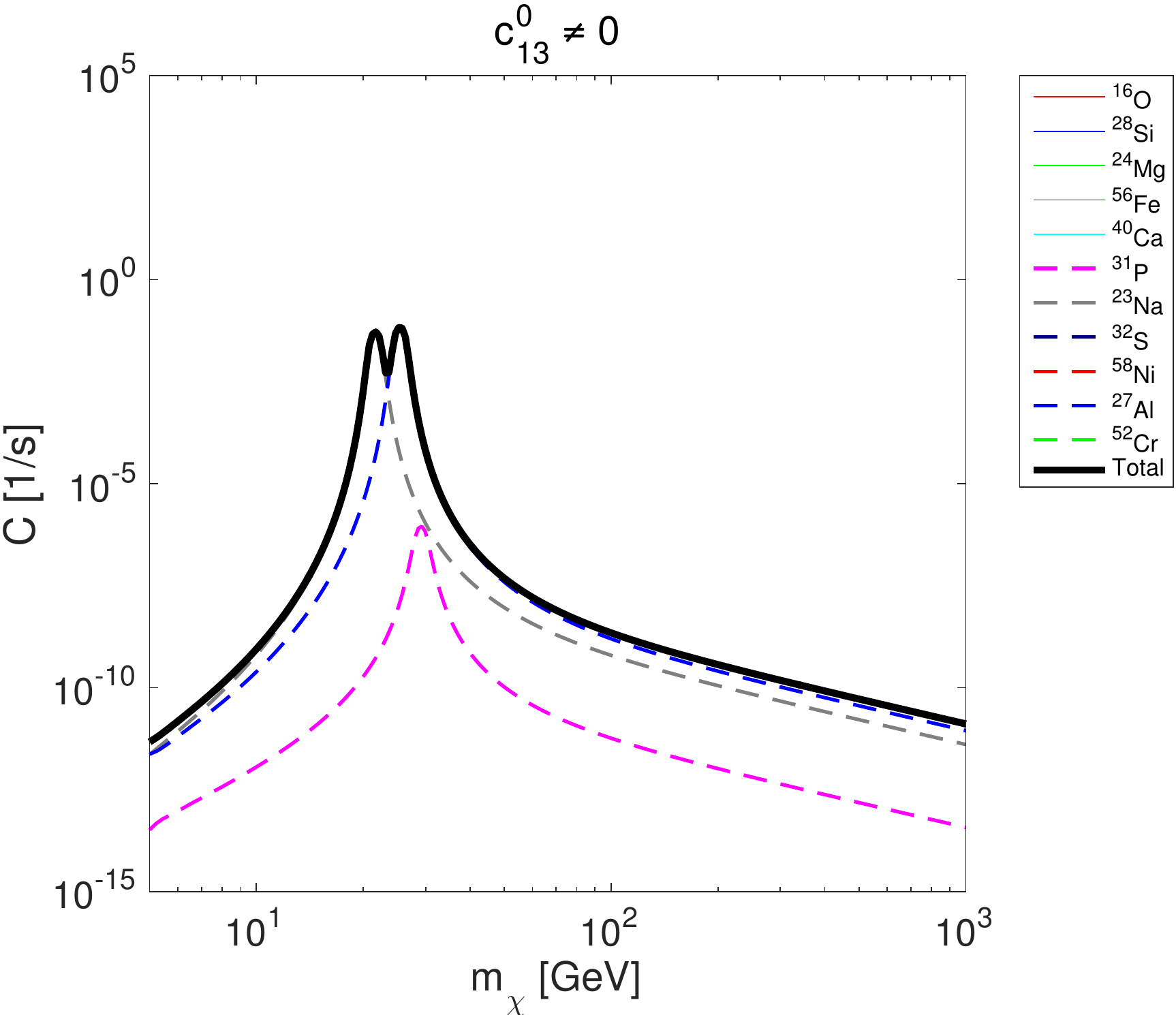}
\end{minipage}
\end{center}
\caption{Same as for Fig.~\ref{fig:C4op1}, but now for the isoscalar component of the interaction operators $\hat{\mathcal{O}}_5 = i{\bf{\hat{S}}}_\chi\cdot({\bf{\hat{q}}}/m_N\times{\bf{\hat{v}}}^{\perp})$, $\hat{\mathcal{O}}_6 = ({\bf{\hat{S}}}_\chi\cdot {\bf{\hat{q}}}/m_N) ({\bf{\hat{S}}}_N\cdot\hat{{\bf{q}}}/m_N)$, $\hat{\mathcal{O}}_8 = {\bf{\hat{S}}}_{\chi}\cdot {\bf{\hat{v}}}^{\perp}$, and $\hat{\mathcal{O}}_{13}=i ({\bf{\hat{S}}}_{\chi}\cdot {\bf{\hat{v}}}^{\perp})({\bf{\hat{S}}}_{N}\cdot {\bf{\hat{q}}}/m_N)$. Capture rates for the remaining operators are in Appendix~\ref{app:figures}.}
\label{fig:C4op2}
\end{figure}
\begin{figure}[t]
\begin{center}
\begin{minipage}[t]{0.49\linewidth}
\centering
\includegraphics[width=\textwidth]{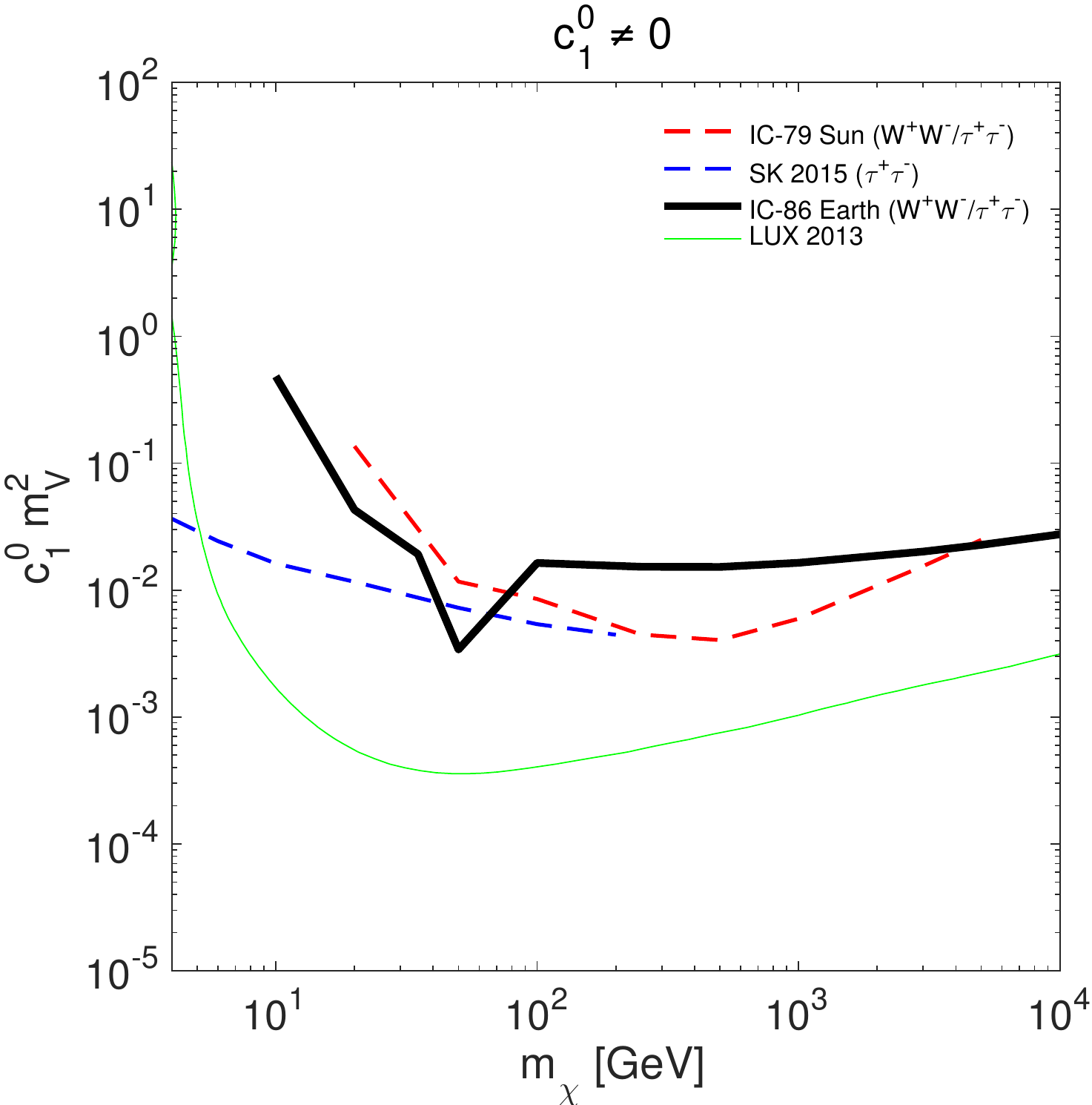}
\end{minipage}
\begin{minipage}[t]{0.49\linewidth}
\centering
\includegraphics[width=\textwidth]{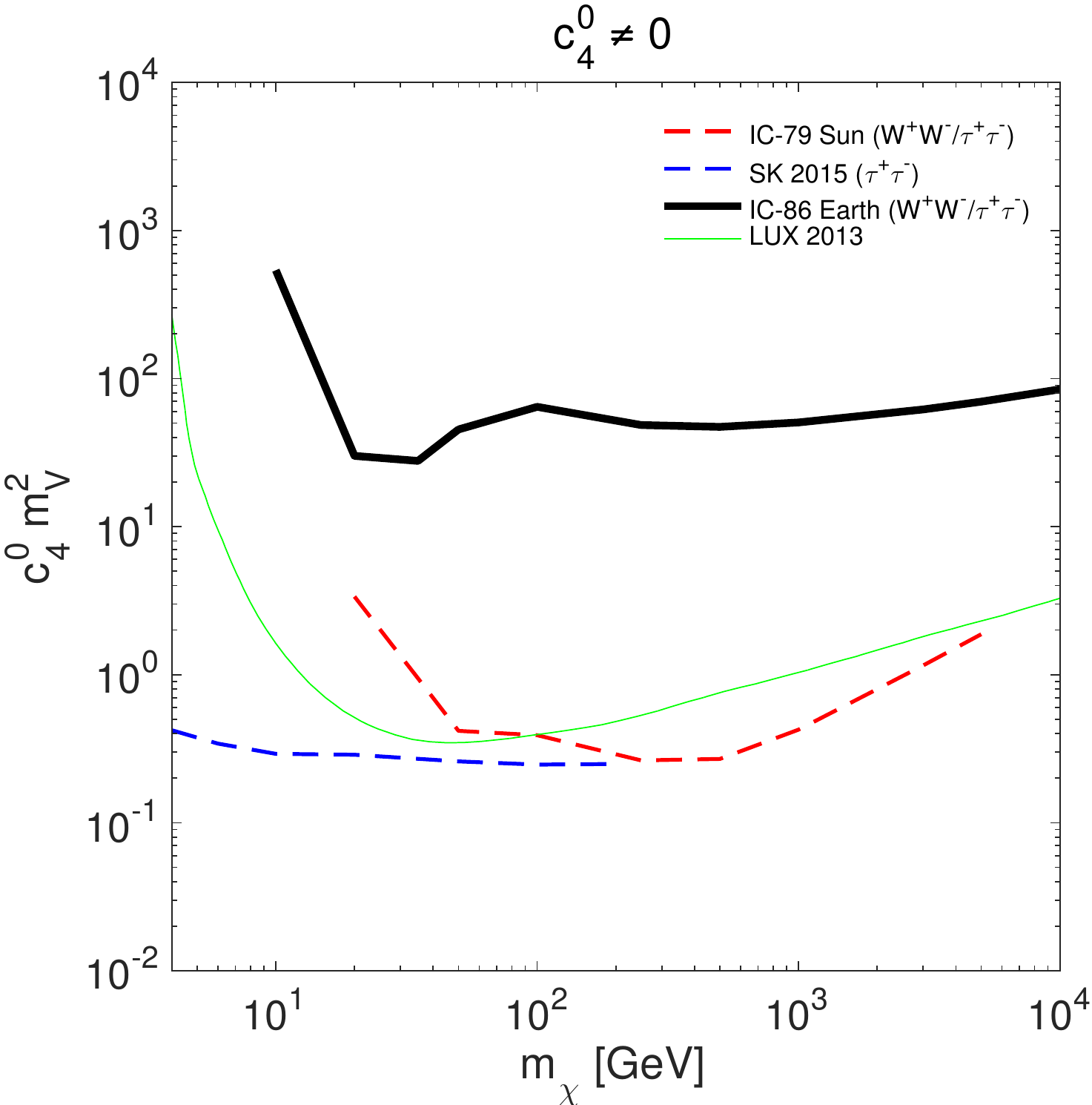}
\end{minipage}
\end{center}
\caption{Exclusion limits in the coupling constant - WIMP mass plane for the operators $\hat{\mathcal{O}}_1$ and $\hat{\mathcal{O}}_4$ (isoscalar components).~Black solid lines are the 90\% CL upper limits on $c_1^0$ and $c_4^0$ found in this work using bounds on the annihilation rate $\Gamma_a$ from a recent WIMP analysis of IceCube in the 86-string configuration.~WIMPs are assumed to annihilate into $W^+ W^-$ for $m_\chi$ larger than the $W$ boson rest mass, and into $\tau^+ \tau^-$ otherwise.~Limits from Super-Kamiokande (SK), IceCube solar WIMP searches, and LUX (2013) are superimposed for comparison.}
\label{fig:limits1}
\end{figure}
\begin{figure}[t]
\begin{center}
\begin{minipage}[t]{0.49\linewidth}
\centering
\includegraphics[width=\textwidth]{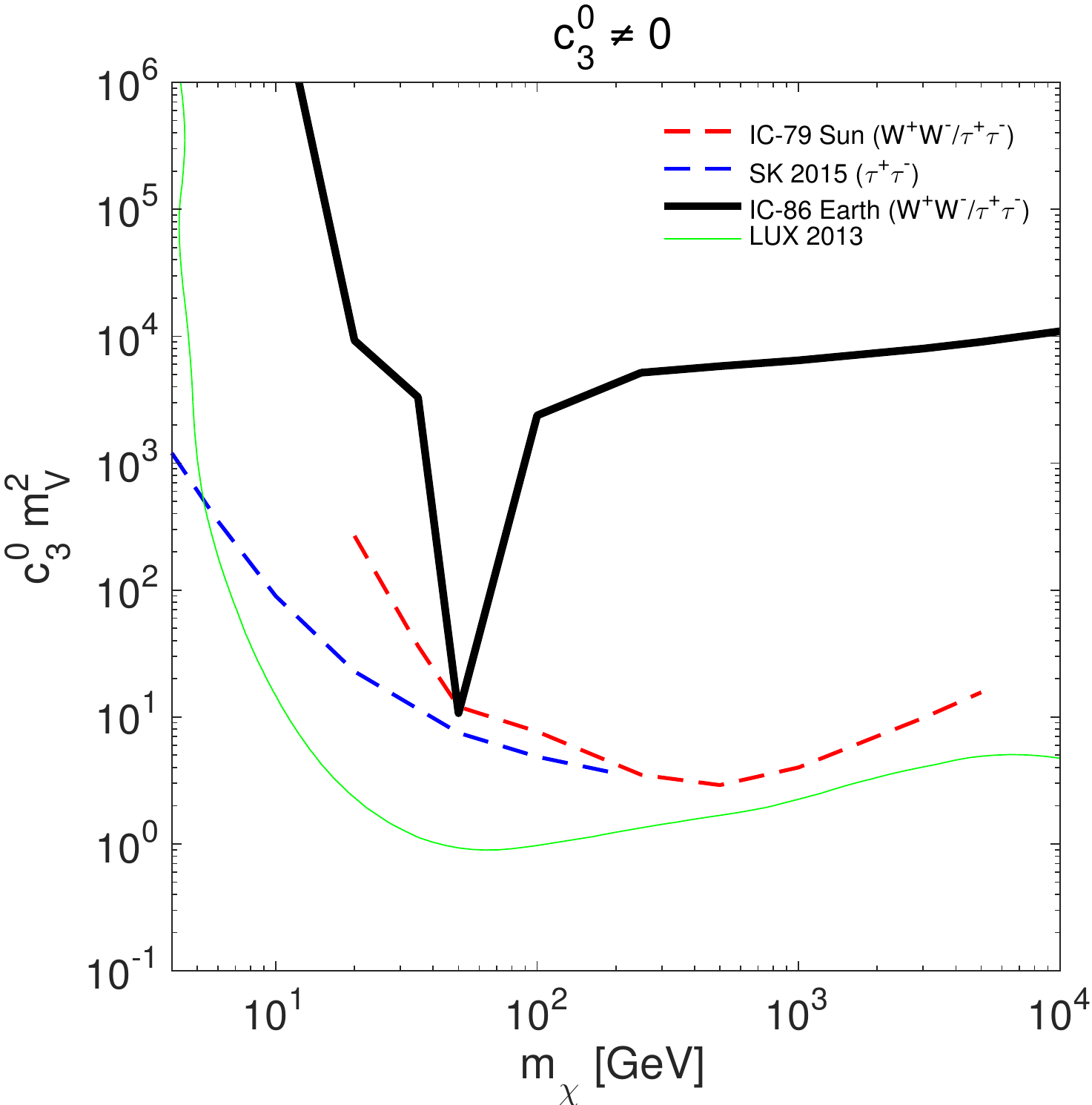}
\end{minipage}
\begin{minipage}[t]{0.49\linewidth}
\centering
\includegraphics[width=\textwidth]{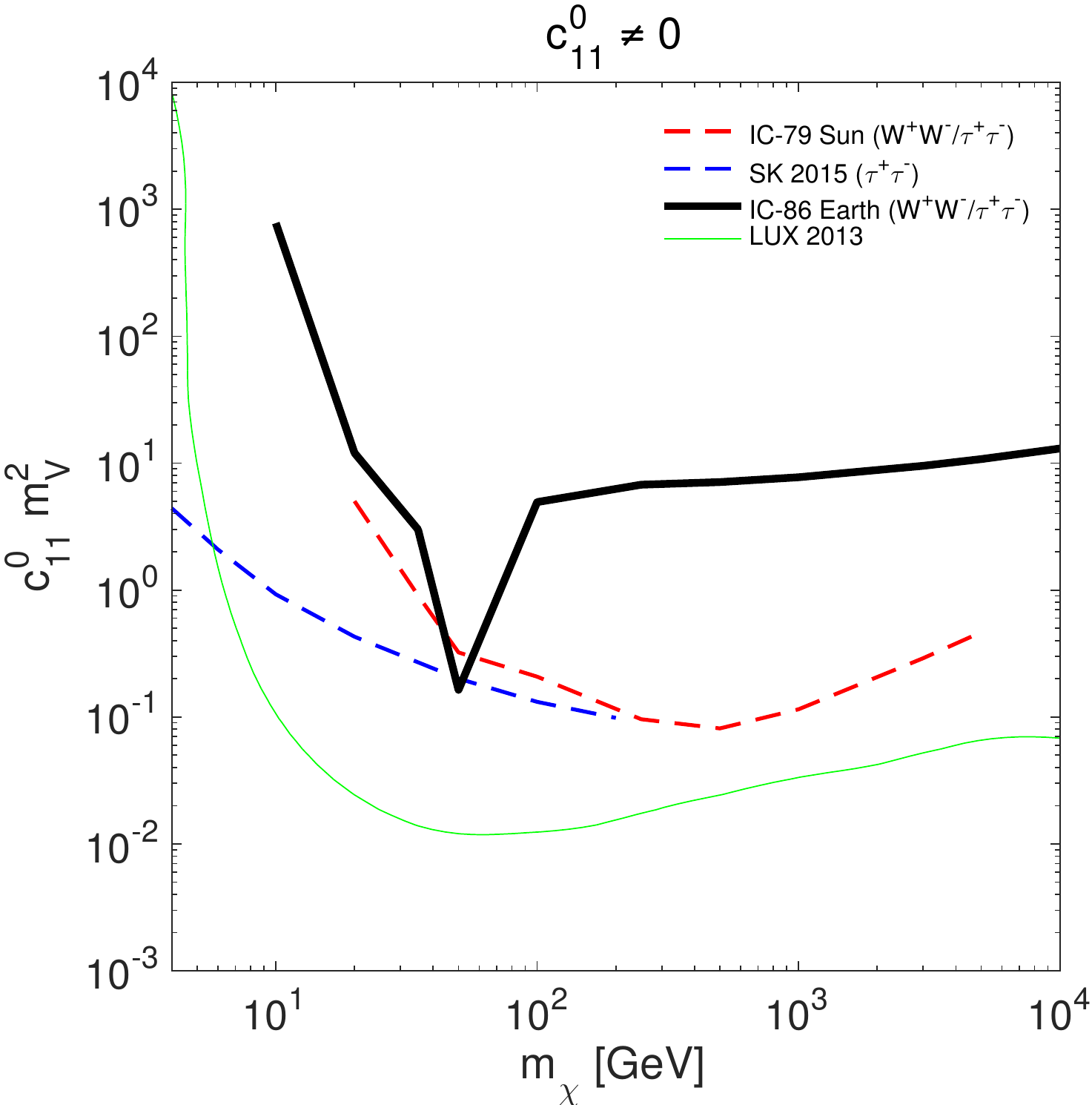}
\end{minipage}
\begin{minipage}[t]{0.49\linewidth}
\centering
\includegraphics[width=\textwidth]{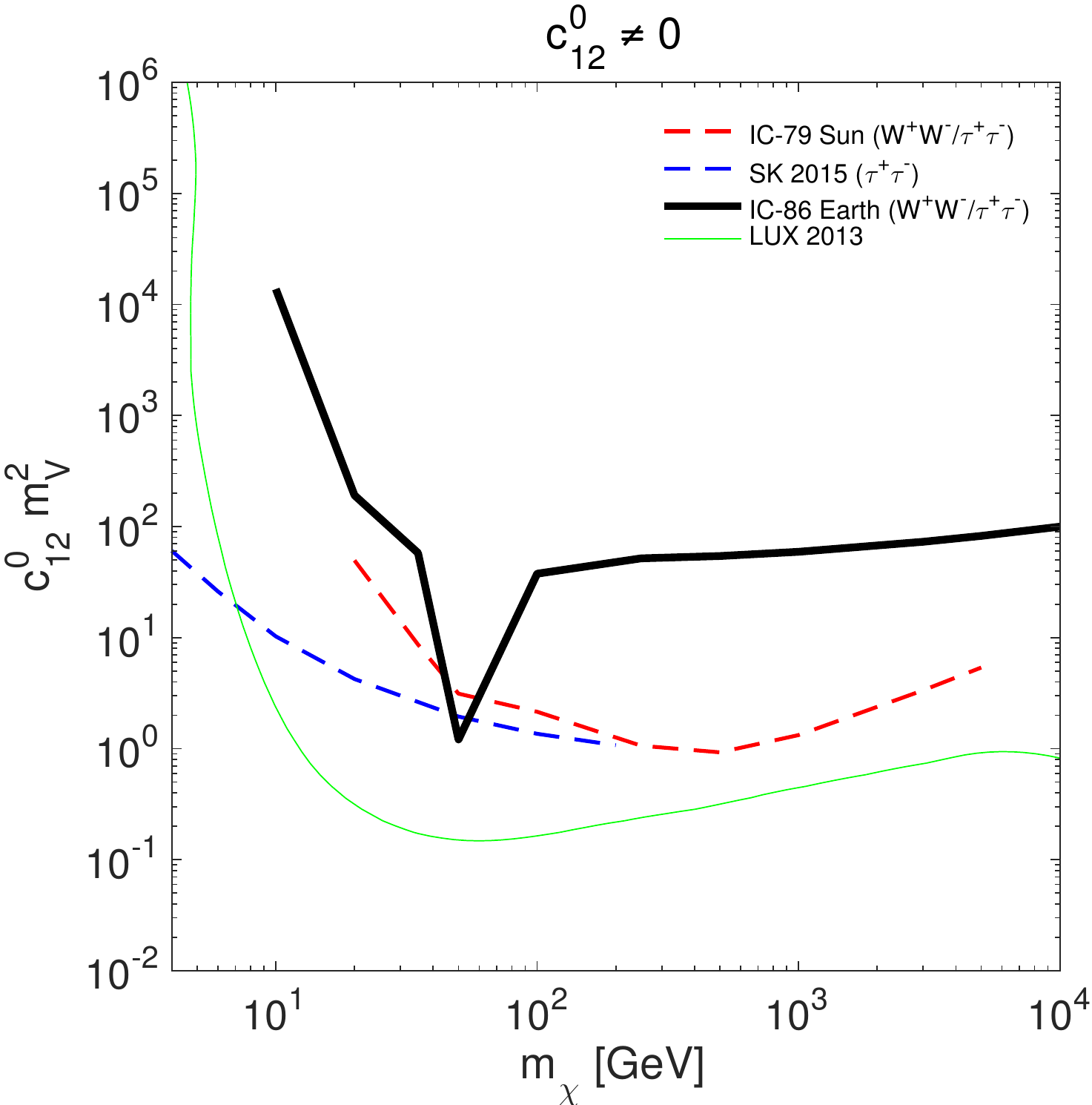}
\end{minipage}
\begin{minipage}[t]{0.49\linewidth}
\centering
\includegraphics[width=\textwidth]{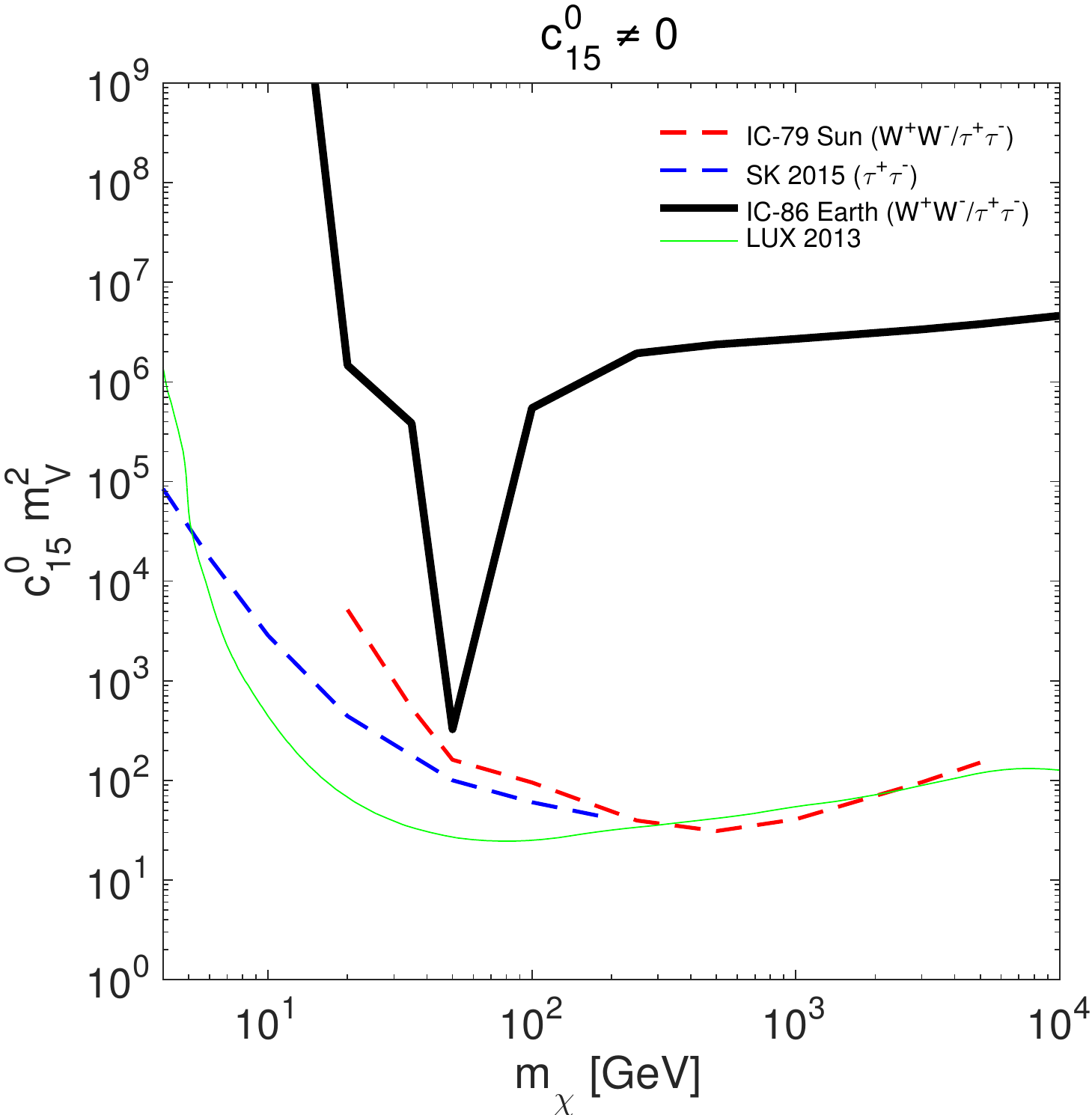}
\end{minipage}
\end{center}
\caption{Same as for Fig.~\ref{fig:limits1} but now for selected interaction operators.~Limits for the remaining operators are in Appendix~\ref{app:figures}.}
\label{fig:limits2}
\end{figure}

\subsection{Capture rate and resonances}
\label{sec:cres}
In this subsection I focus on the rate of WIMP capture by the Earth in the effective theory of Sec.~\ref{sec:scattering}.~Exclusion limits on the coupling constants appearing in Eq.~(\ref{eq:H_chiT}) will be presented in the next subsection.

\noindent In the calculation of the rate of WIMP capture by the Earth, for each interaction operator $\hat{\mathcal{O}}_j$, $j=1,3,\dots,15$, in Tab.~\ref{tab:operators}, I consider two scenarios:~a first one where $c_j^0=10^{-3}/m_V^2$ and $c_j^1=0$; a second one in which $c_j^1=10^{-3}/m_V^2$ and $c_j^0=0$.~Here $m_V=246.2$~GeV is the electroweak scale.~The value $10^{-3}/m_V^2$ is arbitrary and corresponds to the reference cross-section $ (\mu^2_{\chi N}/m_V^4)/(4\pi) \sim 7\times10^{-45}$~cm$^2$, where $\mu_{\chi N}$ is the WIMP-nucleon reduced mass.~Results can trivially be rescaled to other values, since capture rates depend quadratically on the coupling constants.

Fig.~\ref{fig:C4op1} and Fig.~\ref{fig:C4op2} show the rate of WIMP capture in the Earth for selected interaction operators.~Specifically, I consider:~$\hat{\mathcal{O}}_1 = \mathbb{1}_{\chi N}$, $\hat{\mathcal{O}}_3 = i{\bf{\hat{S}}}_N\cdot({\bf{\hat{q}}}/m_N\times{\bf{\hat{v}}}^{\perp})$, $\hat{\mathcal{O}}_4 = {\bf{\hat{S}}}_{\chi}\cdot {\bf{\hat{S}}}_{N}$, $\hat{\mathcal{O}}_5 = i{\bf{\hat{S}}}_\chi\cdot({\bf{\hat{q}}}/m_N\times{\bf{\hat{v}}}^{\perp})$, $\hat{\mathcal{O}}_6 = ({\bf{\hat{S}}}_\chi\cdot {\bf{\hat{q}}}/m_N) ({\bf{\hat{S}}}_N\cdot\hat{{\bf{q}}}/m_N)$, $\hat{\mathcal{O}}_8 = {\bf{\hat{S}}}_{\chi}\cdot {\bf{\hat{v}}}^{\perp}$, and, finally, $\hat{\mathcal{O}}_{13}=i ({\bf{\hat{S}}}_{\chi}\cdot {\bf{\hat{v}}}^{\perp})({\bf{\hat{S}}}_{N}\cdot {\bf{\hat{q}}}/m_N)$.~Results are presented in the ($C,m_\chi$) plane.~I focus on the isoscalar component of all operators and on the isovector component of the $\hat{\mathcal{O}}_1$ operator.~$\hat{\mathcal{O}}_1$ is an important benchmark, since it generates the standard spin-idenpendent WIMP-nucleon scattering cross-section.~Results for all remaining interaction operators are shown in Appendix~\ref{app:figures}.

The peaks in Fig.~\ref{fig:C4op1} and Fig.~\ref{fig:C4op2} are expected, since the small local escape velocity $v(r)$ in Eq.~(\ref{C}) allows for resonant WIMP scattering and capture in the Earth.~Indeed, in the limit $m_\chi \rightarrow m_T$, where $m_T$ is the mass of an element in the Earth, the WIMP total energy, $E=m_\chi w^2/2$, can entirely be transferred in the scattering.~As a result, the Heaviside step function in Eq.~(\ref{C}) is identically one, and the WIMP capture rate develops a local maximum at $m_\chi \rightarrow m_T$.~In this limit, WIMP scattering and capture in the Earth's core and mantle become resonant.~The exact position of the peaks in Figs.~\ref{fig:C4op1} and \ref{fig:C4op2} depends on the Earth's composition.~Including elements not considered here, additional resonances might appear in the mass - capture rate plane.~This is shown in~\cite{visinelli} for the isoscalar component of the $\hat{\mathcal{O}}_4$ operator using Helm form factors. 

The rate of WIMP capture in the Earth and the resonances characterising the isoscalar component of the operator $\hat{\mathcal{O}}_1$ in the top-left panel of Fig.~\ref{fig:C4op1} have extensively been studied in the literature~\cite{Gould:1987ir,Lundberg:2004dn}.~For all other operators, here I present the first detailed calculation of the rate of WIMP capture in the Earth.~The results that I report in Figs.~\ref{fig:C4op1} and \ref{fig:C4op2}, together with the figures in Appendix~\ref{app:figures}, fully characterise the resonant structure of the WIMP capture in the Earth in the general non-relativistic effective theory of dark matter-nucleon interactions~\cite{Fitzpatrick:2012ix,Anand:2013yka}.

I find significantly different resonant patterns in the panels of Figs.~\ref{fig:C4op1} and \ref{fig:C4op2}.~Such differences arise since not all response functions $W_k^{\tau\tau'}$ can be generated in the WIMP scattering by nuclei for a given operator (see~\cite{Catena:2015uha} and Appendix~\ref{sec:nuc}).~For the $\hat{\mathcal{O}}_3$ operator, for instance, the visible resonances are those associated with $^{56}$Fe, $^{40}$Ca (partially), $^{32}$S, $^{28}$Si, $^{24}$Mg, and $^{16}$O.~This pattern is similar to the one found for the isoscalar component of the $\hat{\mathcal{O}}_1$ operator, although the relative size of the peaks is different in the two cases, and $^{56}$Fe is by far the most important element in the WIMP capture for the $\hat{\mathcal{O}}_3$ interaction.~For other operators however, like $\hat{\mathcal{O}}_{13}$, $\hat{\mathcal{O}}_6$ or $\hat{\mathcal{O}}_4$, less peaks are visible.~At the same time, $^{56}$Fe is not the most important element in the capture process.~For these operators the most important elements are $^{27}$Al and $^{23}$Na (or $^{31}$P), producing peaks of comparable high in the ($C,m_\chi$) plane.~The most important element in the WIMP capture is determined by a tradeoff between nuclear abundance in the Earth, powers of momentum transfer in the WIMP-nucleon interaction, and strength of the associated nuclear response.~Similar conclusions were found in~\cite{Catena:2015iea} in the case of WIMP capture by the Sun.

\subsection{Exclusion limits}
\label{sec:lres}
Here I derive 90\% CL upper limits on the strength of the interaction operators in Tab.~\ref{tab:operators} using bounds on the annihilation rate $\Gamma_a$ from a recent Earth WIMP analysis of IceCube in the 86-string configuration~\cite{Aartsen:2016fep}.~For 11 WIMP masses, I use the 90\% CL upper limits on $\Gamma_a$ given in Tab.~2 of~\cite{Aartsen:2016fep}.~For each operator, I compute $\Gamma_a$, Eq.~(\ref{eq:annrate}), using the WIMP capture rates of Sec.~\ref{sec:cres}.

Fig.~\ref{fig:limits1} shows the 90\% CL upper limits on the coupling constants $c_1^0$ and $c_4^0$ that I find in this work from the recent IceCube Earth WIMP search in~\cite{Aartsen:2016fep}.~In the figures, limits are presented as a function of the dark matter particle mass.~For comparison, each panel in Fig.~\ref{fig:limits1} also shows:~limits from data collected in a solar WIMP search at IceCube in the 79-string configuration and interpreted in~\cite{Catena:2015iea}; limits from data collected at Super-Kamiokande~\cite{Choi:2015ara} and interpreted in this work (as explained below); exclusion limits from LUX 2013 data derived in~\cite{Catena:2015iea}.~Limits from Super-Kamiokande have been derived by imposing that the neutrino flux from WIMP annihilation in the Sun, Eq.~(\ref{eq:nuflux}), is less than its 90\% CL upper limit~\cite{Choi:2015ara} for 7 WIMP masses (see Tab.~1 in~\cite{Choi:2015ara}).~Under the assumption $c_1^p=c_1^n$, the upper limit in the left panel of Fig.~\ref{fig:limits1} can be translated into a limit on the spin-independent WIMP-nucleon scattering cross-section, $\sigma_{\rm SI}=(\mu_{\chi N}^2/\pi)|c_1^0|^2/4$.~Notably, at the Iron resonance, that is for $m_\chi\sim 50$~GeV, limits on $c_1^0$ from present Earth WIMP searches at IceCube are stronger than the limits on $c_1^0$ from the search for solar WIMPs at neutrino telescopes.~The limits on $c_4^0$ found here are significantly weaker than those from solar WIMP searches, since the only spin-dependent elements in the Earth (considered here) are $^{31}$P, $^{27}$Al, and $^{23}$Na, which have relatively low mass fractions, ranging from 0 to 0.002 in the core, and from 0.00009 to 0.0235 in the mantle~\cite{raddens}.

Fig.~\ref{fig:limits2} shows the upper limits that I find for the isoscalar coupling constants of the interaction operators $\hat{\mathcal{O}}_3 = i{\bf{\hat{S}}}_N\cdot({\bf{\hat{q}}}/m_N\times{\bf{\hat{v}}}^{\perp})$, $\hat{\mathcal{O}}_{11} = i{\bf{\hat{S}}}_\chi\cdot{\bf{\hat{q}}}/m_N$,  $\hat{\mathcal{O}}_{12} = {\bf{\hat{S}}}_{\chi}\cdot({\bf{\hat{S}}}_{N} \times{\bf{\hat{v}}}^{\perp} )$ and $\hat{\mathcal{O}}_{15} = -({\bf{\hat{S}}}_{\chi}\cdot {\bf{\hat{q}}}/m_N)[ ({\bf{\hat{S}}}_{N}\times {\bf{\hat{v}}}^{\perp} ) \cdot {\bf{\hat{q}}}/m_N] $.~For all operators in Fig.~\ref{fig:limits2}, in the $m_\chi~\sim 50$~GeV region, my limits from data collected in a WIMP analysis of IceCube in the 86-string configuration are comparable or stronger than those I obtain in this work from Super-Kamiokande data, and in~\cite{Catena:2015iea} from a solar WIMP search at IceCube in the 79-string configuration.~For the operators in Fig.~\ref{fig:limits2}, the leading nuclear response operators are $M_{LM;0}$ or $\Phi''_{LM;0}$, Eq.~(\ref{eq:resop}).~In the small momentum transfer limit, the former measures the number of nucleons in the nucleus, and is large for Iron.~Limits for the remaining operators and coupling constants are reported in Appendix~\ref{app:figures}.

\section{Conclusions}
\label{sec:conclusions}
I have studied the capture and annihilation of WIMP dark matter in the Earth in the effective theory of dark matter-nucleon interactions.~It is the first time that the neutrino signal from WIMP annihilation in the Earth's interior is investigated in this general theoretical framework.

Computing the rate of WIMP capture in the Earth I have used nuclear response functions derived through numerical shell model calculations partly in~\cite{Catena:2015uha}, and partly in this work (i.e.~for $^{31}$P and $^{52}$Cr).~For all operators and coupling constants in the effective theory, I have computed the position and shape of the predicted resonances in the corresponding WIMP mass - capture rate plane.~I have found that Iron is not the most important element in the capture process for many interaction operators.~The number of resonances, and their relative high also drastically depend on the interaction operator in analysis.~A variety of factors are relevant in this calculation, ranging from the dependence on the momentum transfer of the WIMP-nucleon interaction to the Earth's composition and associated nuclear physics inputs.

Next, I have calculated the rate of WIMP annihilation in the Earth in the effective theory of dark matter-nucleon interactions.~I have compared this prediction with the 90\% CL upper limits on the same rate from a WIMP analysis of IceCube in the 86-string configuration~\cite{Aartsen:2016fep}.~Through this comparison, I have set 90\% CL upper limits on all isoscalar and isovector coupling constants in Eq.~(\ref{eq:H_chiT}).~For comparison, I have also derived limits on the same coupling constants by demanding that the predicted neutrino flux from WIMP annihilation in the Sun is not larger than the corresponding 90\% CL upper limit from observations performed at Super-Kamiokande~\cite{Choi:2015ara}.~For WIMPs with a mass of about 50 GeV, I find that present Earth WIMP searches at IceCube in the 86-string configuration place comparable or even stronger constraints on the strength of the $\hat{\mathcal{O}}_1$, $\hat{\mathcal{O}}_3$, $\hat{\mathcal{O}}_{11}$, $\hat{\mathcal{O}}_{12}$ and $\hat{\mathcal{O}}_{15}$ interactions than current searches for solar WIMPs at neutrino telescopes in general.~This is in particular true for interaction operators that can generate a large nuclear response for WIMP-Iron scattering.

\acknowledgments It is a pleasure to thank Sebastian Baum, Katherine Freese and Luca Visinelli for sharing a preliminary version of their work~\cite{visinelli}, and for important remarks on the Earth's composition adopted in a first version of this paper.

\appendix
\section{Dark matter response functions}
\label{sec:appDM}
Dark matter response functions appearing in Eq.~(\ref{eq:dsigma}): 
\begin{eqnarray}
 R_{M}^{\tau \tau^\prime}\left(v_T^{\perp 2}, {q^2 \over m_N^2}\right) &=& c_1^\tau c_1^{\tau^\prime } + {J_\chi (J_\chi+1) \over 3} \left[ {q^2 \over m_N^2} v_T^{\perp 2} c_5^\tau c_5^{\tau^\prime }+v_T^{\perp 2}c_8^\tau c_8^{\tau^\prime }
+ {q^2 \over m_N^2} c_{11}^\tau c_{11}^{\tau^\prime } \right] \nonumber \\
 R_{\Phi^{\prime \prime}}^{\tau \tau^\prime}\left(v_T^{\perp 2}, {q^2 \over m_N^2}\right) &=& {q^2 \over 4 m_N^2} c_3^\tau c_3^{\tau^\prime } + {J_\chi (J_\chi+1) \over 12} \left( c_{12}^\tau-{q^2 \over m_N^2} c_{15}^\tau\right) \left( c_{12}^{\tau^\prime }-{q^2 \over m_N^2}c_{15}^{\tau^\prime} \right)  \nonumber \\
 R_{\Phi^{\prime \prime} M}^{\tau \tau^\prime}\left(v_T^{\perp 2}, {q^2 \over m_N^2}\right) &=&  c_3^\tau c_1^{\tau^\prime } + {J_\chi (J_\chi+1) \over 3} \left( c_{12}^\tau -{q^2 \over m_N^2} c_{15}^\tau \right) c_{11}^{\tau^\prime } \nonumber \\
  R_{\tilde{\Phi}^\prime}^{\tau \tau^\prime}\left(v_T^{\perp 2}, {q^2 \over m_N^2}\right) &=&{J_\chi (J_\chi+1) \over 12} \left[ c_{12}^\tau c_{12}^{\tau^\prime }+{q^2 \over m_N^2}  c_{13}^\tau c_{13}^{\tau^\prime}  \right] \nonumber \\
   R_{\Sigma^{\prime \prime}}^{\tau \tau^\prime}\left(v_T^{\perp 2}, {q^2 \over m_N^2}\right)  &=&{q^2 \over 4 m_N^2} c_{10}^\tau  c_{10}^{\tau^\prime } +
  {J_\chi (J_\chi+1) \over 12} \left[ c_4^\tau c_4^{\tau^\prime} + \right.  \nonumber \\
 && \left. {q^2 \over m_N^2} ( c_4^\tau c_6^{\tau^\prime }+c_6^\tau c_4^{\tau^\prime })+
 {q^4 \over m_N^4} c_{6}^\tau c_{6}^{\tau^\prime } +v_T^{\perp 2} c_{12}^\tau c_{12}^{\tau^\prime }+{q^2 \over m_N^2} v_T^{\perp 2} c_{13}^\tau c_{13}^{\tau^\prime } \right] \nonumber \\
    R_{\Sigma^\prime}^{\tau \tau^\prime}\left(v_T^{\perp 2}, {q^2 \over m_N^2}\right)  &=&{1 \over 8} \left[ {q^2 \over  m_N^2}  v_T^{\perp 2} c_{3}^\tau  c_{3}^{\tau^\prime } + v_T^{\perp 2}  c_{7}^\tau  c_{7}^{\tau^\prime }  \right]
       + {J_\chi (J_\chi+1) \over 12} \left[ c_4^\tau c_4^{\tau^\prime} +  \right.\nonumber \\
       &&\left. {q^2 \over m_N^2} c_9^\tau c_9^{\tau^\prime }+{v_T^{\perp 2} \over 2} \left(c_{12}^\tau-{q^2 \over m_N^2}c_{15}^\tau \right) \left( c_{12}^{\tau^\prime }-{q^2 \over m_N^2}c_{15}^{\tau \prime} \right) +{q^2 \over 2 m_N^2} v_T^{\perp 2}  c_{14}^\tau c_{14}^{\tau^\prime } \right] \nonumber \\
     R_{\Delta}^{\tau \tau^\prime}\left(v_T^{\perp 2}, {q^2 \over m_N^2}\right)&=&  {J_\chi (J_\chi+1) \over 3} \left[ {q^2 \over m_N^2} c_{5}^\tau c_{5}^{\tau^\prime }+ c_{8}^\tau c_{8}^{\tau^\prime } \right] \nonumber \\
 R_{\Delta \Sigma^\prime}^{\tau \tau^\prime}\left(v_T^{\perp 2}, {q^2 \over m_N^2}\right)&=& {J_\chi (J_\chi+1) \over 3} \left[c_{5}^\tau c_{4}^{\tau^\prime }-c_8^\tau c_9^{\tau^\prime} \right].
 \label{eq:R}
\end{eqnarray}
In all numerical application I set $J_\chi=1/2$, where $J_\chi$ is the dark matter particle spin.

\section{Phosphorus and Chromium nuclear response functions}
\label{sec:nuc}
In this section I describe the shell model calculation that I have performed to obtain the $W_k^{\tau\tau'}$ functions for Phosphorus and Chromium.~These response functions have never been computed in the literature before.
\subsection{Nuclear response functions}
\label{sec:nuc2}
The nuclear response functions $W_k^{\tau\tau'}$ in Eq.~(\ref{eq:dsigma}), with $k=M,\Sigma',\Sigma'',\Phi'', \Phi'' M, \tilde{\Phi}', \Delta, \Delta \Sigma'$ are defined as follows~\cite{Fitzpatrick:2012ix}
\begin{align}
W_{M}^{\tau \tau^\prime}\left(q^2\right) &= \sum_{L=0,2,\dots}  \langle J,T,M_T ||~ M_{L;\tau} (q)~ || J,T,M_T \rangle \langle J,T,M_T ||~ M_{L;\tau^\prime} (q)~ || J,T,M_T \rangle \nonumber\\
W_{\Sigma'}^{\tau \tau^\prime}\left(q^2\right) &= \sum_{L=1,3,\dots}  \langle J,T,M_T ||~ \Sigma'_{L;\tau} (q)~ || J,T,M_T \rangle \langle J,T,M_T ||~ \Sigma'_{L;\tau^\prime} (q)~ || J,T,M_T \rangle \nonumber\\
W_{\Sigma'}^{\tau \tau^\prime}\left(q^2\right) &= \sum_{L=1,3,\dots}  \langle J,T,M_T ||~ \Sigma''_{L;\tau} (q)~ || J,T,M_T \rangle \langle J,T,M_T ||~ \Sigma''_{L;\tau^\prime} (q)~ || J,T,M_T \rangle \nonumber\\
W_{\Phi''}^{\tau \tau^\prime}\left(q^2\right) &= \sum_{L=0,2,\dots}  \langle J,T,M_T ||~ \Phi''_{L;\tau} (q)~ || J,T,M_T \rangle \langle J,T,M_T ||~ \Phi''_{L;\tau^\prime} (q)~ || J,T,M_T \rangle \nonumber\\
W_{\Phi'' M}^{\tau \tau^\prime}\left(q^2\right) &= \sum_{L=0,2,\dots}  \langle J,T,M_T ||~ \Phi''_{L;\tau} (q)~ || J,T,M_T \rangle \langle J,T,M_T ||~ M_{L;\tau^\prime} (q)~ || J,T,M_T \rangle \nonumber\\
W_{\tilde{\Phi}'}^{\tau \tau^\prime}\left(q^2\right) &= \sum_{L=2,4,\dots}  \langle J,T,M_T ||~ \tilde{\Phi}'_{L;\tau} (q)~ || J,T,M_T \rangle \langle J,T,M_T ||~ \tilde{\Phi}'_{L;\tau^\prime} (q)~ || J,T,M_T \rangle \nonumber\\
W_{\Delta}^{\tau \tau^\prime}\left(q^2\right) &= \sum_{L=1,3,\dots}  \langle J,T,M_T ||~ \Delta_{L;\tau} (q)~ || J,T,M_T \rangle \langle J,T,M_T ||~ \Delta_{L;\tau^\prime} (q)~ || J,T,M_T \rangle \nonumber\\
W_{\Delta \Sigma'}^{\tau \tau^\prime}\left(q^2\right) &= \sum_{L=1,3,\dots}  \langle J,T,M_T ||~ \Delta_{L;\tau} (q)~ || J,T,M_T \rangle \langle J,T,M_T ||~ \Sigma'_{L;\tau^\prime} (q)~ || J,T,M_T \rangle \,.
\label{eq:W}
\end{align}
All nuclear matrix elements in Eq.~(\ref{eq:W}) are reduced in the spin magnetic quantum number $M_J$ via the Wigner-Eckart theorem, e.g.
\begin{equation}
\langle J,M_J |\,{M}_{LM;\tau}\,|J,M_J\rangle =(-1)^{J-M_J}\left(
\begin{array}{ccc} J&L&J\\
-M_J&M&M_J 
\end{array} 
\right)
\langle  J  ||\,{M}_{L;\tau}\,|| J  \rangle \,.
\label{eq:red}
\end{equation}
In Eqs.~(\ref{eq:red}) and (\ref{eq:W}), $T$ and $M_T$ are the nuclear isospin and associated magnetic quantum number, respectively.~The nuclear response operators in Eq.~(\ref{eq:W}) arise from the multipole expansion of nuclear charges and currents generated in the scattering of WIMPs by nuclei (see, e.g., \cite{Anand:2013yka,Catena:2015uha} for further details).~The response operator $M_{LM;\tau}$ arises from the multiple expansion of the nuclear vector charge, $\Sigma'_{LM;\tau}$ and $\Sigma''_{LM;\tau}$ from the expansion of the nuclear spin current, $\Delta_{LM;\tau}$ from the nuclear convection current, and, finally, $\tilde{\Phi}'_{LM;\tau}$ and $\Phi''_{LM;\tau}$ from the nuclear spin-velocity current.~In all equations, $L$ is the operator multipolar decomposition index and is restricted by the requirement of nuclear wave functions of definite P and CP, and by the constraint $L\le 2J$.~The nuclear response operators in Eq.~(\ref{eq:W}) admit the following representation
\begin{eqnarray}
M_{LM;\tau}(q) &=& \sum_{i=1}^{A} M_{LM}(q {\bf{r}}_i) t^{\tau}_{(i)}\nonumber\\
\Sigma'_{LM;\tau}(q) &=& -i \sum_{i=1}^{A} \left[ \frac{1}{q} \overrightarrow{\nabla}_{{\bf{r}}_i} \times {\bf{M}}_{LL}^{M}(q {\bf{r}}_i)  \right] \cdot \vec{\sigma}_{(i)} t^{\tau}_{(i)}\nonumber\\
\Sigma''_{LM;\tau}(q) &=&\sum_{i=1}^{A} \left[ \frac{1}{q} \overrightarrow{\nabla}_{{\bf{r}}_i} M_{LM}(q {\bf{r}}_i)  \right] \cdot \vec{\sigma}_{(i)} t^{\tau}_{(i)}\nonumber\\
\Delta_{LM;\tau}(q) &=&\sum_{i=1}^{A}  {\bf{M}}_{LL}^{M}(q {\bf{r}}_i) \cdot \frac{1}{q}\overrightarrow{\nabla}_{{\bf{r}}_i} t^{\tau}_{(i)} \nonumber\\
\tilde{\Phi}^{\prime}_{LM;\tau}(q) &=& \sum_{i=1}^A \left[ \left( {1 \over q} \overrightarrow{\nabla}_{{\bf{r}}_i} \times {\bf{M}}_{LL}^M(q {\bf{r}}_i) \right) \cdot \left(\vec{\sigma}_{(i)} \times {1 \over q} \overrightarrow{\nabla}_{{\bf{r}}_i} \right) + {1 \over 2} {\bf{M}}_{LL}^M(q {\bf{r}}_i) \cdot \vec{\sigma}_{(i)} \right]~t^\tau_{(i)} \nonumber \\
\Phi^{\prime \prime}_{LM;\tau}(q ) &=& i  \sum_{i=1}^A\left( {1 \over q} \overrightarrow{\nabla}_{{\bf{r}}_i}  M_{LM}(q {\bf{r}}_i) \right) \cdot \left(\vec{\sigma}_{(i)} \times \
{1 \over q} \overrightarrow{\nabla}_{{\bf{r}}_i}  \right)~t^\tau_{(i)} \,,
\label{eq:resop}
\end{eqnarray}
where ${\bf{r}}_i$ and $\vec{\sigma}_{(i)}$ are the $i$-th nucleon position vector in the nucleus centre of mass frame and spin Pauli matrices, respectively.~In Eq.~(\ref{eq:resop}) I have introduced the notation ${\bf{M}}_{LL}^{M}(q {\bf{r}}_i)=j_{L}(q r_i){\bf Y}^M_{LL1}(\Omega_{{\bf{r}}_i})$ and $M_{LM}(q {\bf{r}}_i)=j_{L}(q r_i)Y_{LM}(\Omega_{{\bf{r}}_i})$, where $Y_{LM}$ and ${\bf Y}^M_{LL1}$ are spherical harmonics and vector spherical harmonics, respectively.~The spherical Bessel functions are denoted by $j_{L}$.

For simplicity, in the following I denote by $A_{LM;\tau}$ all nuclear response operators.~With this notation, I now expand the nuclear response operators in Eq.~(\ref{eq:resop}) in a complete set of spherically symmetric single-particle states, $|\alpha\rangle$, assuming the nuclear harmonic oscillator model for the radial part of the nucleon wave functions.~Such states are labelled by their principal quantum number, $n_\alpha$, angular momentum, $l_\alpha$, and spin $s_\alpha$, and by their total spin and isospin, $j_{\alpha}$ and $t_{\alpha}$ respectively:~$|\alpha\rangle=|n_\alpha,l_\alpha, s_\alpha=1/2,j_\alpha, m_{j_\alpha};t_{\alpha}=1/2,m_{t_\alpha}\rangle$, where $m_{j_\alpha}$ and $m_{t_\alpha}$ are the total spin and isospin magnetic quantum numbers.~Analogously, $|\alpha\rangle\equiv ||\alpha|,m_{j_\alpha};m_{t_{\alpha}}\rangle$.~The expansion of the operators in Eq.~(\ref{eq:resop}) in a basis of single-particle states gives
\begin{eqnarray}
  A_{LM;\tau} &=&\;  \sum_{\alpha\beta}\langle \alpha|~A_{LM;\tau}~|\beta\rangle \,a_{\alpha}^{\dagger}a_{\beta} \nonumber\\
&=&\; \sum_{|\alpha||\beta|}\langle\left|\alpha\right|\vdots\vdots A_{L;\tau}\vdots\vdots \left|\beta\right|\rangle\frac{[a_{|\alpha|}^{\dagger}\otimes\tilde{a}_{|\beta|}]_{LM;\tau}}{\sqrt{(2L+1)(2\tau+1)}} \,,
\label{eq:keyeq}
\end{eqnarray}  
where $\tilde{a}_{|\beta|,m_{j_\beta},m_{t_{\beta}}}\equiv (-1)^{j_\beta-m_{j_\beta}+1/2-m_{t_\beta}}\,a_{|\beta|,-m_{j_\beta},-m_{t_{\beta}}}$, and the symbol $\vdots\vdots$ appears in matrix elements reduced in spin and isospin according to Eq.~(\ref{eq:red}).~The creation and annihilation operators $a^{\dagger}_{\alpha}$ and $\tilde{a}_\beta$ transform as tensors under spin and isospin transformations.~Their tensor product can be decomposed as follows
\begin{eqnarray}
[a_{|\alpha|}^{\dagger}\otimes\tilde{a}_{|\beta|}]_{LM;\tau} &=&  \sqrt{(2L+1)(2\tau+1)}\sum_{m_{j_\alpha}m_{t_{\alpha}}m_{j_\beta}m_{t_{\beta}}}(-1)^{\,j_{\alpha}-m_{j_\alpha}+t_{j_\alpha}-m_{t_{\alpha}}}\nonumber\\
&\times&\begin{pmatrix}
    j_{\alpha}&L&j_{\beta}\\
    -m_{\alpha}&M&m_{\beta}
  \end{pmatrix}\begin{pmatrix}
    t_{\alpha}&\tau&t_{\beta}\\
    -m_{t_{\alpha}}&0&m_{t_{\beta}}
  \end{pmatrix}a_{\alpha}^{\dagger}a_{\beta}
\,.
\end{eqnarray}
Combining Eqs.~(\ref{eq:keyeq}) and (\ref{eq:red}), I find the following matrix elements for the nuclear response operators $A_{LM;\tau}$
\begin{eqnarray}
\langle J, T, M_T  ||~ A_{LM;\tau} ~ || J, T, M_T \rangle &=& (-1)^{T-M_T}
\begin{pmatrix}
    T&\tau&T\\
    -M_T&0&M_T
  \end{pmatrix} \nonumber\\
&\times& \sum_{|\alpha||\beta|}\langle\left|\alpha\right|\vdots\vdots A_{L;\tau}\vdots\vdots \left|\beta\right|\rangle\frac{\langle J, T \vdots\vdots ~[a_{|\alpha|}^{\dagger}\otimes\tilde{a}_{|\beta|}]_{L;\tau}
~\vdots\vdots J, T\rangle}{\sqrt{(2L+1)(2\tau+1)}} \,.
\label{eq:matel}
\end{eqnarray}
In Eq.~(\ref{eq:matel}) one can recognise the definition of ground-state to ground-state one-body density matrix elements (OBDME):
\begin{equation}
\psi^{L;\tau}_{|\alpha||\beta|} \equiv \frac{\langle J, T\vdots\vdots ~[a_{|\alpha|}^{\dagger}\otimes\tilde{a}_{|\beta|}]_{L;\tau}
~\vdots\vdots J, T\rangle}{\sqrt{(2L+1)(2\tau+1)}} \,,
\label{eq:OBDME}
\end{equation}
which can be used to rewrite Eq.~(\ref{eq:matel}) as follows
\begin{eqnarray}
\langle J, T, M_T  ||~ A_{LM;\tau} ~ || J, T, M_T \rangle 
&=& (-1)^{T-M_T}
\begin{pmatrix}
    T&\tau&T\\
    -M_T&0&M_T
  \end{pmatrix} 
   \sum_{|\alpha||\beta|}\psi^{L;\tau}_{|\alpha||\beta|} \,\langle \left|\alpha\right|\vdots\vdots A_{L;\tau}\vdots\vdots \left|\beta\right|\rangle \,. \nonumber\\
   \label{eq:master}
\end{eqnarray}
The doubly reduced matrix elements in Eq.~(\ref{eq:master}) can be further simplified, since the nuclear operators $A_{LM;\tau}$ depend on isospin through the matrices $t^{\tau}_{(i)}$ only.~As a result, Eq.~(\ref{eq:master}) can be factorised using
\begin{equation}
\langle \left|\alpha\right|\vdots\vdots A_{L;\tau}\vdots\vdots \left|\beta\right|\rangle = \sqrt{2(2\tau+1)} \,\langle n_{\alpha},l_\alpha,1/2,j_\alpha ||\,A_{L}\,|| n_{\beta},l_\beta,1/2,j_\beta  \rangle\,,
\label{eq:math}
\end{equation}
where $A_L$ is the factor in $A_{L;\tau}$ acting on nuclear spin and space coordinates.~I use the {\sffamily Mathematica} package in \cite{Anand:2013yka} to evaluate the matrix elements in Eq.~(\ref{eq:math}) in the harmonic oscillator basis.~In this basis, the $W_k^{\tau\tau'}$ functions depend on $q^2$ through the variable $y=(bq/2)^2$, where $b=\sqrt{41.467/(45 A^{-1/3}-25A^{-2/3})}$~fm is the harmonic oscillator basis length parameter. 

\begin{table}
  \centering
  \begin{tabular}[!h]{|c|c|c|c|c|c|r|c|}
\hline
    Element&$2J$&$2T$&P&core-orbits&valence-orbits&Hamiltonian&restrictions\\
\hline
\hline
${}^{31}$P&1&1&+&s-p&sd&w~\cite{Warburton:1992rh}&none\\
\hline
${}^{52}$Cr&0&4&+&s-p-sd&pf&gx1~\cite{Honma:2004xk}&$1p_{1/2}0f_{5/2}$\\
\hline
  \end{tabular}
  \caption{Input parameters for OBDME calculation via {\sffamily Nushell@MSU}.~The notation in~\cite{Brown:2001zz} is used to define core and valence orbits, interaction Hamiltonians, and restrictions in valence space.~Not allowed single-particle states are listed in the last column.}
  \label{tab:inputs}
\end{table}

In order to complete the calculation of the matrix elements in Eq.~(\ref{eq:master}), and therefore of the nuclear response functions in Eq.~(\ref{eq:W}), the OBDME $\psi^{L;\tau}_{|\alpha||\beta|}$ in Eq.~(\ref{eq:OBDME}) must be computed numerically.~In this analysis, I use the shell model code {\sffamily Nushell@MSU}~\cite{NuShell,Brown:2001zz}.~This code relies on three sets of inputs: the target nucleus spin, isospin and parity; the Hamiltonian for valence nucleon interactions; and the model space, including restrictions on the allowed single-particle states.~The assumptions made for Phosphorus and Chromium are listed in Tab.~\ref{tab:inputs}, and are based on guidelines provided in Ref.~\cite{Brown:2001zz} and references therein.~{\sffamily Nushell@MSU} first calculates the nuclear many-body ground-state wave function, and then evaluates the overlap of this wave function with the single-particle states $|\alpha \rangle$, according to the definition of OBDME in Eq.~(\ref{eq:OBDME}).~As a test of the {\sffamily Nushell@MSU} code, I have compared the OBDME found for $^{23}$Na, $^{28}$Si and $^{19}$F using this programme and its implementation of the w-interaction~\cite{Warburton:1992rh} with the OBDME independently obtained in~\cite{Anand:2013yka} for the same elements.~The result of the two calculations differ negligibly.~In this test I have also included $^{19}$F, although it does not enter the WIMP capture calculation.~Regarding the gx1-interaction~\cite{Honma:2004xk} used in the present calculation of the $^{52}$Cr OBDME, it has been found that in the full $pf$ model space gx1 can successfully describe binding energies, electro-magnetic transitions, and excitation spectra of Iron, and of various Nickel isotopes~\cite{Honma:2004xk}.~The major limitation of my numerical OBDME calculation thus resides in the use of model space restrictions for $^{52}$Cr.~On the other hand, such restrictions allow me to complete the calculation of the $^{52}$Cr OBDME with limited computing resources.~At the same time, corrections to the binding energies of nuclei with about 35 neutrons or less due to similar model space restrictions are expected to be at the percent level when the gx1-interaction is used~\cite{Honma:2004xk}.

\subsection{Phosphorus}
Below I list the non-zero nuclear response functions that I find for $^{31}$P proceeding as described in the previous subsection.~Notice that in general $W_k^{\tau\tau'} = W_k^{\tau'\tau}$, with the exception of the nuclear response functions $W^{\tau\tau'}_{\Phi'' M}$ and $W^{\tau\tau'}_{\Delta \Sigma'}$, which are not symmetric under the transformation of indexes $\tau \rightarrow \tau'$:
\begin{align}
W_{M}^{00}(y)&=e^{-2 y} \left(0.982271 y^4-11.0707 y^3+43.4498 y^2-69.0704 y+38.2357\right) \nonumber \\
W_{M}^{11}(y)&=e^{-2 y} \left(0.0076155 y^4-0.0464186 y^3+0.105547 y^2-0.1061 y+0.0397876\right)  \nonumber \\
W_{M}^{10}(y)&=e^{-2 y} \left(-0.0864898 y^4+0.75098 y^3-2.2227 y^2+2.75859 y-1.23341\right) \nonumber \\
W_{M}^{01}(y)&=e^{-2 y} \left(-0.0864898 y^4+0.75098 y^3-2.2227 y^2+2.75859 y-1.23341\right) \nonumber \\
W_{\Sigma''}^{00}(y)&= e^{-2 y} \left(0.00277379 y^4-0.00269138 y^3+0.00743249 y^2-0.0032891 y+0.00414265\right)\nonumber \\
W_{\Sigma''}^{11}(y)&=e^{-2 y} \left(0.0016697 y^4-0.000679619 y^3+0.00467858 y^2-0.000938087
   y+0.00318122\right)\nonumber \\
W_{\Sigma''}^{10}(y)&=e^{-2 y} \left(0.00215207 y^4-0.00148204 y^3+0.00581303 y^2-0.00197639
   y+0.00363025\right)\nonumber \\
W_{\Sigma''}^{01}(y)&=e^{-2 y} \left(0.00215207 y^4-0.00148204 y^3+0.00581303 y^2-0.00197639
   y+0.00363025\right)\nonumber \\
W_{\Sigma'}^{00}(y)&=e^{-2 y} \left(0.0117227 y^4-0.0355087 y^3+0.0465999 y^2-0.0298521 y+0.0082853\right)\nonumber \\
W_{\Sigma'}^{11}(y)&=e^{-2 y} \left(0.00919363 y^4-0.0294649 y^3+0.0389045 y^2-0.0245117 y+0.00636245\right)\nonumber \\
W_{\Sigma'}^{10}(y)&=e^{-2 y} \left(0.0103814 y^4-0.0323588 y^3+0.0425594 y^2-0.0270656 y+0.0072605\right)\nonumber \\
W_{\Sigma'}^{01}(y)&=e^{-2 y} \left(0.0103814 y^4-0.0323588 y^3+0.0425594 y^2-0.0270656 y+0.0072605\right)\nonumber \\
W_{\Phi''}^{00}(y)&=e^{-2 y} \left(0.191514 y^2-0.957567 y+1.19696\right)\nonumber \\
W_{\Phi''}^{11}(y)&=e^{-2 y} \left(0.00132747 y^2-0.00663733 y+0.00829666\right)\nonumber \\
W_{\Phi''}^{10}(y)&=e^{-2 y} \left(0.0159445 y^2-0.0797226 y+0.0996532\right)\nonumber \\
W_{\Phi''}^{01}(y)&=e^{-2 y} \left(0.0159445 y^2-0.0797226 y+0.0996532\right)\nonumber \\
W_{\Delta}^{00}(y)&=0.000365072 e^{-2 y} (2.5\, -y)^2\nonumber \\
W_{\Delta}^{11}(y)&=0.000026963 e^{-2 y} (2.5\, -y)^2\nonumber \\
W_{\Delta}^{10}(y)&=0.0000992141 e^{-2 y} (2.5\, -y)^2\nonumber \\
W_{\Delta}^{01}(y)&=0.0000992141 e^{-2 y} (2.5\, -y)^2\nonumber \\
W_{M \Phi''}^{00}(y)&=e^{-2 y} \left(0.433726 y^3-3.52846 y^2+8.81641 y-6.7651\right)\nonumber \\
W_{M \Phi''}^{11}(y)&=e^{-2 y} \left(-0.00317951 y^3+0.0176388 y^2-0.0314925 y+0.0181688\right)\nonumber \\
W_{M \Phi''}^{10}(y)&=e^{-2 y} \left(-0.0381899 y^3+0.211864 y^2-0.378265 y+0.21823\right)\nonumber \\
W_{M \Phi''}^{01}(y)&=e^{-2 y} \left(0.03611 y^3-0.293763 y^2+0.734013 y-0.563231\right)\nonumber \\
W_{\Sigma' \Delta}^{00}(y)&=e^{-2 y} \left(0.00206873 y^3-0.00830496 y^2+0.00957203 y-0.00434794\right)\nonumber \\
W_{\Sigma' \Delta}^{11}(y)&=e^{-2 y} \left(0.000497883 y^3-0.00204255 y^2+0.00240878 y-0.00103547\right)\nonumber \\
W_{\Sigma' \Delta}^{10}(y)&=e^{-2 y} \left(0.00183203 y^3-0.00751584 y^2+0.00886346 y-0.00381015\right)\nonumber \\
W_{\Sigma' \Delta}^{01}(y)&=e^{-2 y} \left(0.000562209 y^3-0.002257 y^2+0.00260135 y-0.00118162\right) \,. \nonumber \\
\end{align}

\subsection{Chromium}
Below I list the non-zero nuclear response functions that I find for $^{52}$Cr as described in Sec.~\ref{sec:nuc2}:
\begin{align}
W_{M}^{00}(y)&=e^{-2 y} \left(0.0178903 y^6-0.668565 y^5+8.66093 y^4-47.0832 y^3+118.148 y^2\right.\nonumber \\ 
&\left.  -132.417 y+53.7944\right)\nonumber \\
W_{M}^{11}(y)&=e^{-2 y} \left(0.00204138 y^6-0.0414801 y^5+0.312679 y^4-1.08692 y^3+1.7912 y^2\right.\nonumber \\ 
&\left.    -1.27324
 y+0.318309\right)\nonumber \\
W_{M}^{10}(y)&=e^{-2 y} \left(-0.00604326 y^6+0.174317 y^5-1.70602 y^4+7.37066 y^3-14.9627 y^2\right.\nonumber \\ 
&\left.  +13.369
  y-4.13803\right)\nonumber \\
W_{M}^{01}(y)&=e^{-2 y} \left(-0.00604326 y^6+0.174317 y^5-1.70602 y^4+7.37066 y^3-14.9627 y^2 \right.\nonumber \\  
 &\left.  +13.369
   y-4.13803\right)\nonumber \\
W_{\Phi''}^{00}(y)&=e^{-2 y} \left(0.0376034 y^4-0.519936 y^3+2.44718 y^2-4.49316 y+2.80823\right)\nonumber \\
W_{\Phi''}^{11}(y)&=e^{-2 y} \left(0.00418588 y^4-0.0576078 y^3+0.270215 y^2-0.495514 y+0.309696\right)\nonumber \\
W_{\Phi''}^{10}(y)&=e^{-2 y} \left(-0.012546 y^4+0.173068 y^3-0.813183 y^2+1.49212 y-0.932576\right)\nonumber \\
W_{\Phi''}^{01}(y)&=e^{-2 y} \left(-0.012546 y^4+0.173068 y^3-0.813183 y^2+1.49212 y-0.932576\right)\nonumber \\
W_{M \Phi''}^{00}(y)&=e^{-2 y} \left(0.0259371 y^5-0.663953 y^4+5.32514 y^3-17.7122 y^2+24.96 y\right.\nonumber \\ 
&\left.-12.2909\right)\nonumber \\
W_{M \Phi''}^{11}(y)&=e^{-2 y} \left(0.00292318 y^5-0.0498139 y^4+0.302513 y^3-0.794315 y^2+0.879125
   y \right.\nonumber \\ 
&\left.-0.313973\right)\nonumber \\
W_{M \Phi''}^{10}(y)&=e^{-2 y} \left(-0.00876145 y^5+0.149586 y^4-0.909919 y^3+2.39138 y^2-2.64727
   y\right.\nonumber \\ 
&\left.+0.945455\right)\nonumber \\
W_{M \Phi''}^{01}(y)&=e^{-2 y} \left(-0.0086537 y^5+0.221244 y^4-1.77113 y^3+5.88422 y^2-8.2889
   y\right.\nonumber \\ 
&\left.+4.08166\right) \,.
\end{align}

\section{Further capture rates and exclusion limits}
\label{app:figures}
\begin{figure}[t]
\begin{center}
\begin{minipage}[t]{0.49\linewidth}
\centering
\includegraphics[width=\textwidth]{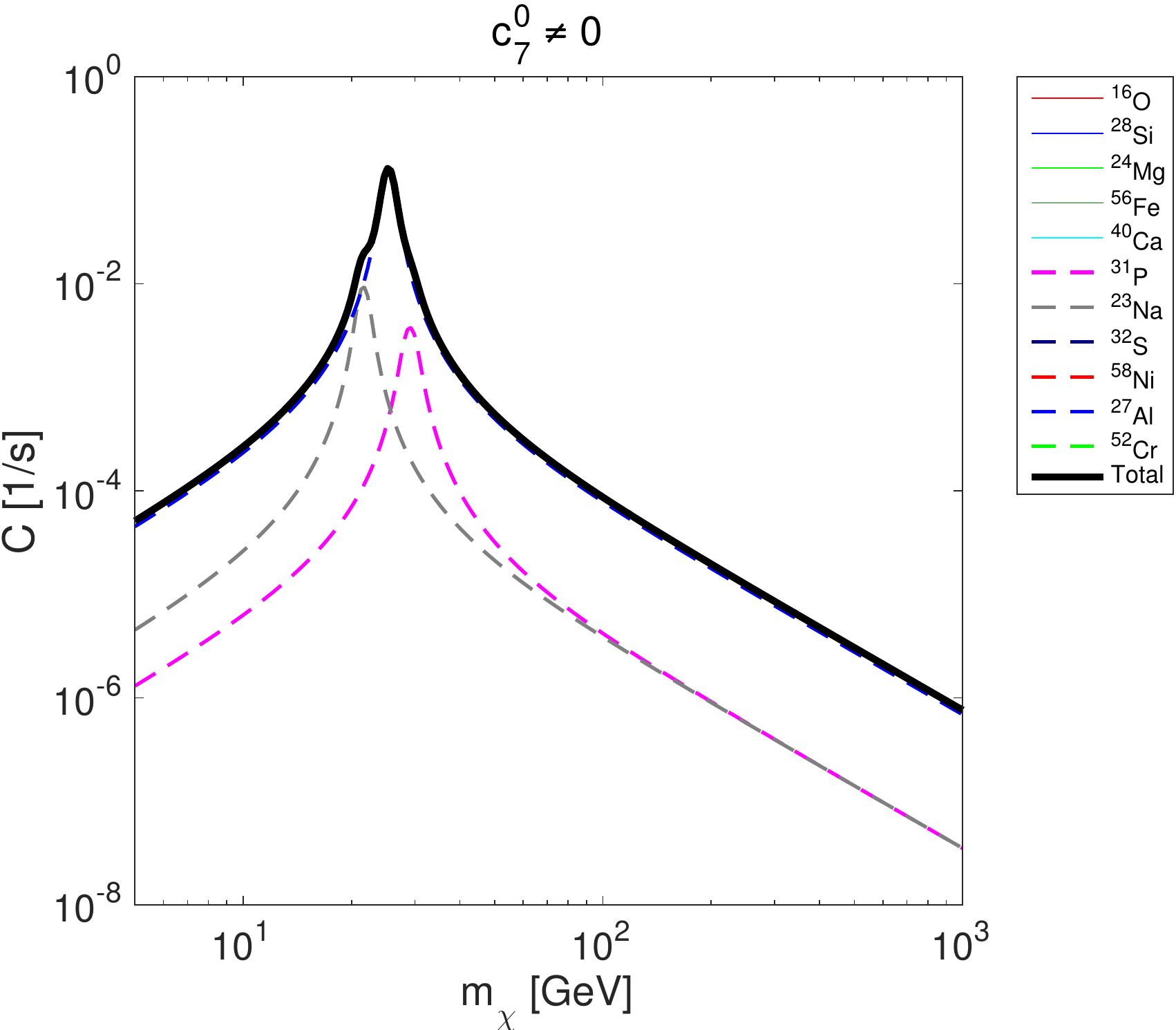}
\end{minipage}
\begin{minipage}[t]{0.49\linewidth}
\centering
\includegraphics[width=\textwidth]{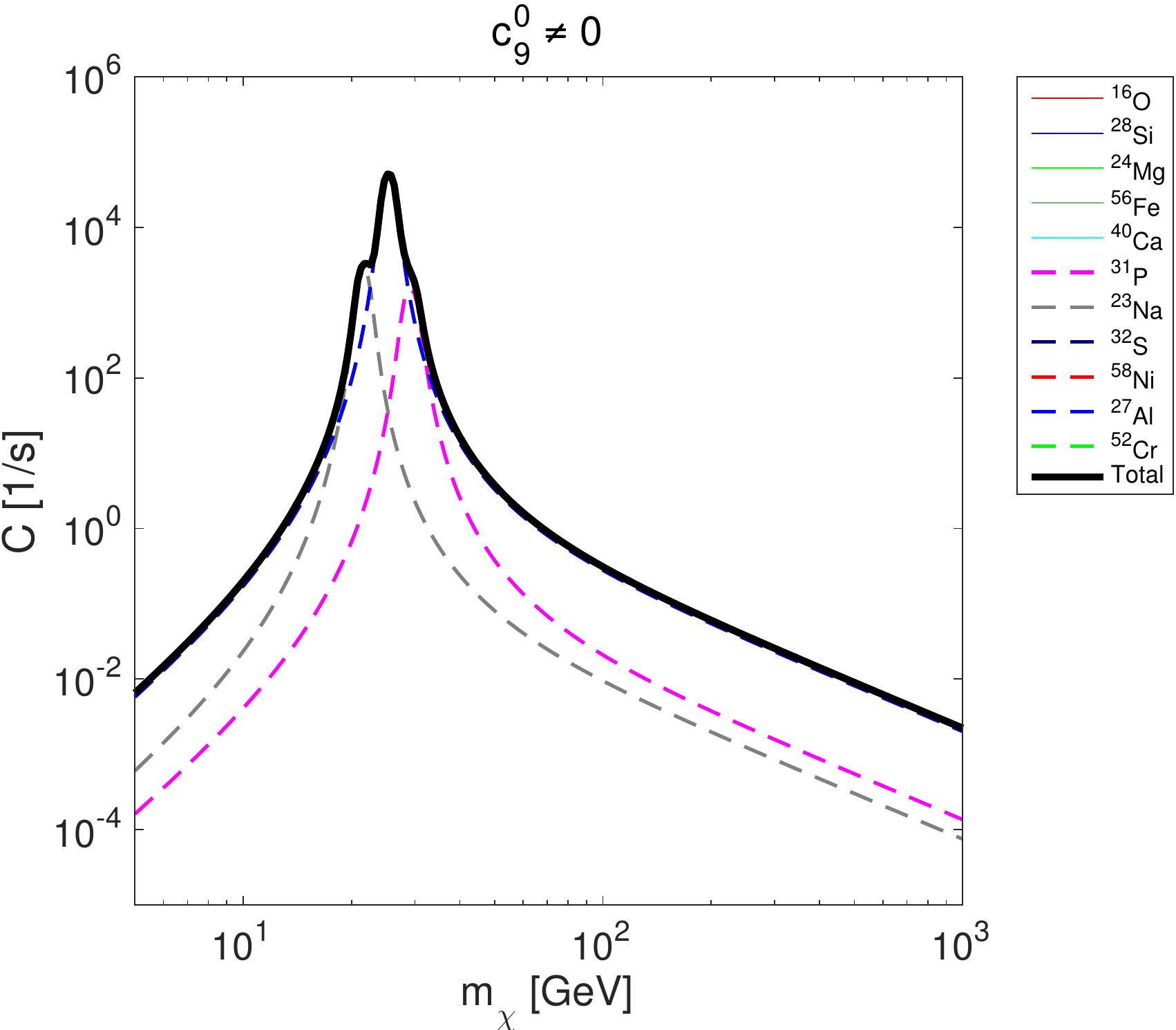}
\end{minipage}
\begin{minipage}[t]{0.49\linewidth}
\centering
\includegraphics[width=\textwidth]{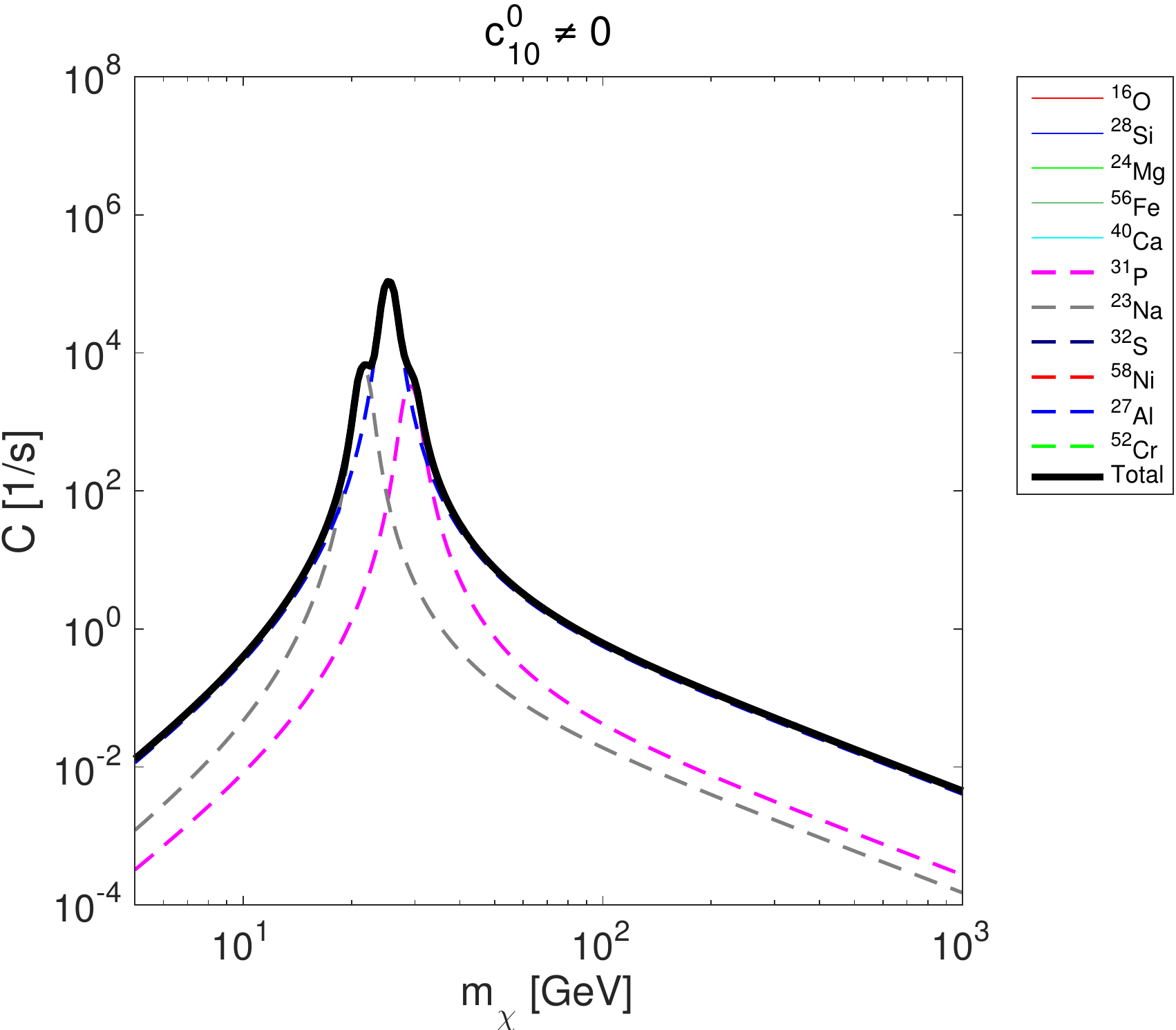}
\end{minipage}
\begin{minipage}[t]{0.49\linewidth}
\centering
\includegraphics[width=\textwidth]{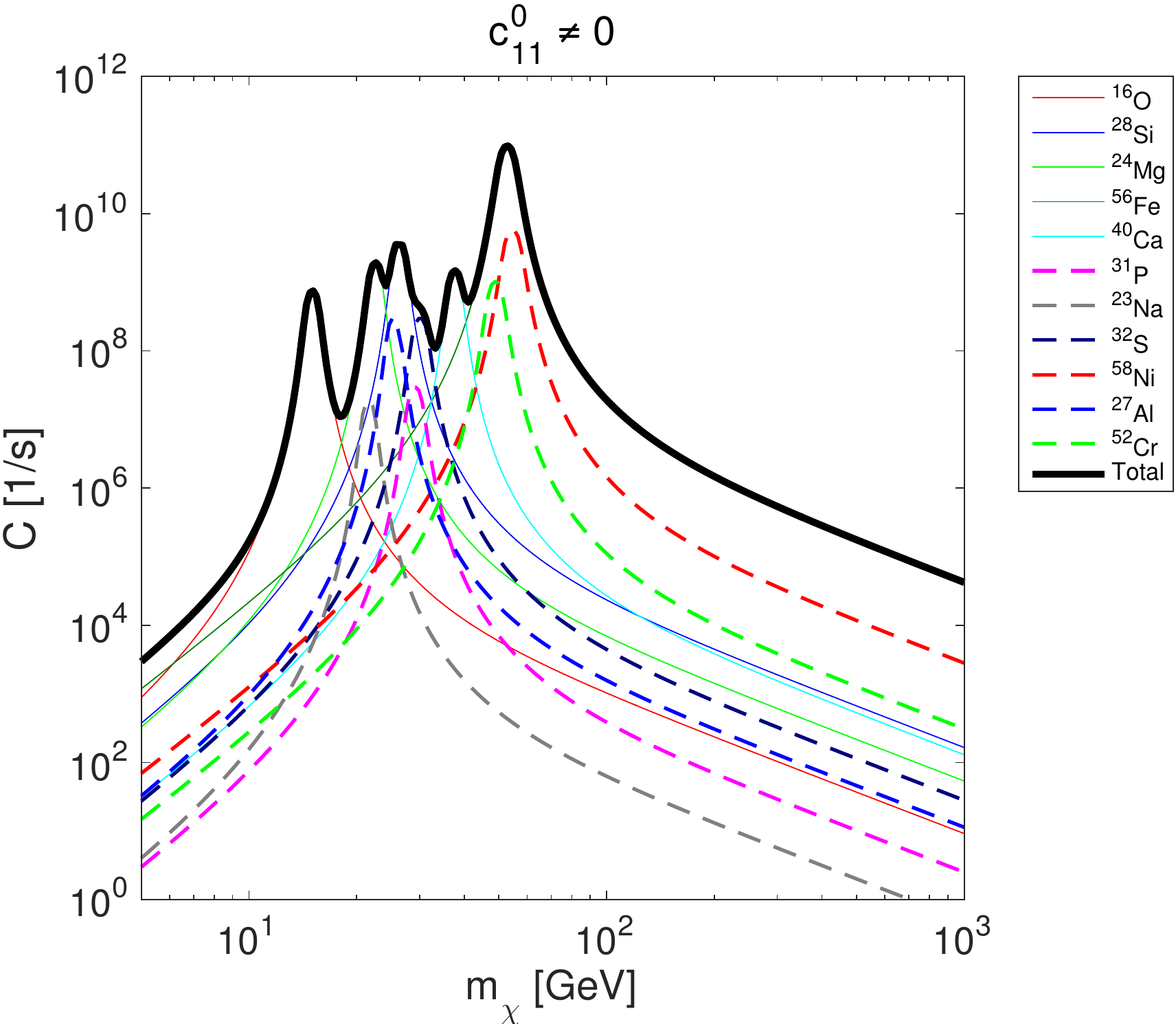}
\end{minipage}
\begin{minipage}[t]{0.49\linewidth}
\centering
\includegraphics[width=\textwidth]{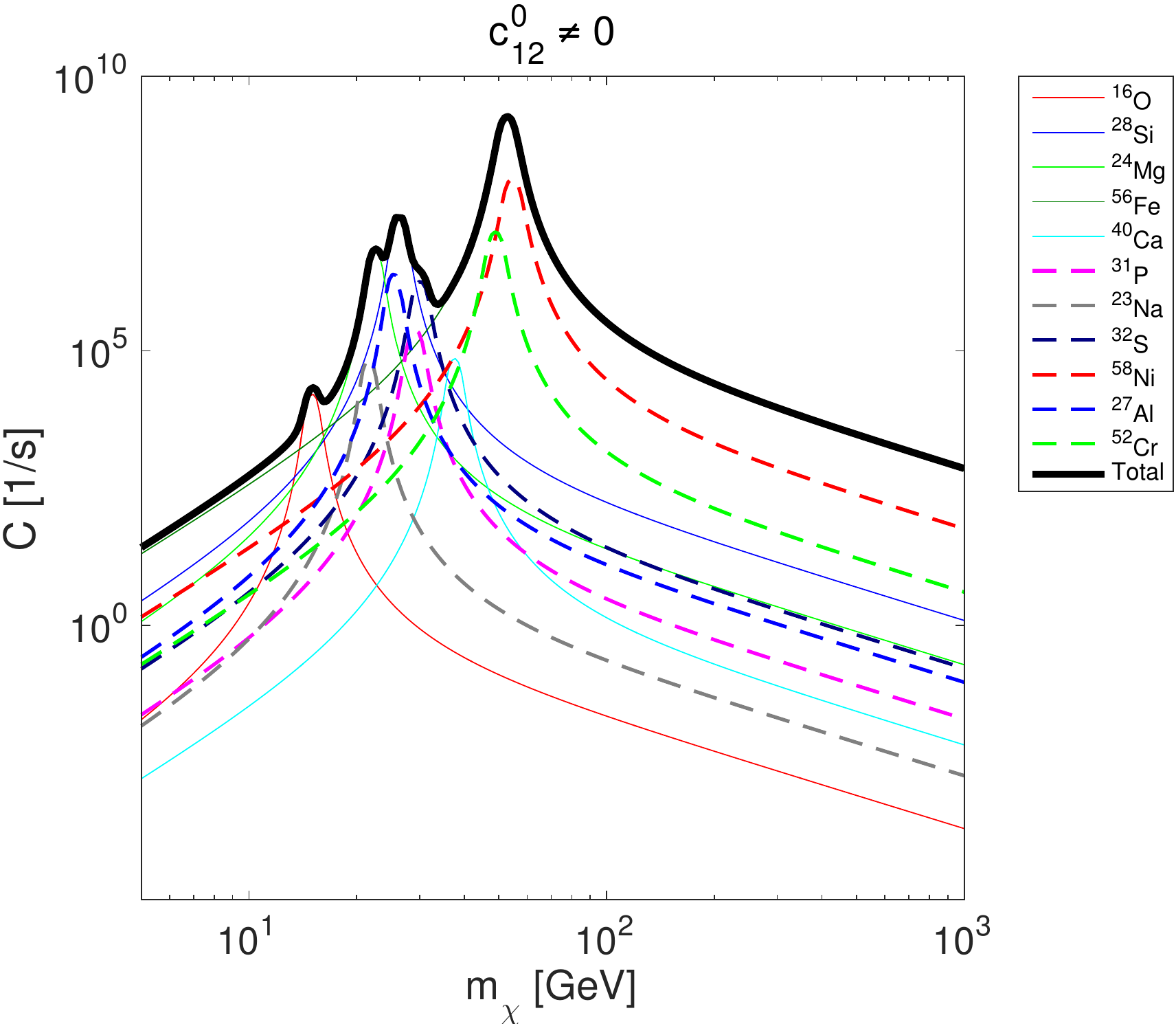}
\end{minipage}
\begin{minipage}[t]{0.49\linewidth}
\centering
\includegraphics[width=\textwidth]{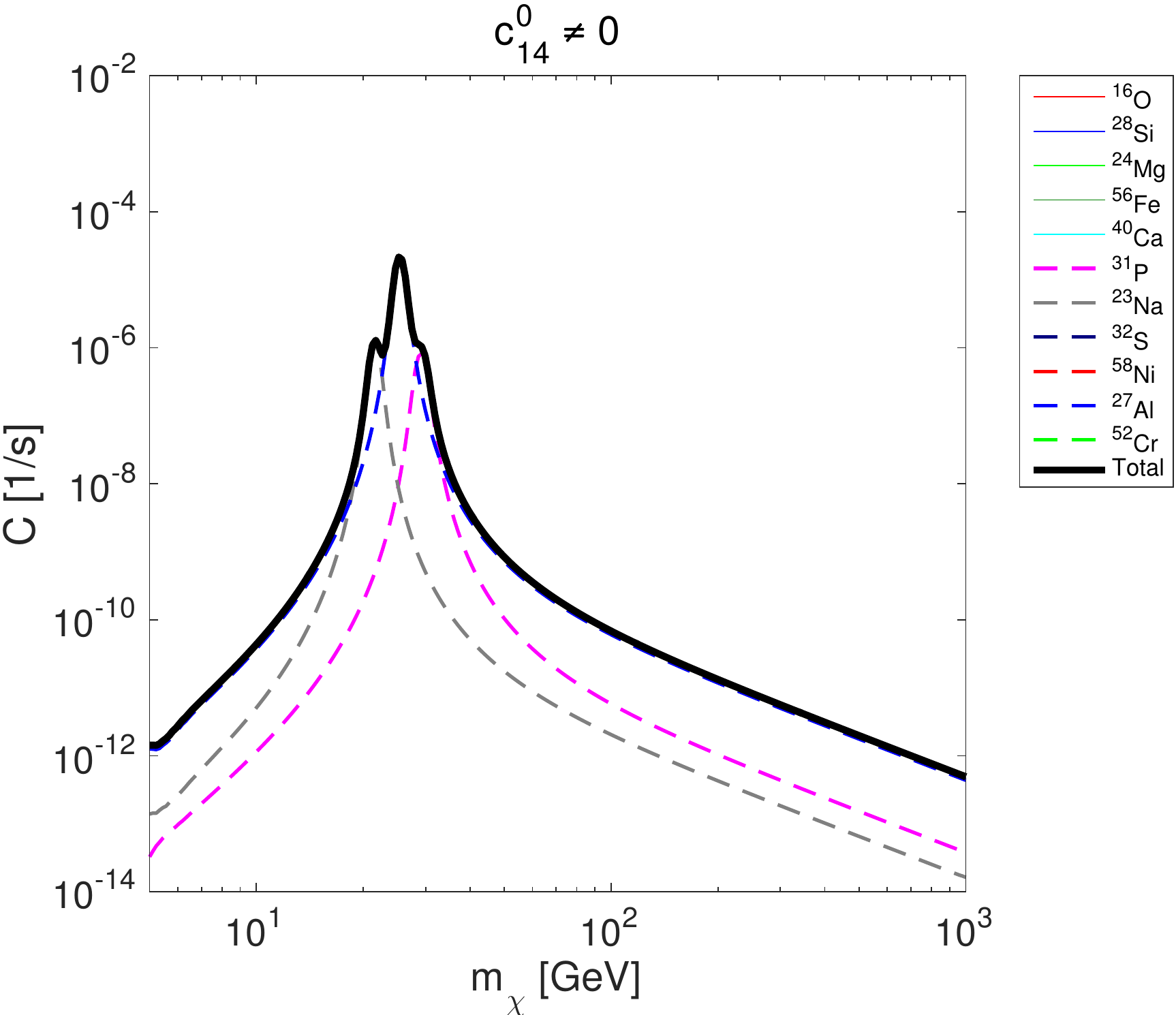}
\end{minipage}
\end{center}
\caption{Rate of WIMP capture in the Earth for the isoscalar component of the interaction operators $\hat{\mathcal{O}}_7$, $\hat{\mathcal{O}}_9$, $\hat{\mathcal{O}}_{10}$, $\hat{\mathcal{O}}_{11}$, $\hat{\mathcal{O}}_{12}$, $\hat{\mathcal{O}}_{14}$.}
\label{fig:ac1}
\end{figure}

\begin{figure}[t]
\begin{center}
\begin{minipage}[t]{0.49\linewidth}
\centering
\includegraphics[width=\textwidth]{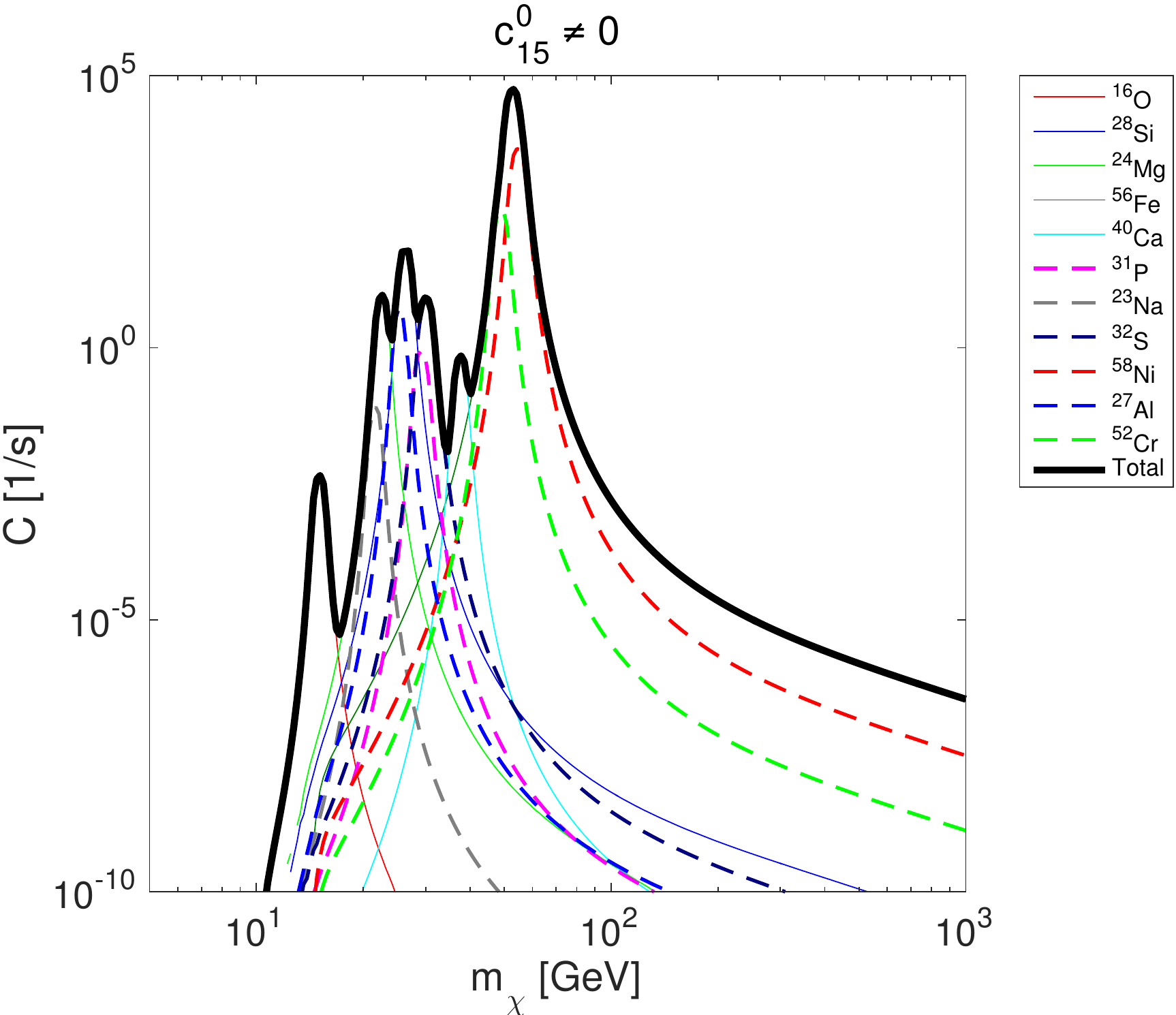}
\end{minipage}
\begin{minipage}[t]{0.49\linewidth}
\centering
\includegraphics[width=\textwidth]{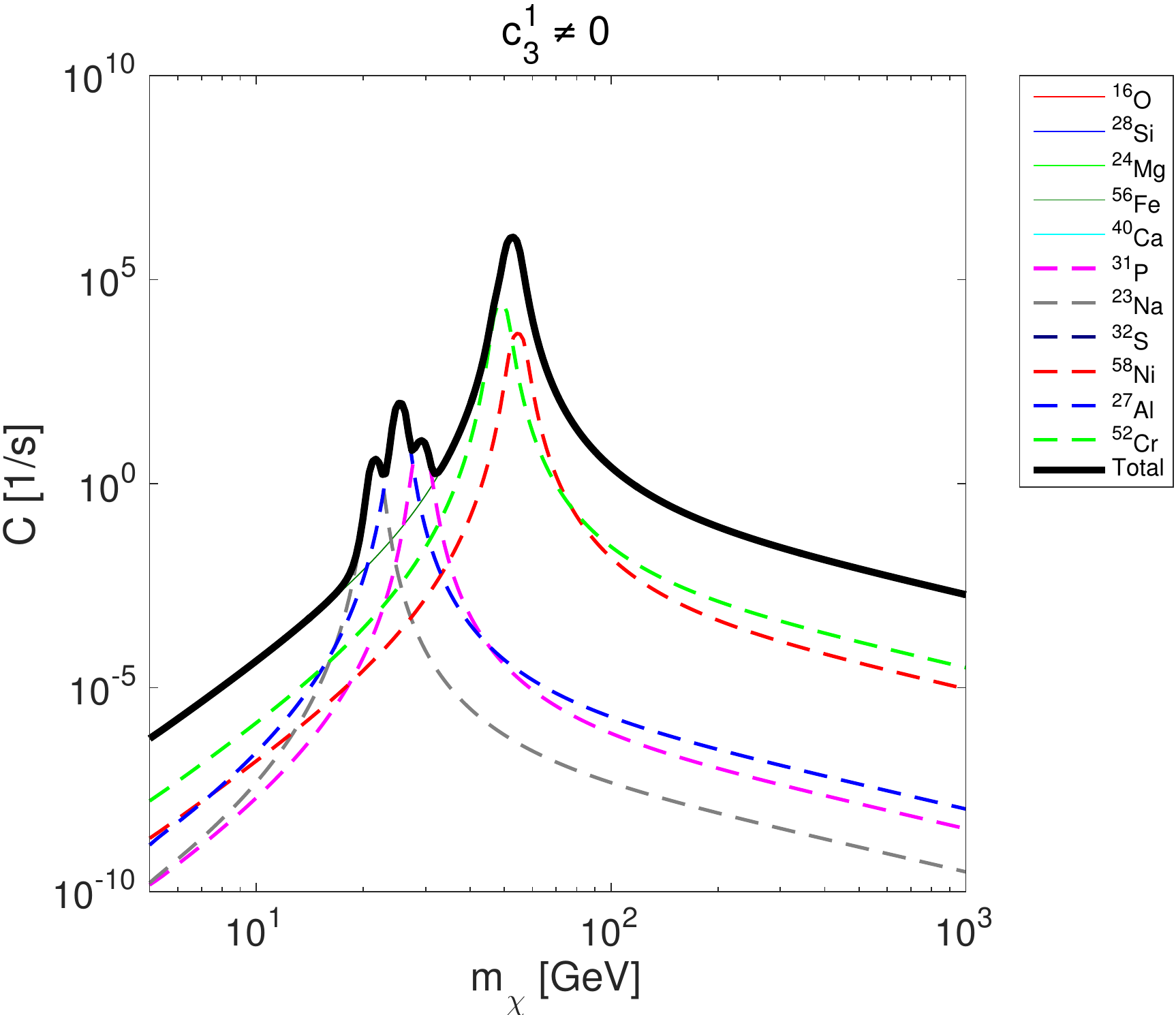}
\end{minipage}
\begin{minipage}[t]{0.49\linewidth}
\centering
\includegraphics[width=\textwidth]{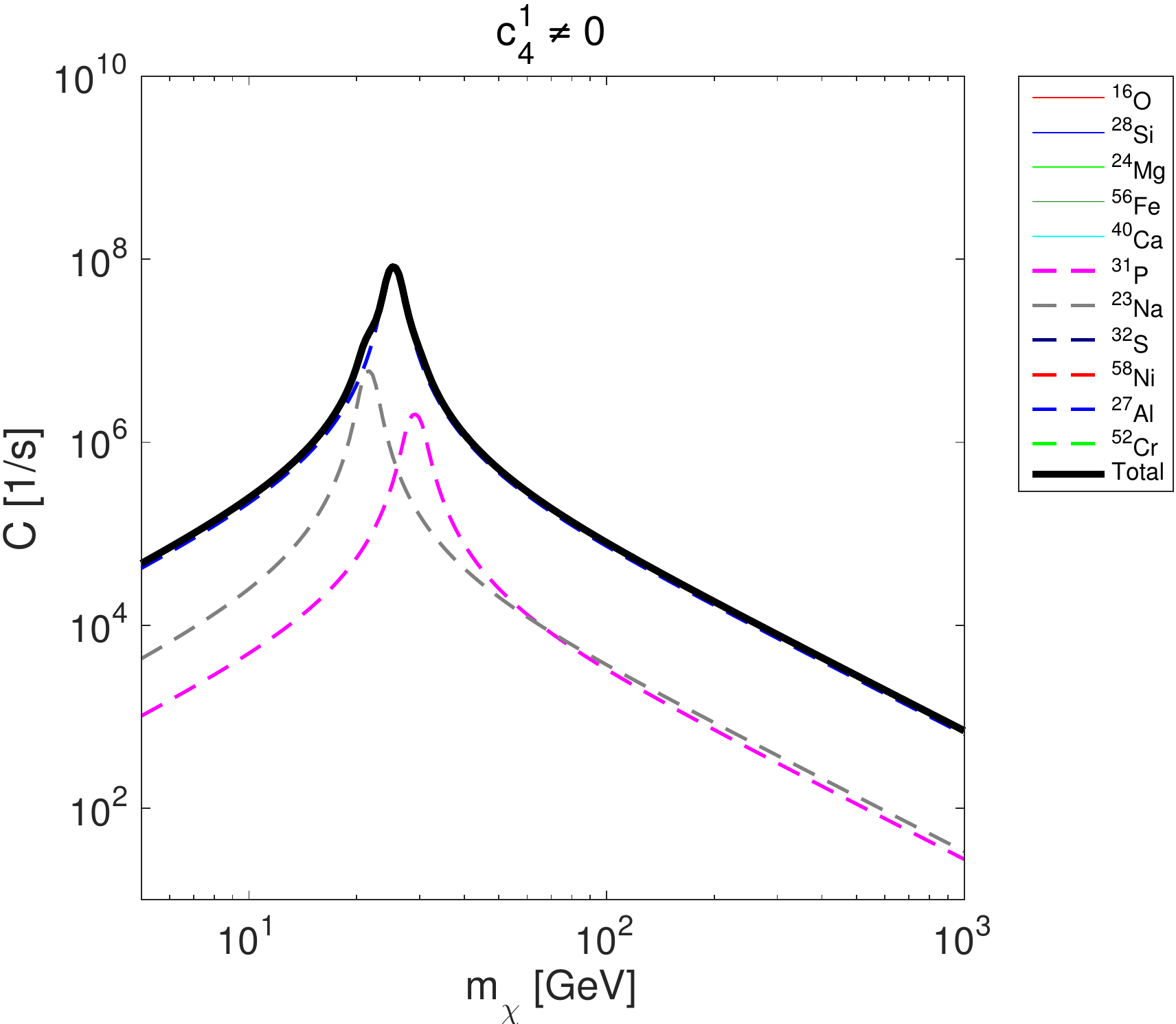}
\end{minipage}
\begin{minipage}[t]{0.49\linewidth}
\centering
\includegraphics[width=\textwidth]{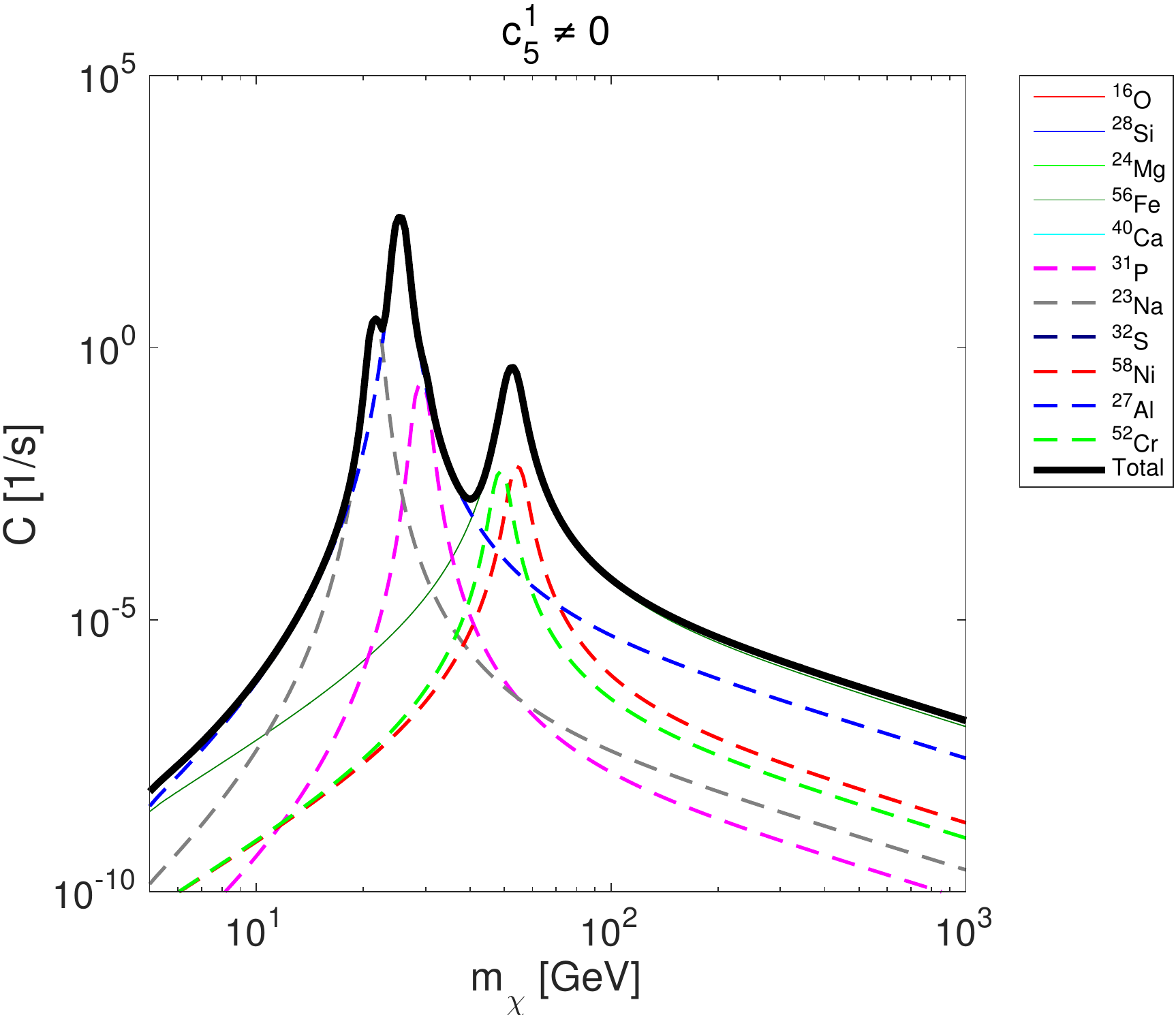}
\end{minipage}
\begin{minipage}[t]{0.49\linewidth}
\centering
\includegraphics[width=\textwidth]{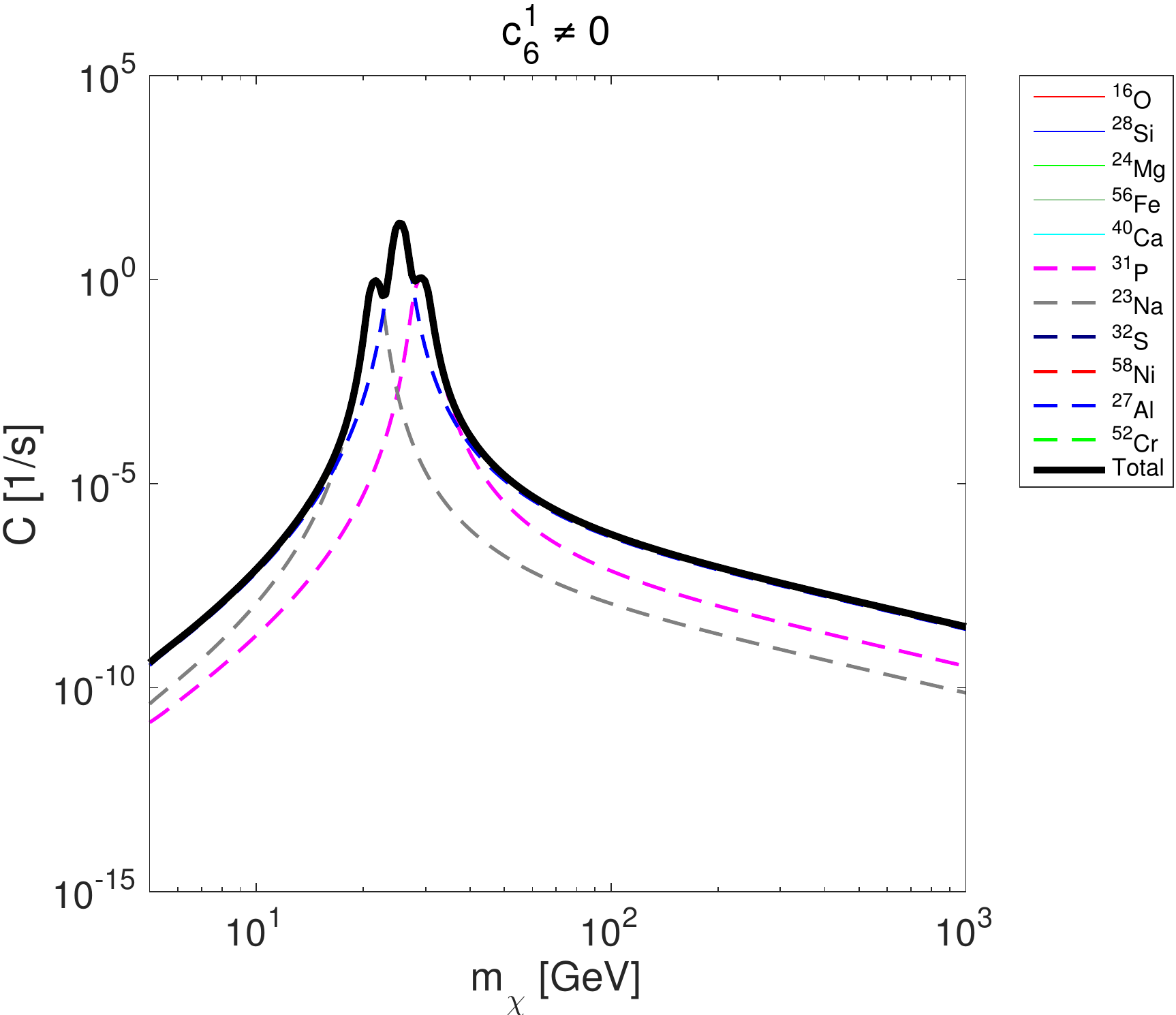}
\end{minipage}
\begin{minipage}[t]{0.49\linewidth}
\centering
\includegraphics[width=\textwidth]{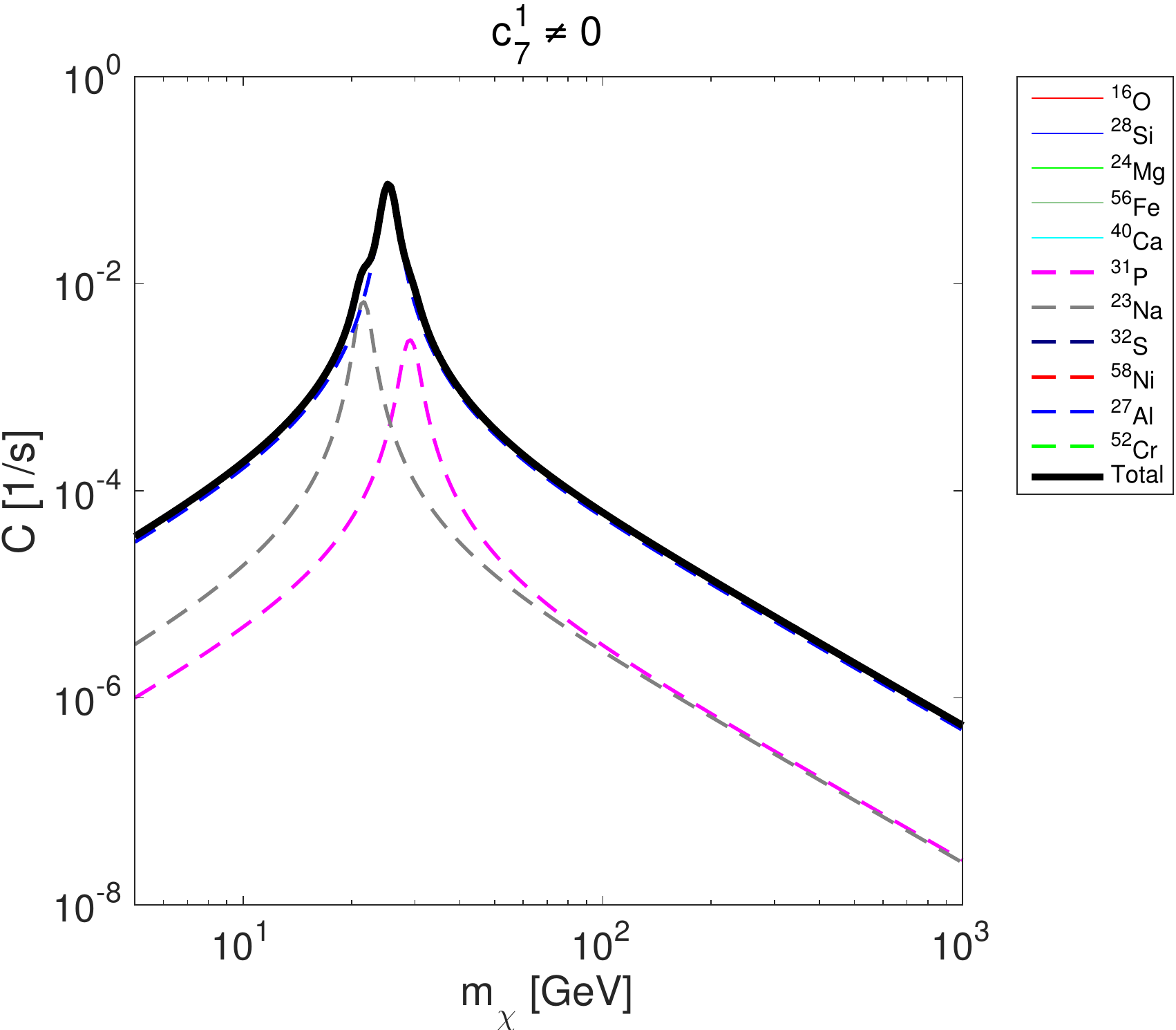}
\end{minipage}
\end{center}
\caption{Rate of WIMP capture in the Earth for the interaction operators $\hat{\mathcal{O}}_{15}$ (isoscalar), and $\hat{\mathcal{O}}_9$, $\hat{\mathcal{O}}_{3}$, $\hat{\mathcal{O}}_{4}$, $\hat{\mathcal{O}}_{5}$, $\hat{\mathcal{O}}_{6}$ and $\hat{\mathcal{O}}_{7}$ (isovector).}
\label{fig:ac2}
\end{figure}

\begin{figure}[t]
\begin{center}
\begin{minipage}[t]{0.49\linewidth}
\centering
\includegraphics[width=\textwidth]{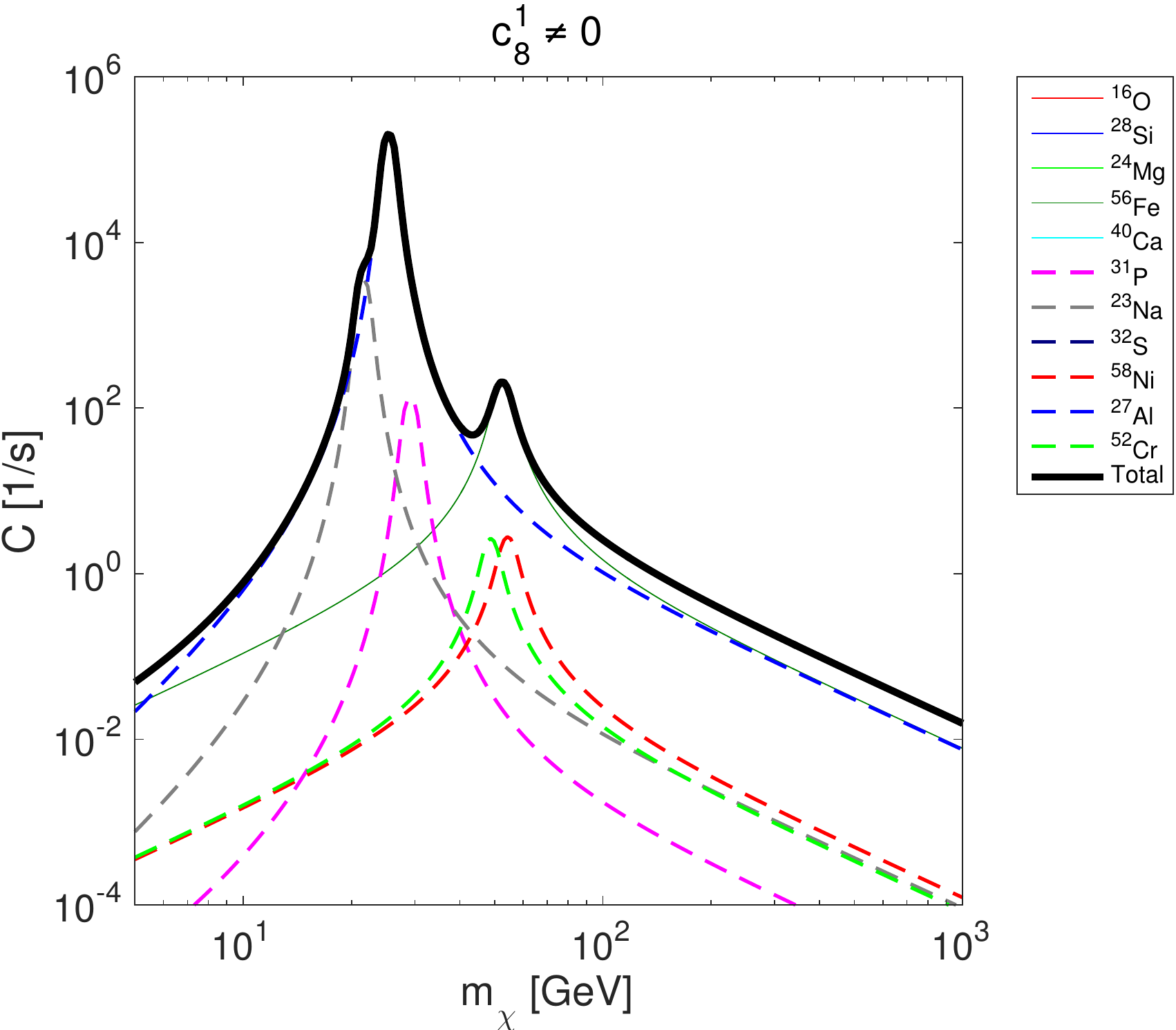}
\end{minipage}
\begin{minipage}[t]{0.49\linewidth}
\centering
\includegraphics[width=\textwidth]{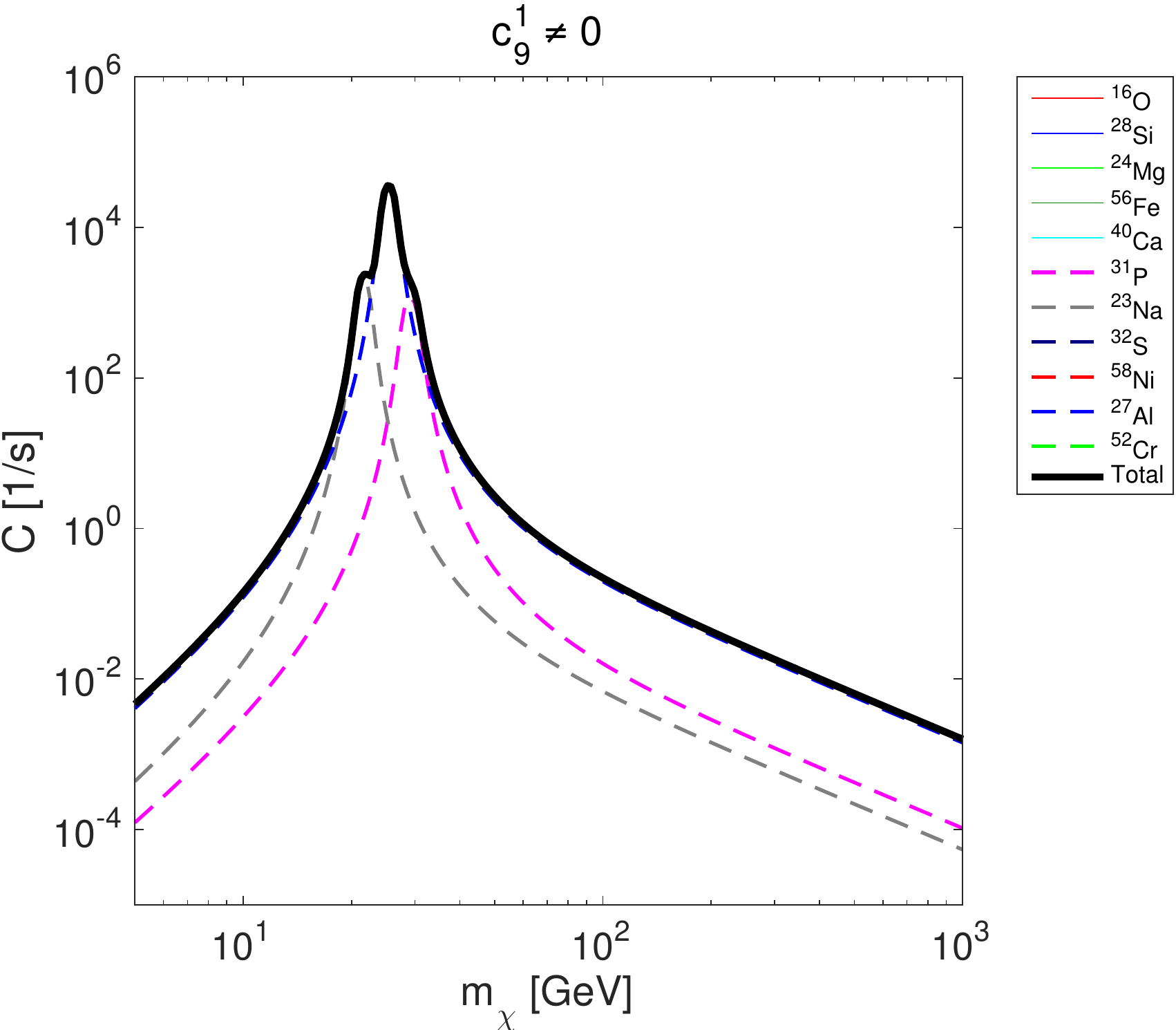}
\end{minipage}
\begin{minipage}[t]{0.49\linewidth}
\centering
\includegraphics[width=\textwidth]{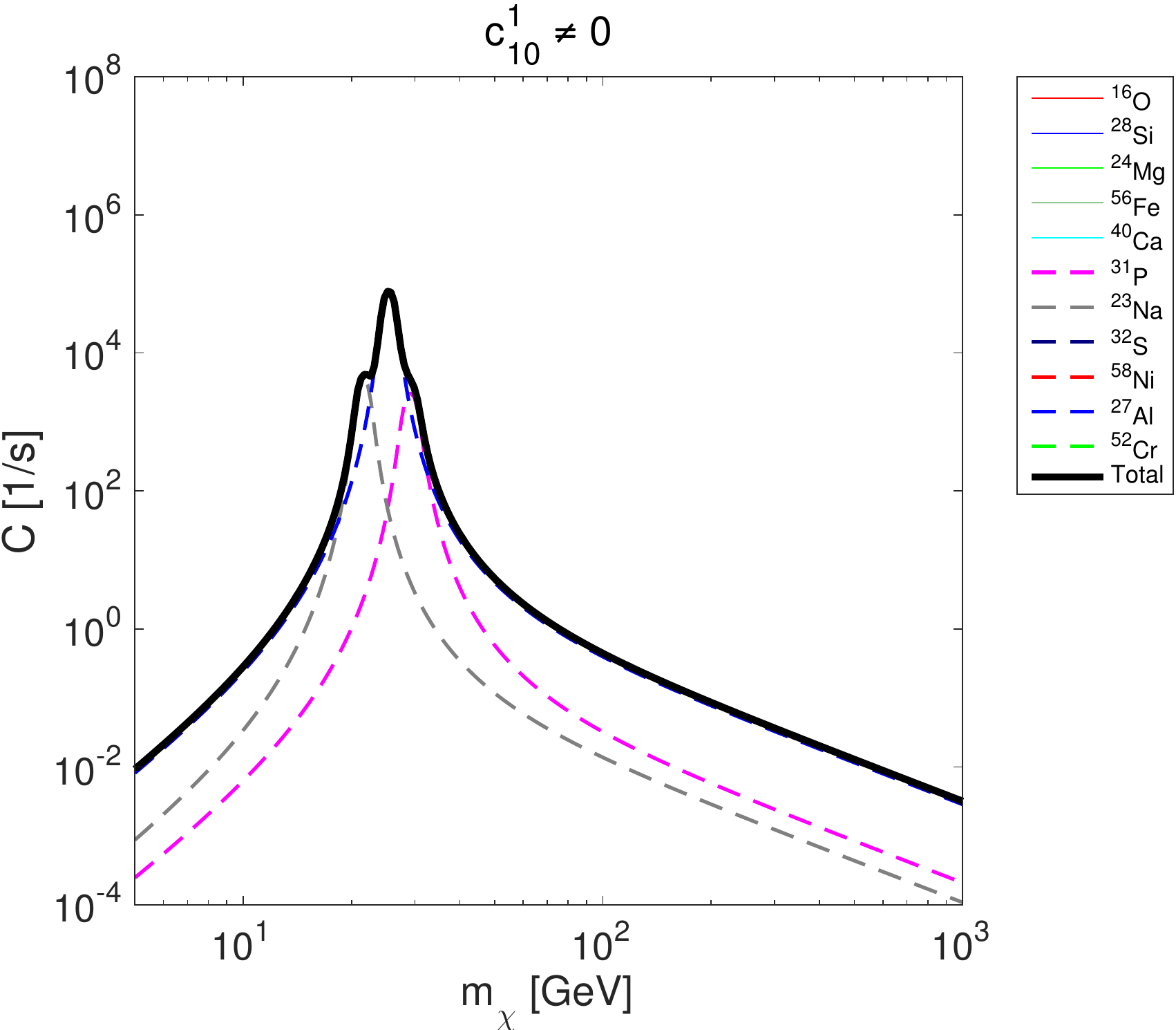}
\end{minipage}
\begin{minipage}[t]{0.49\linewidth}
\centering
\includegraphics[width=\textwidth]{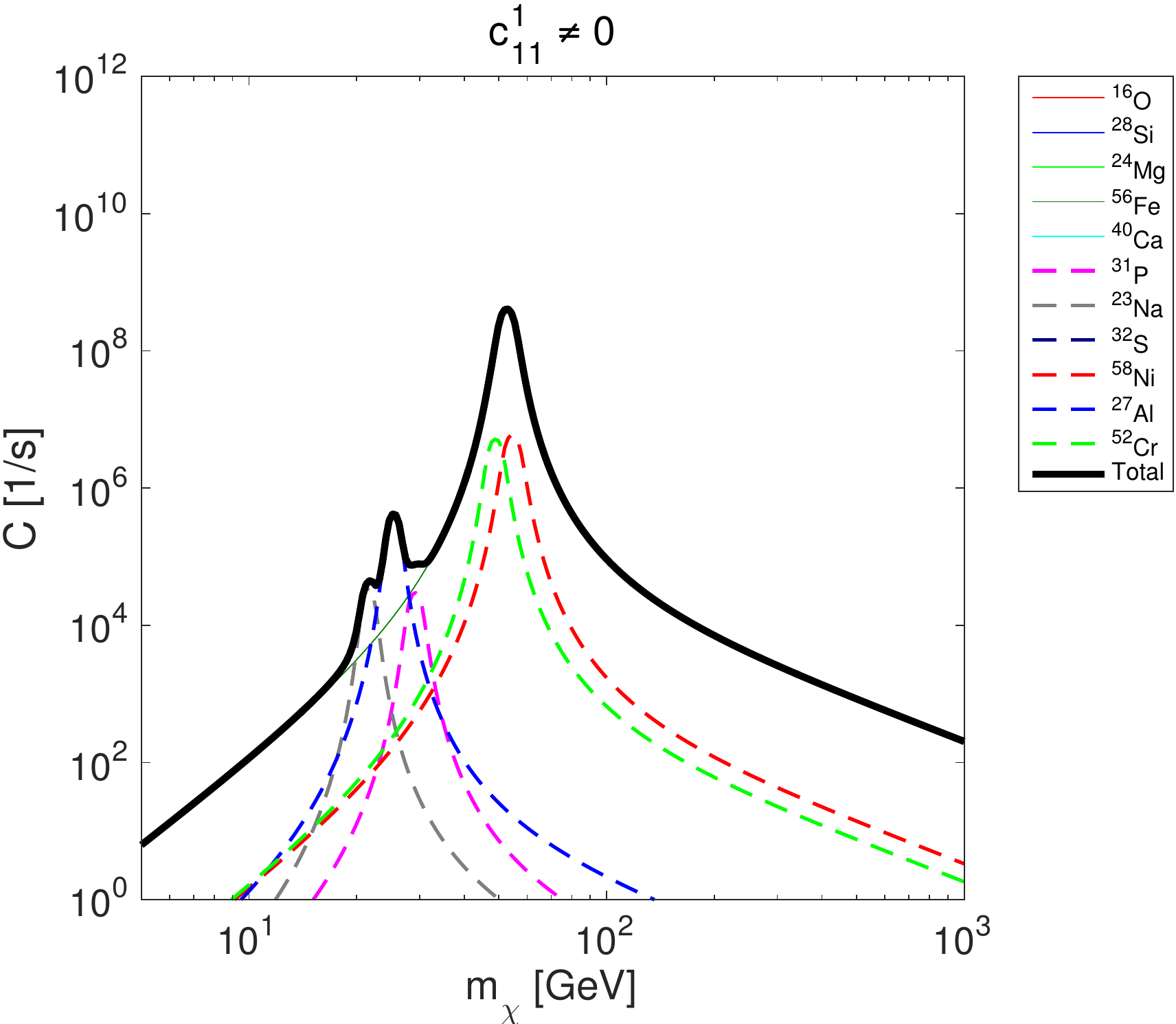}
\end{minipage}
\begin{minipage}[t]{0.49\linewidth}
\centering
\includegraphics[width=\textwidth]{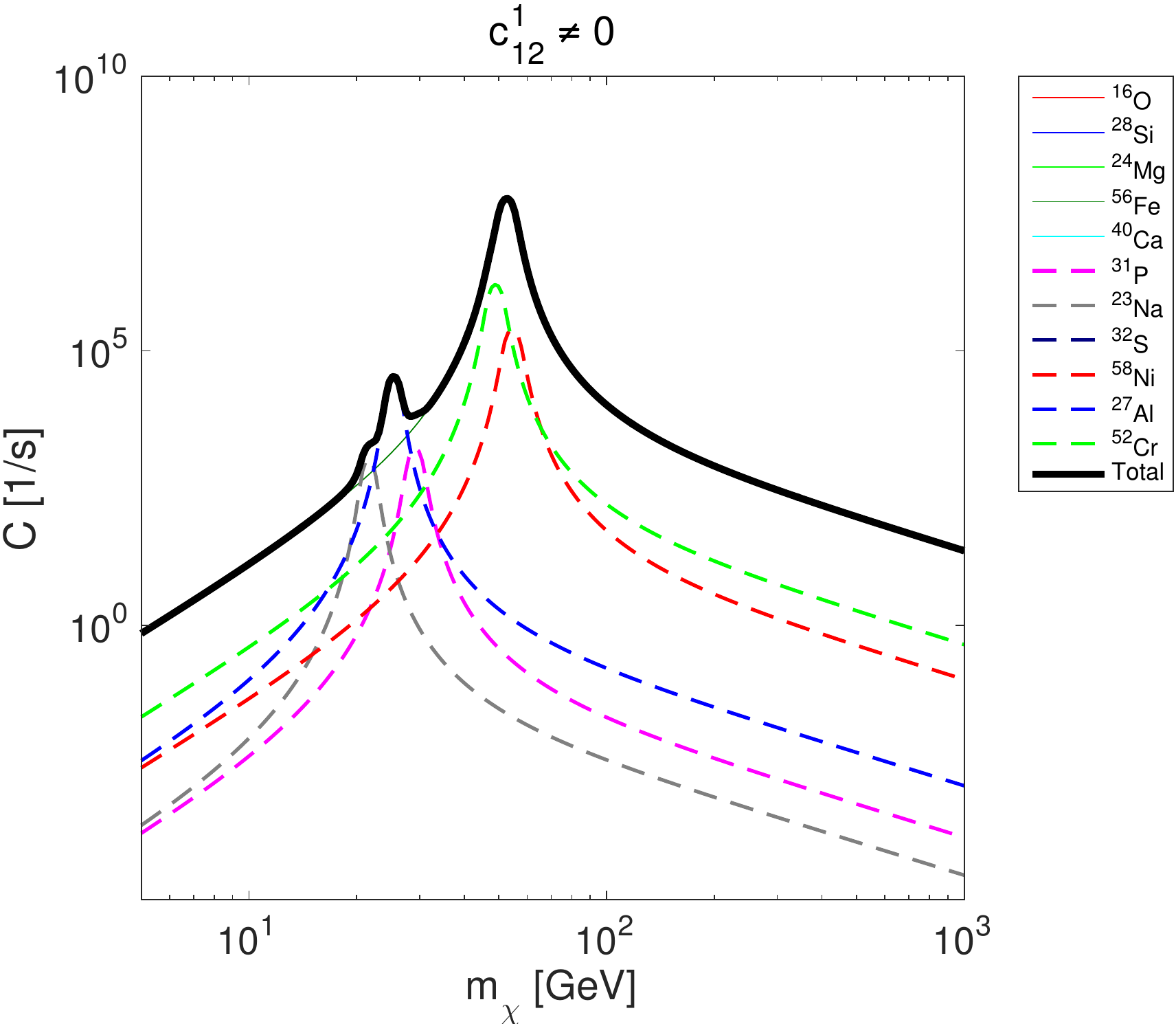}
\end{minipage}
\begin{minipage}[t]{0.49\linewidth}
\centering
\includegraphics[width=\textwidth]{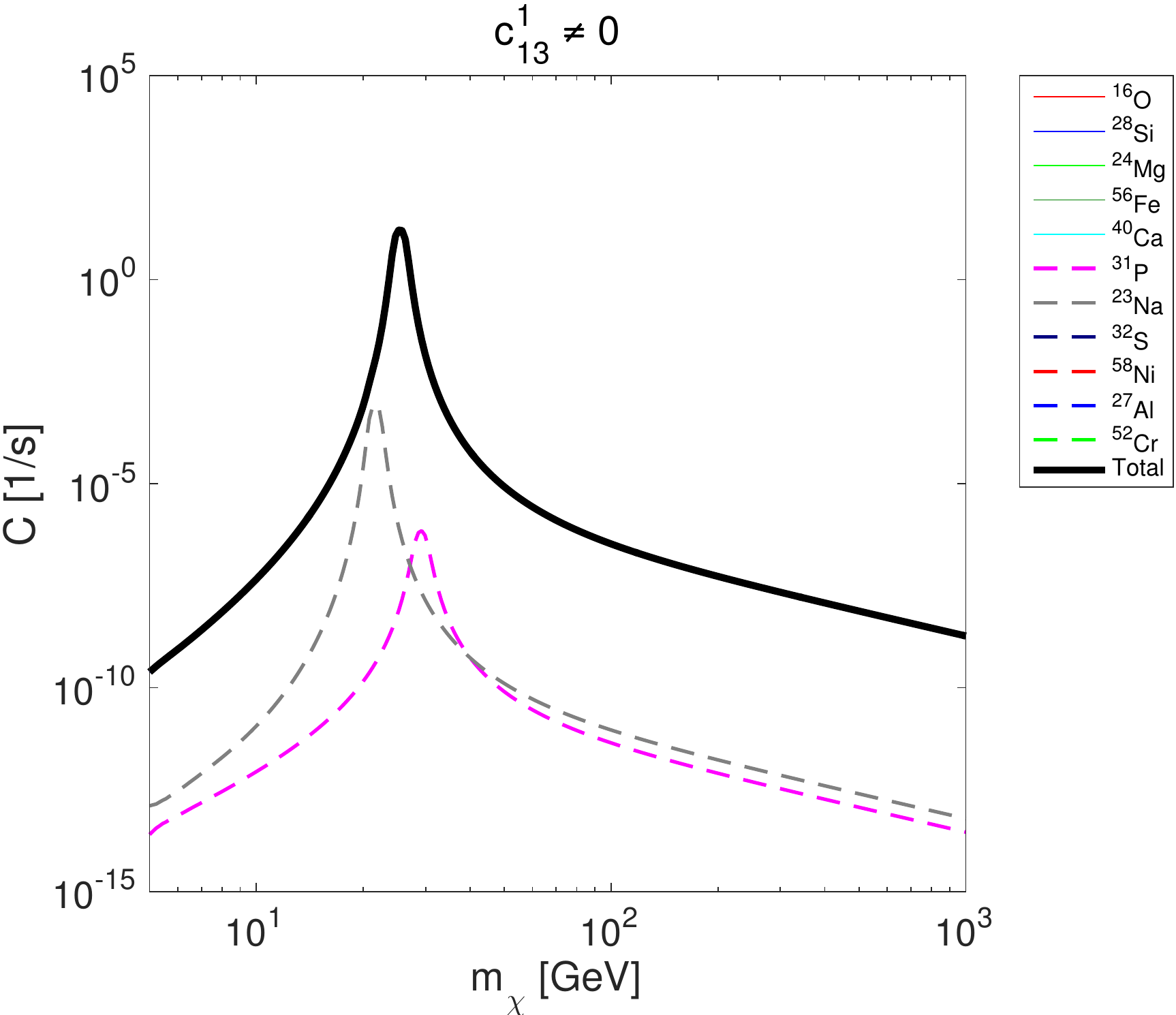}
\end{minipage}
\end{center}
\caption{Rate of WIMP capture in the Earth for the isovector component of the interaction operators $\hat{\mathcal{O}}_8$, $\hat{\mathcal{O}}_9$, $\hat{\mathcal{O}}_{10}$, $\hat{\mathcal{O}}_{11}$, $\hat{\mathcal{O}}_{12}$, $\hat{\mathcal{O}}_{13}$.}
\label{fig:ac3}
\end{figure}

\begin{figure}[t]
\begin{center}
\begin{minipage}[t]{0.45\linewidth}
\centering
\includegraphics[width=\textwidth]{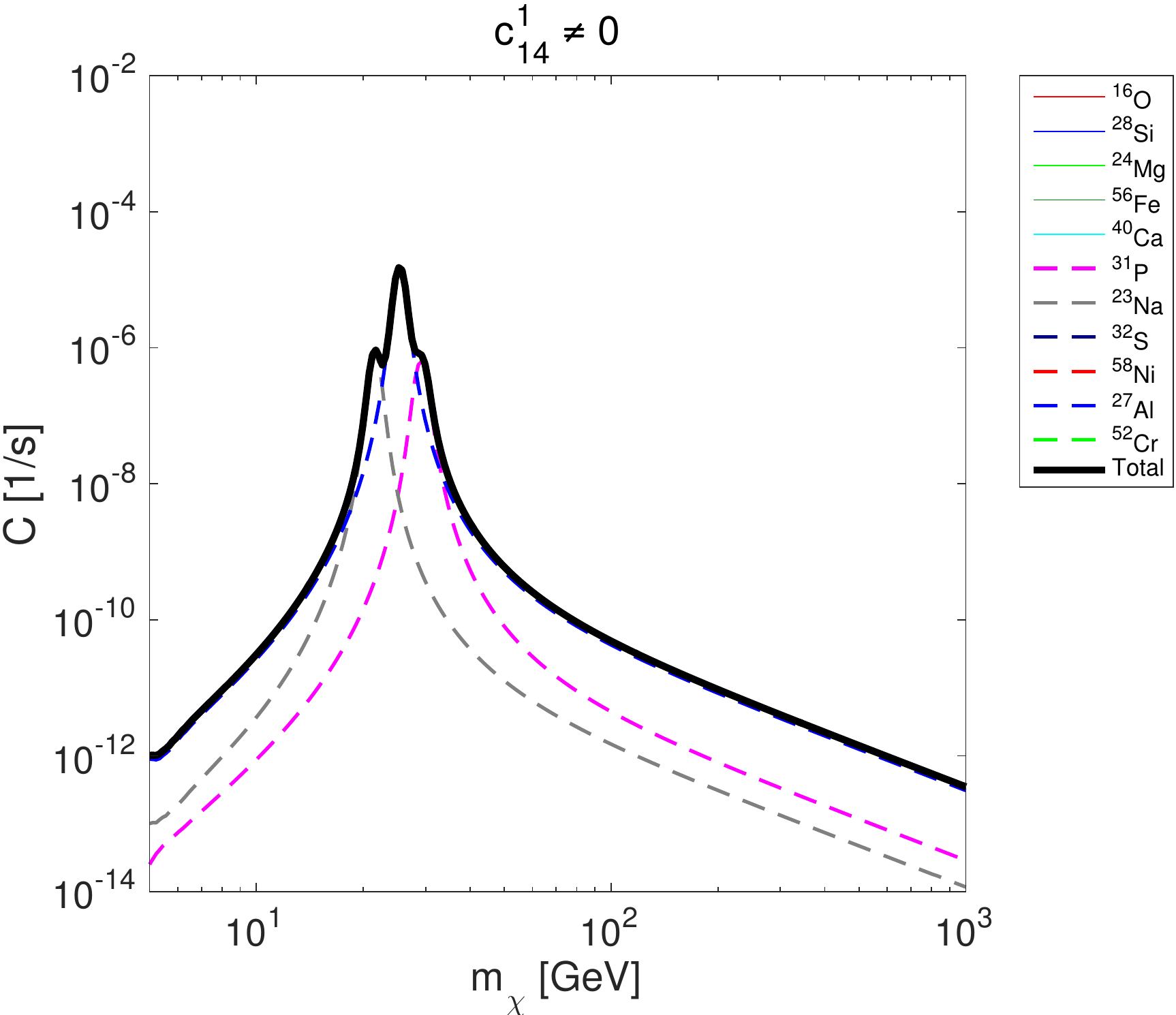}
\end{minipage}
\begin{minipage}[t]{0.45\linewidth}
\centering
\includegraphics[width=\textwidth]{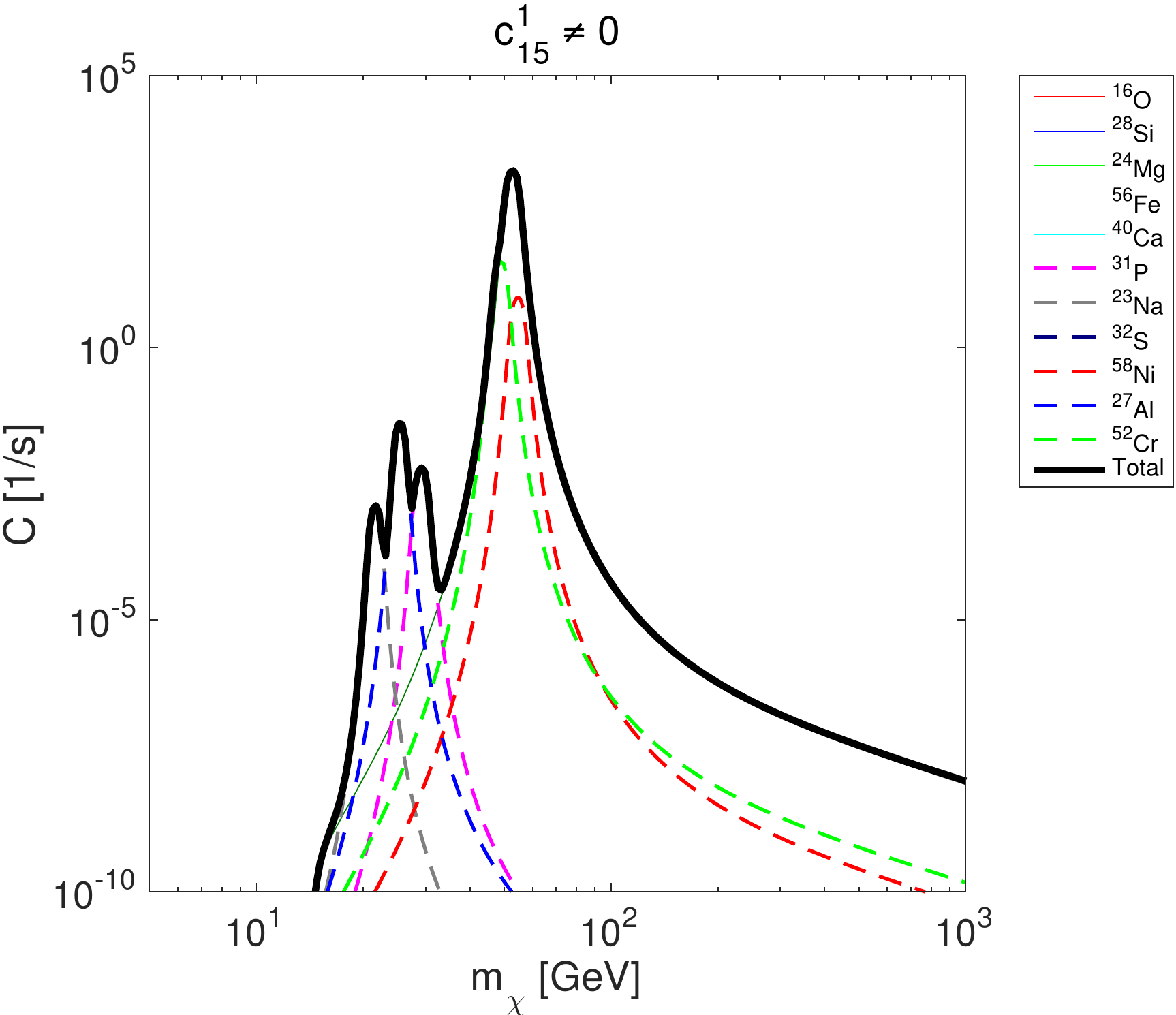}
\end{minipage}
\end{center}
\caption{Rate of WIMP capture in the Earth for the isovector component of the interaction operators $\hat{\mathcal{O}}_{14}$ and $\hat{\mathcal{O}}_{15}$.}
\label{fig:ac4}
\end{figure}

In this last section of the paper I collect figures for capture rates and exclusion limits relative to interaction operators which -- for brevity -- were not considered in Sec.~\ref{sec:results}.~I list them here for completeness.~More specifically:~Fig.~\ref{fig:ac1} shows the rate of WIMP capture in the Earth for the isoscalar component of the interaction operators $\hat{\mathcal{O}}_7$, $\hat{\mathcal{O}}_9$, $\hat{\mathcal{O}}_{10}$, $\hat{\mathcal{O}}_{11}$, $\hat{\mathcal{O}}_{12}$, $\hat{\mathcal{O}}_{14}$;~Fig.~\ref{fig:ac2} shows the rate of WIMP capture in the Earth for the isoscalar component of the interaction operator $\hat{\mathcal{O}}_{15}$, and for the isovector component of the interaction operators $\hat{\mathcal{O}}_9$, $\hat{\mathcal{O}}_{3}$, $\hat{\mathcal{O}}_{4}$, $\hat{\mathcal{O}}_{5}$, $\hat{\mathcal{O}}_{6}$ and $\hat{\mathcal{O}}_{7}$;~Fig.~\ref{fig:ac3} shows the rate of WIMP capture in the Earth for the isovector component of the interaction operators $\hat{\mathcal{O}}_8$, $\hat{\mathcal{O}}_9$, $\hat{\mathcal{O}}_{10}$, $\hat{\mathcal{O}}_{11}$, $\hat{\mathcal{O}}_{12}$, $\hat{\mathcal{O}}_{13}$; and, finally, Fig.~\ref{fig:ac4} shows the rate of WIMP capture in the Earth for the isovector component of the interaction operators $\hat{\mathcal{O}}_{14}$ and $\hat{\mathcal{O}}_{15}$.

\begin{figure}[t]
\begin{center}
\begin{minipage}[t]{0.4\linewidth}
\centering
\includegraphics[width=\textwidth]{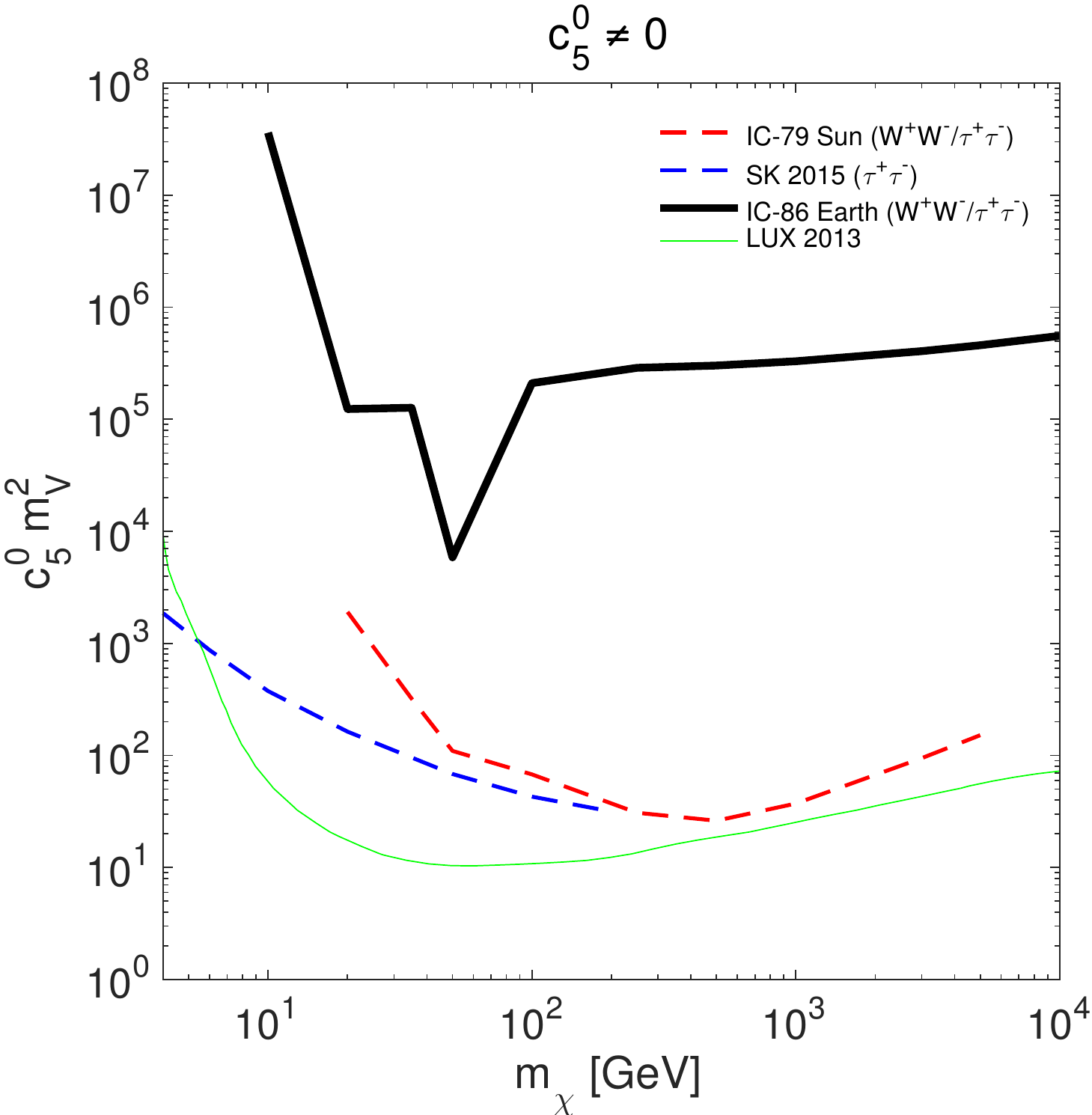}
\end{minipage}
\begin{minipage}[t]{0.4\linewidth}
\centering
\includegraphics[width=\textwidth]{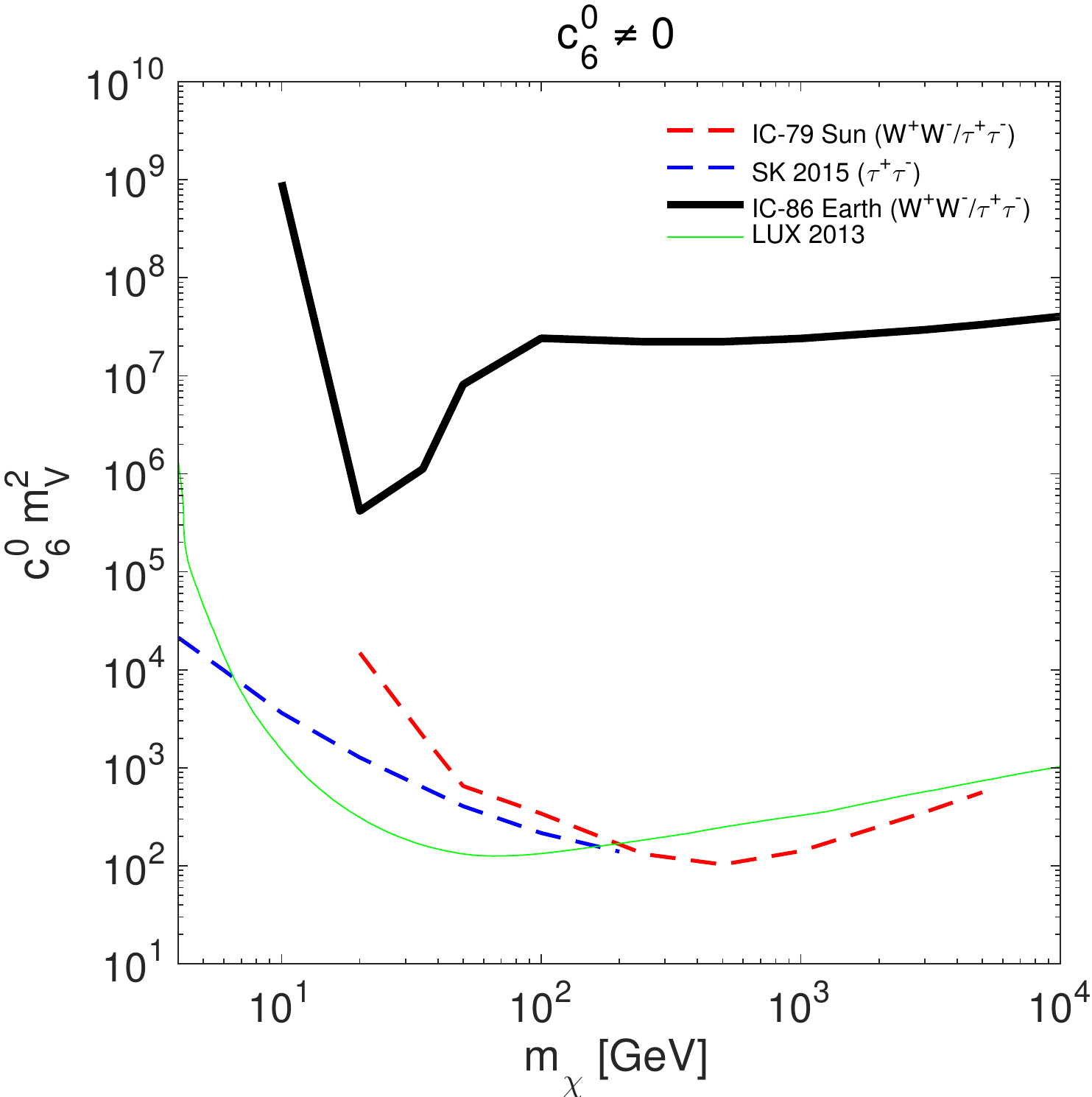}
\end{minipage}
\begin{minipage}[t]{0.4\linewidth}
\centering
\includegraphics[width=\textwidth]{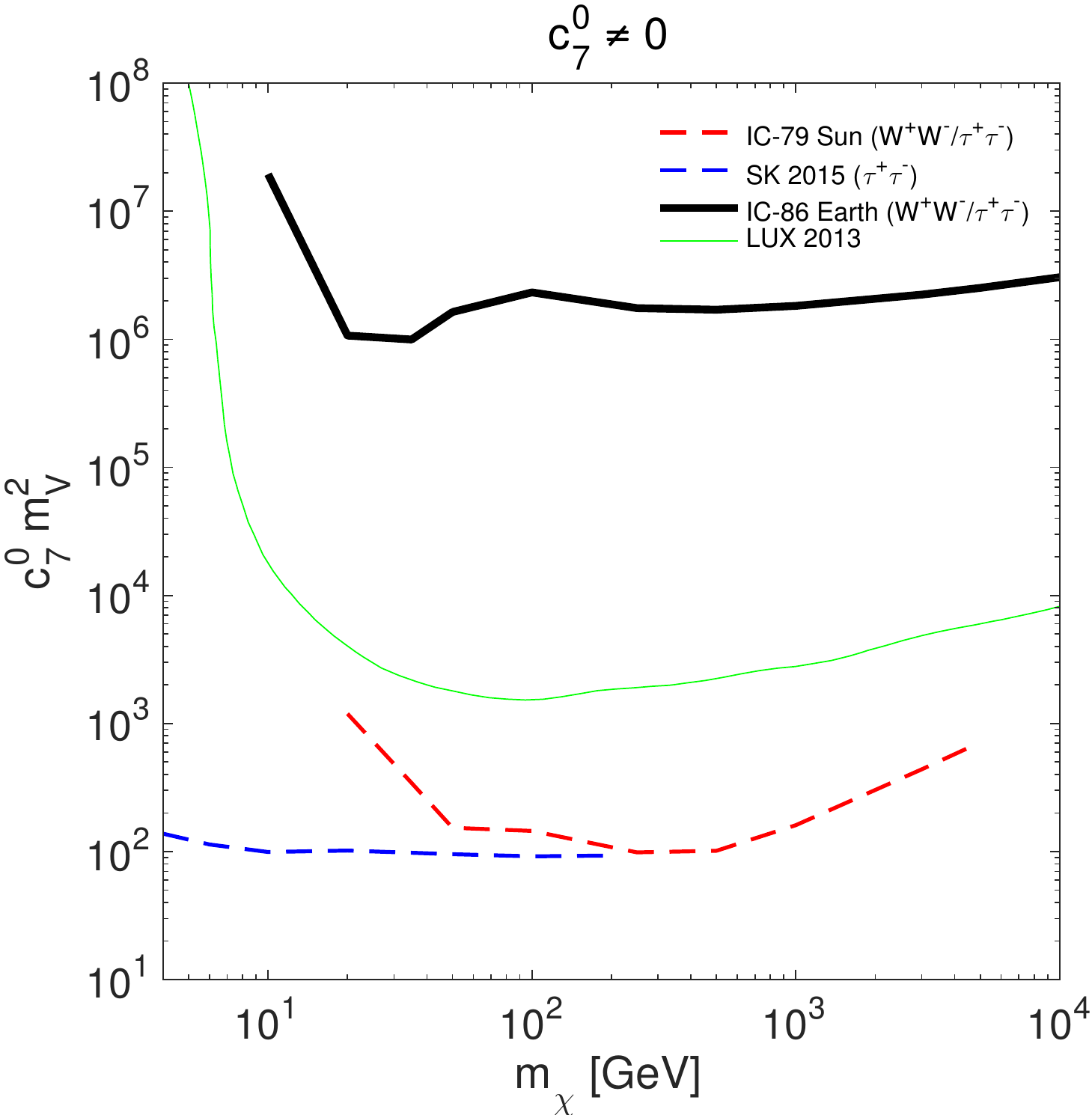}
\end{minipage}
\begin{minipage}[t]{0.4\linewidth}
\centering
\includegraphics[width=\textwidth]{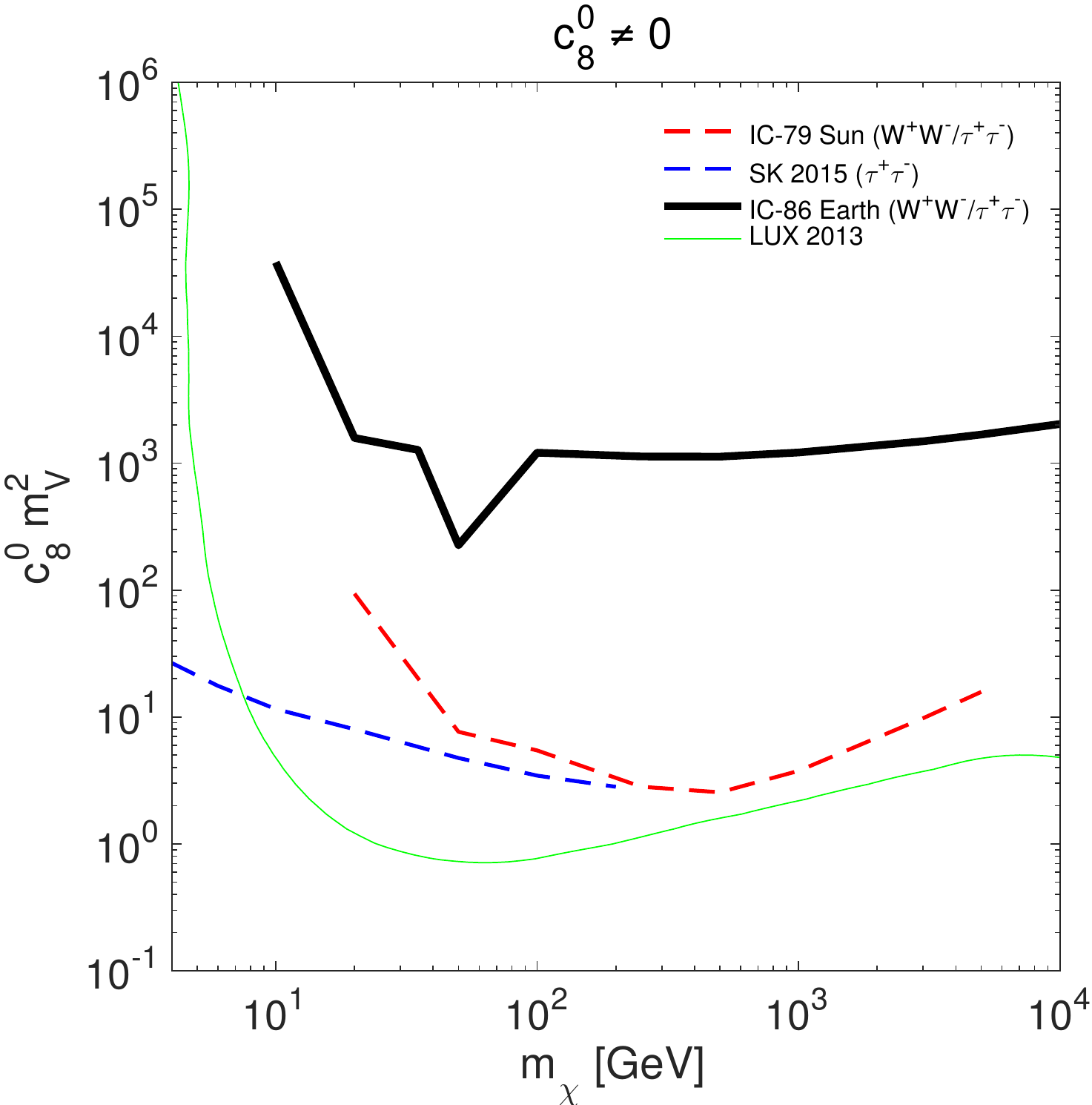}
\end{minipage}
\end{center}
\caption{90\% CL upper limits on the isoscalar coupling constants of the interaction operators $\hat{\mathcal{O}}_5$, $\hat{\mathcal{O}}_6$, $\hat{\mathcal{O}}_7$, and $\hat{\mathcal{O}}_8$.}
\label{fig:bc1}
\end{figure}

\begin{figure}[t]
\begin{center}
\begin{minipage}[t]{0.45\linewidth}
\centering
\includegraphics[width=\textwidth]{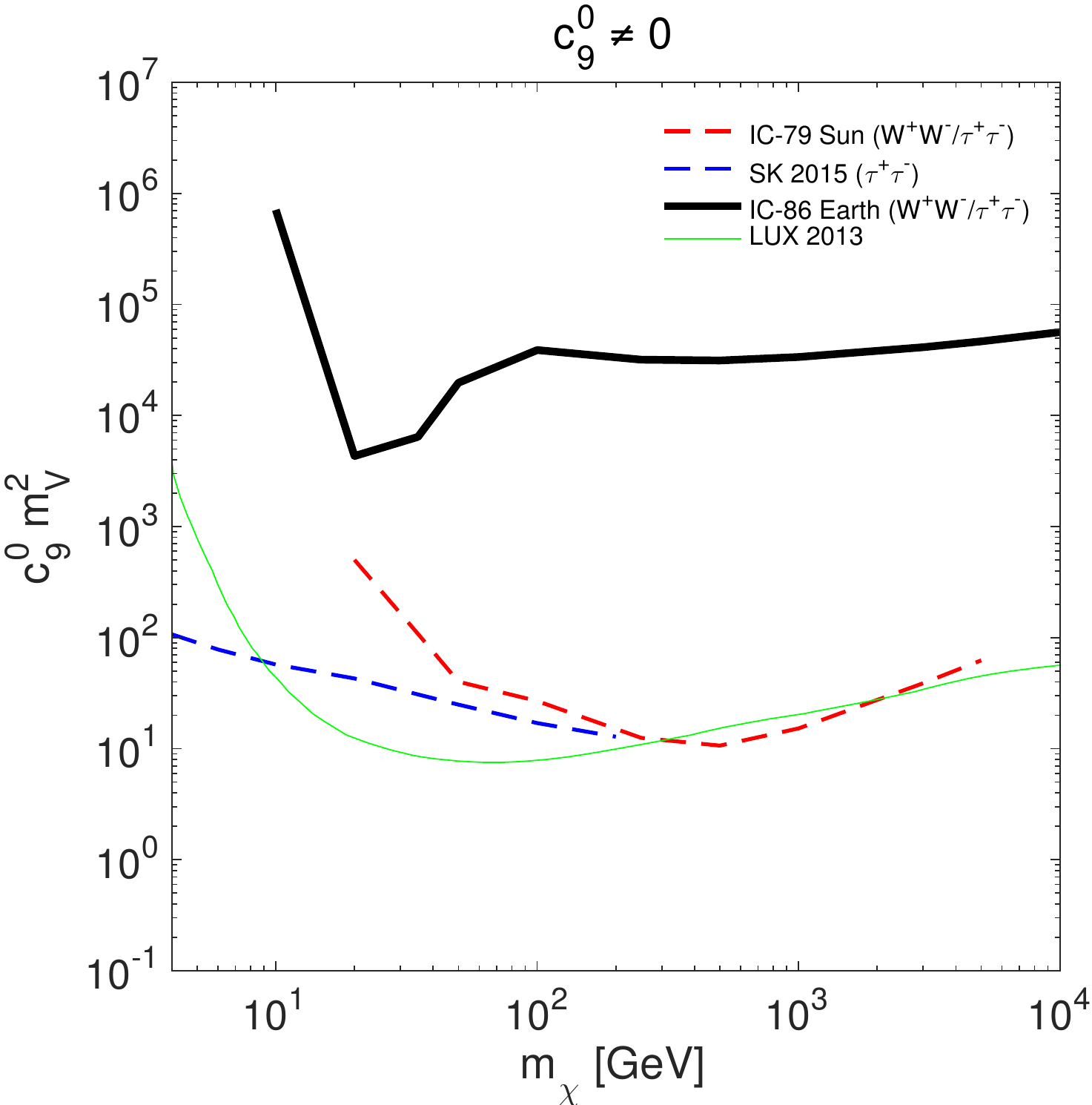}
\end{minipage}
\begin{minipage}[t]{0.45\linewidth}
\centering
\includegraphics[width=\textwidth]{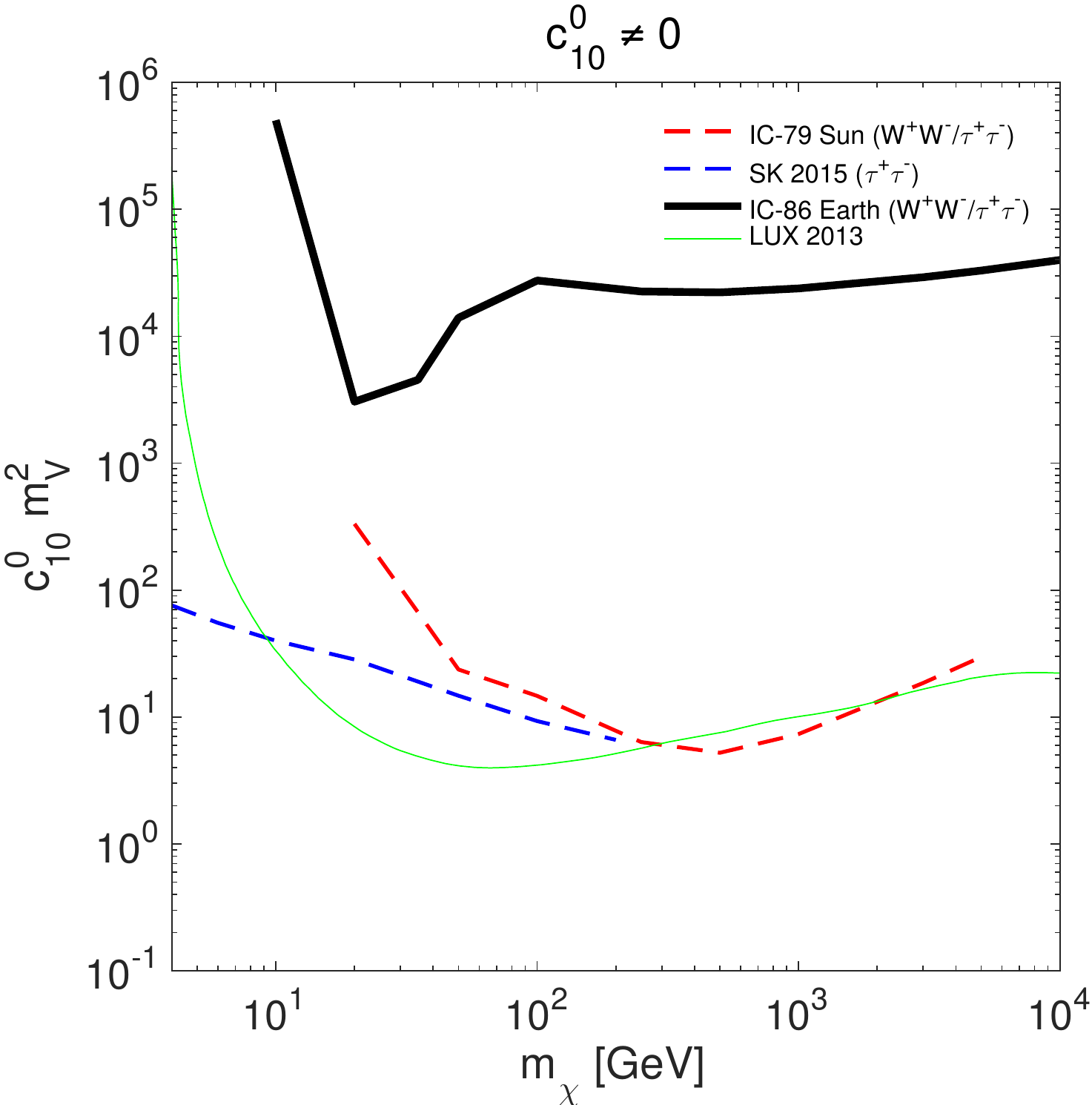}
\end{minipage}
\begin{minipage}[t]{0.45\linewidth}
\centering
\includegraphics[width=\textwidth]{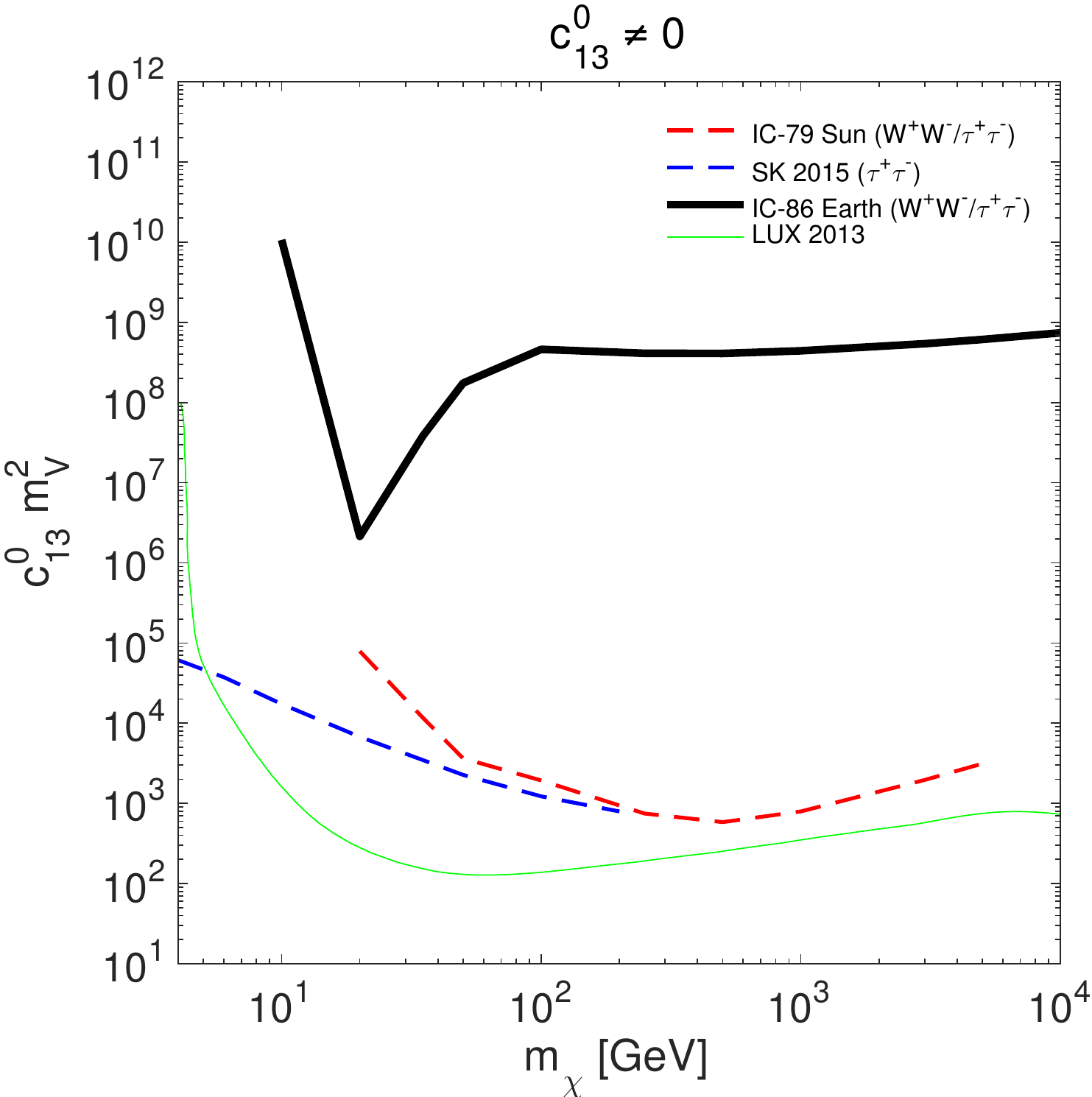}
\end{minipage}
\begin{minipage}[t]{0.45\linewidth}
\centering
\includegraphics[width=\textwidth]{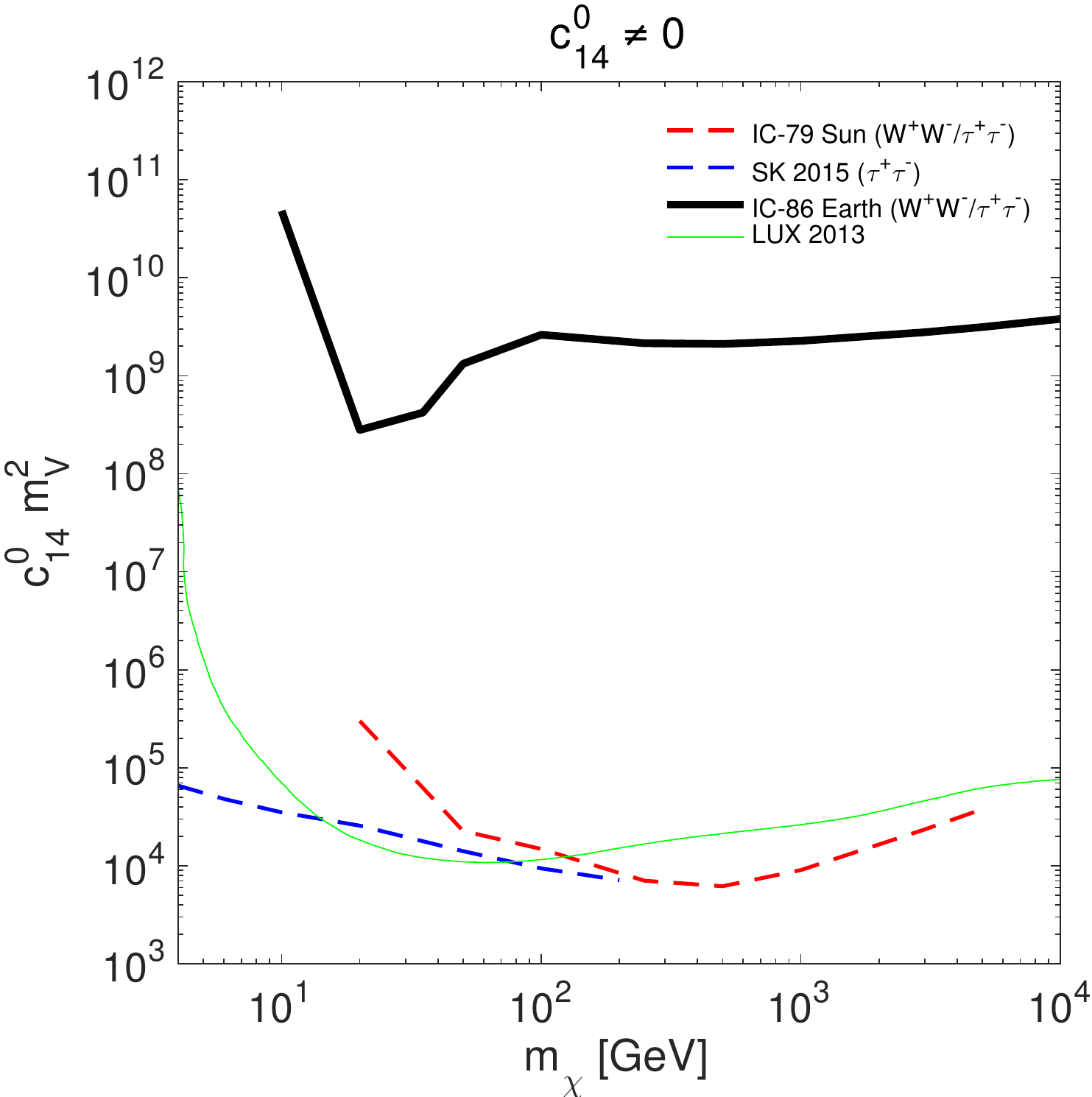}
\end{minipage}
\begin{minipage}[t]{0.45\linewidth}
\centering
\includegraphics[width=\textwidth]{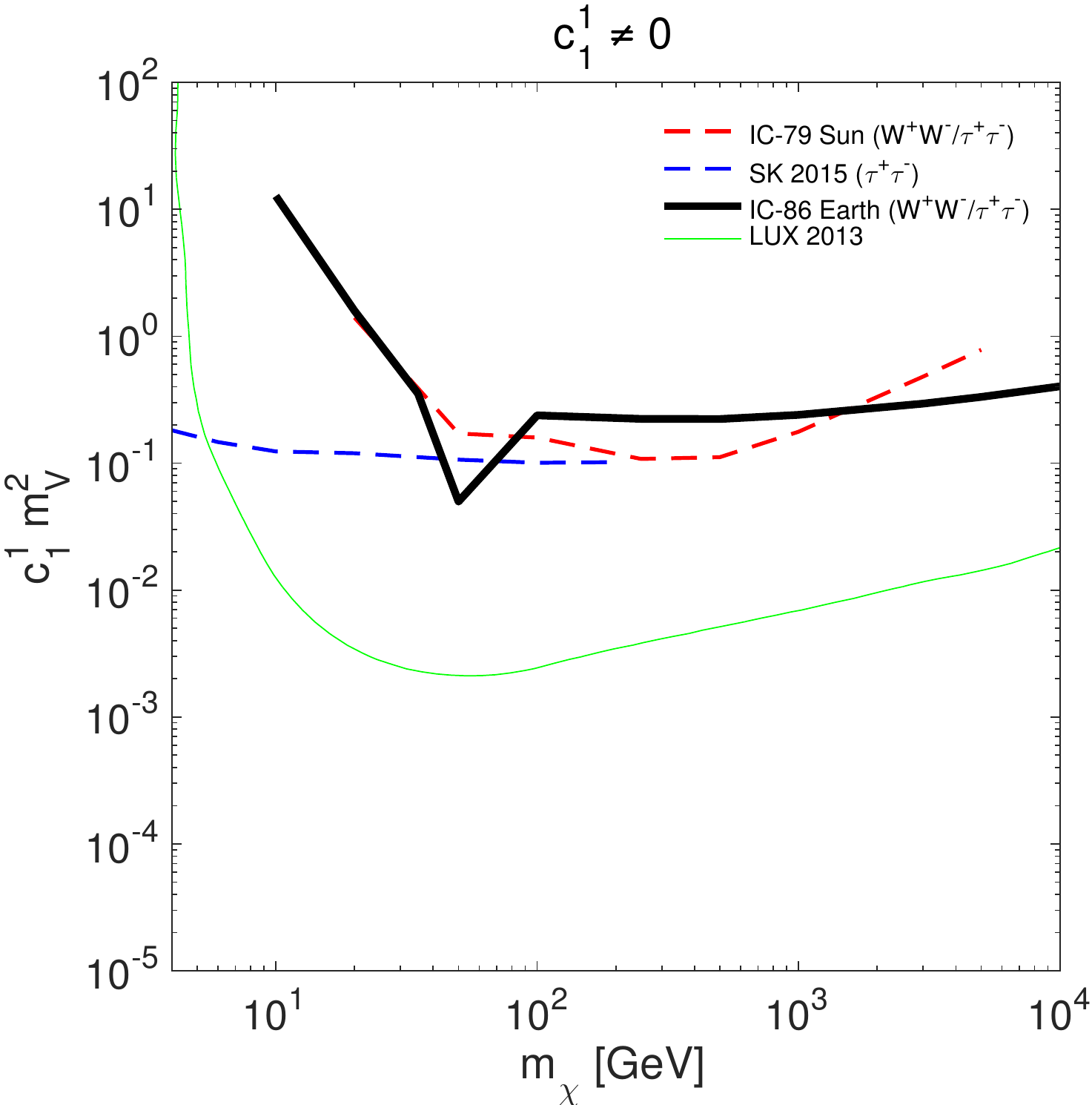}
\end{minipage}
\begin{minipage}[t]{0.45\linewidth}
\centering
\includegraphics[width=\textwidth]{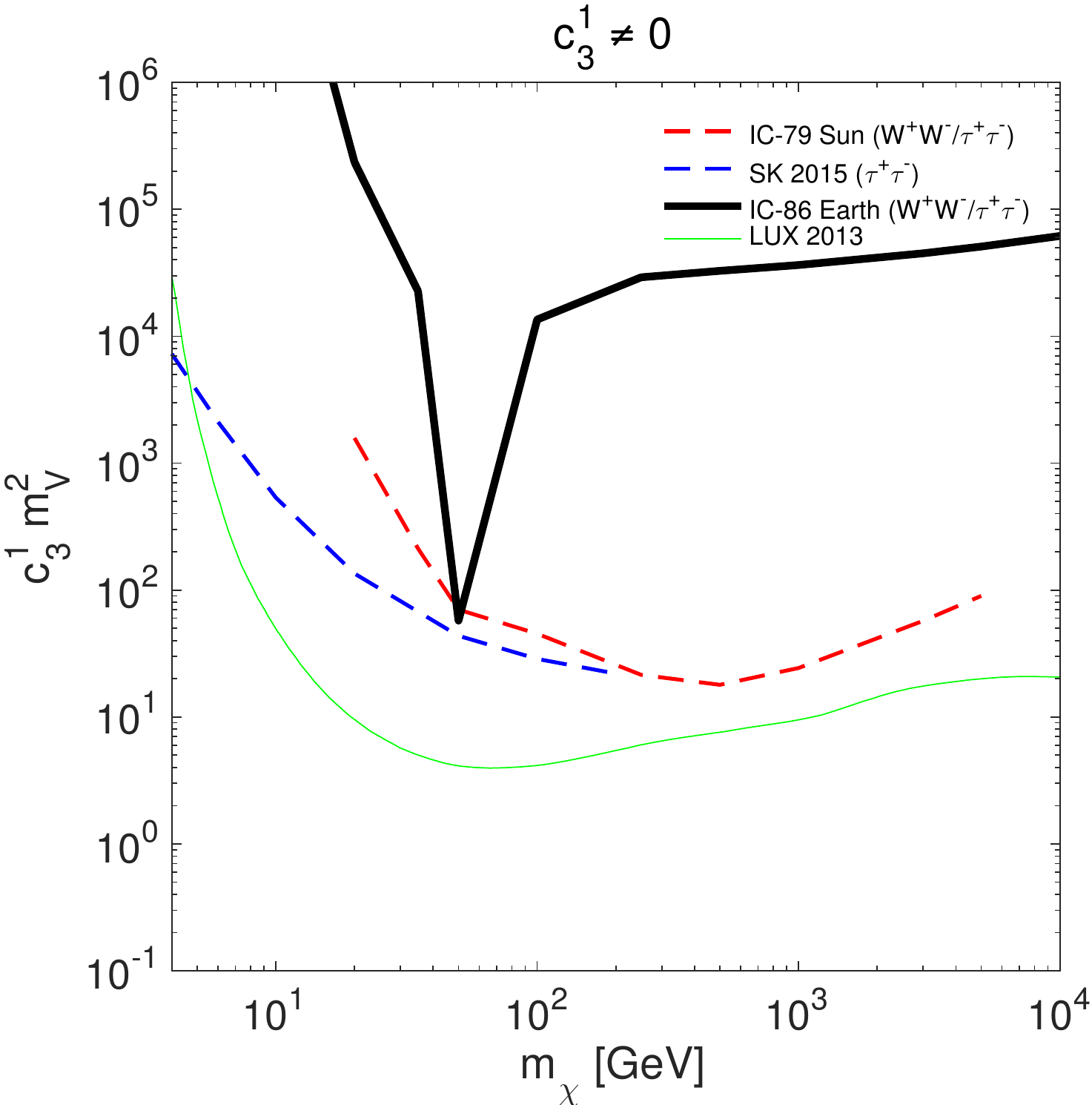}
\end{minipage}
\end{center}
\caption{90\% CL upper limits on the isoscalar coupling constants of $\hat{\mathcal{O}}_9$, $\hat{\mathcal{O}}_{10}$, $\hat{\mathcal{O}}_{13}$, and $\hat{\mathcal{O}}_{14}$, and on the isovector coupling constants of $\hat{\mathcal{O}}_{1}$ and $\hat{\mathcal{O}}_{3}$.}
\label{fig:bc2}
\end{figure}

\begin{figure}[t]
\begin{center}
\begin{minipage}[t]{0.45\linewidth}
\centering
\includegraphics[width=\textwidth]{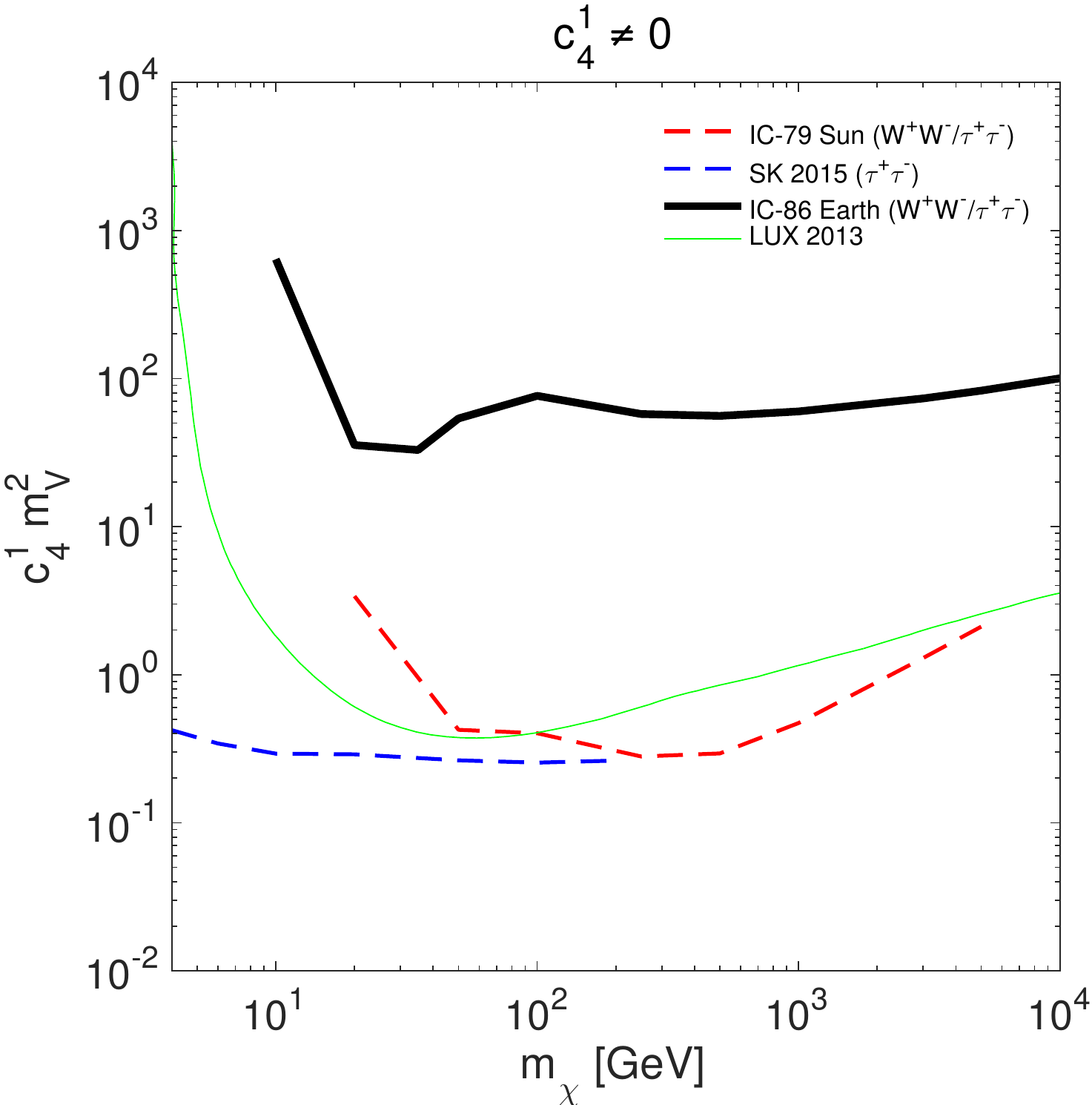}
\end{minipage}
\begin{minipage}[t]{0.45\linewidth}
\centering
\includegraphics[width=\textwidth]{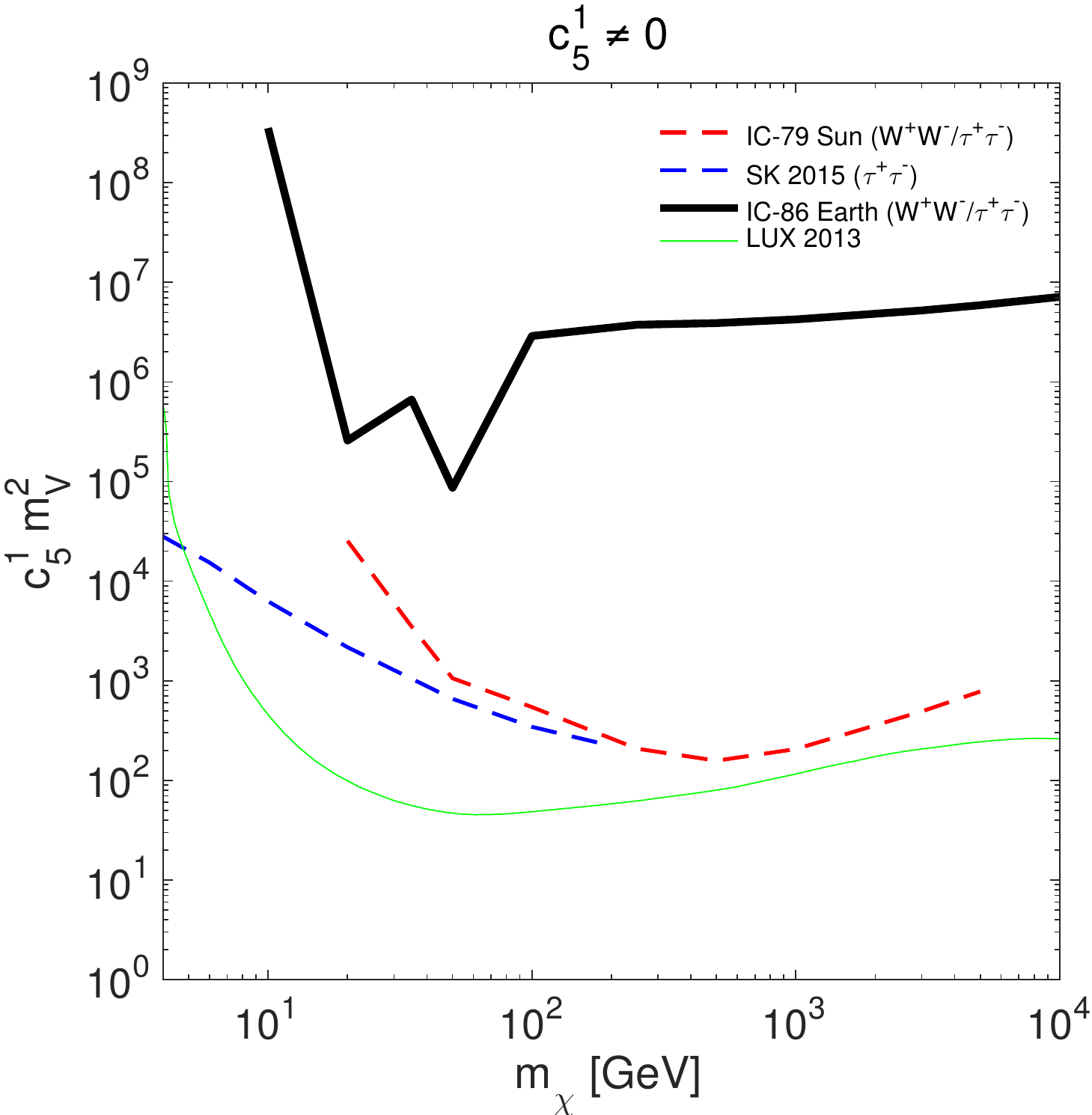}
\end{minipage}
\begin{minipage}[t]{0.45\linewidth}
\centering
\includegraphics[width=\textwidth]{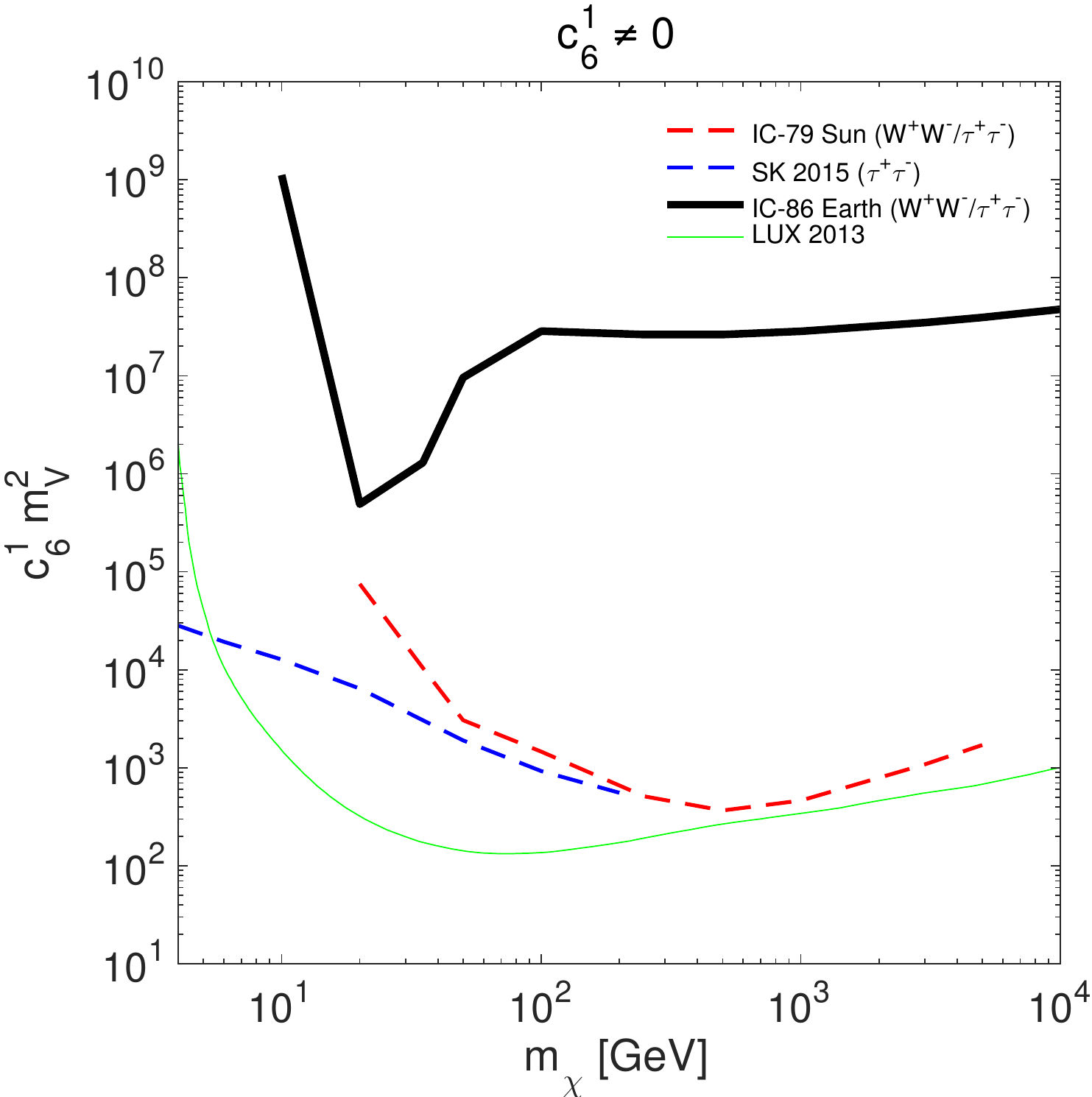}
\end{minipage}
\begin{minipage}[t]{0.45\linewidth}
\centering
\includegraphics[width=\textwidth]{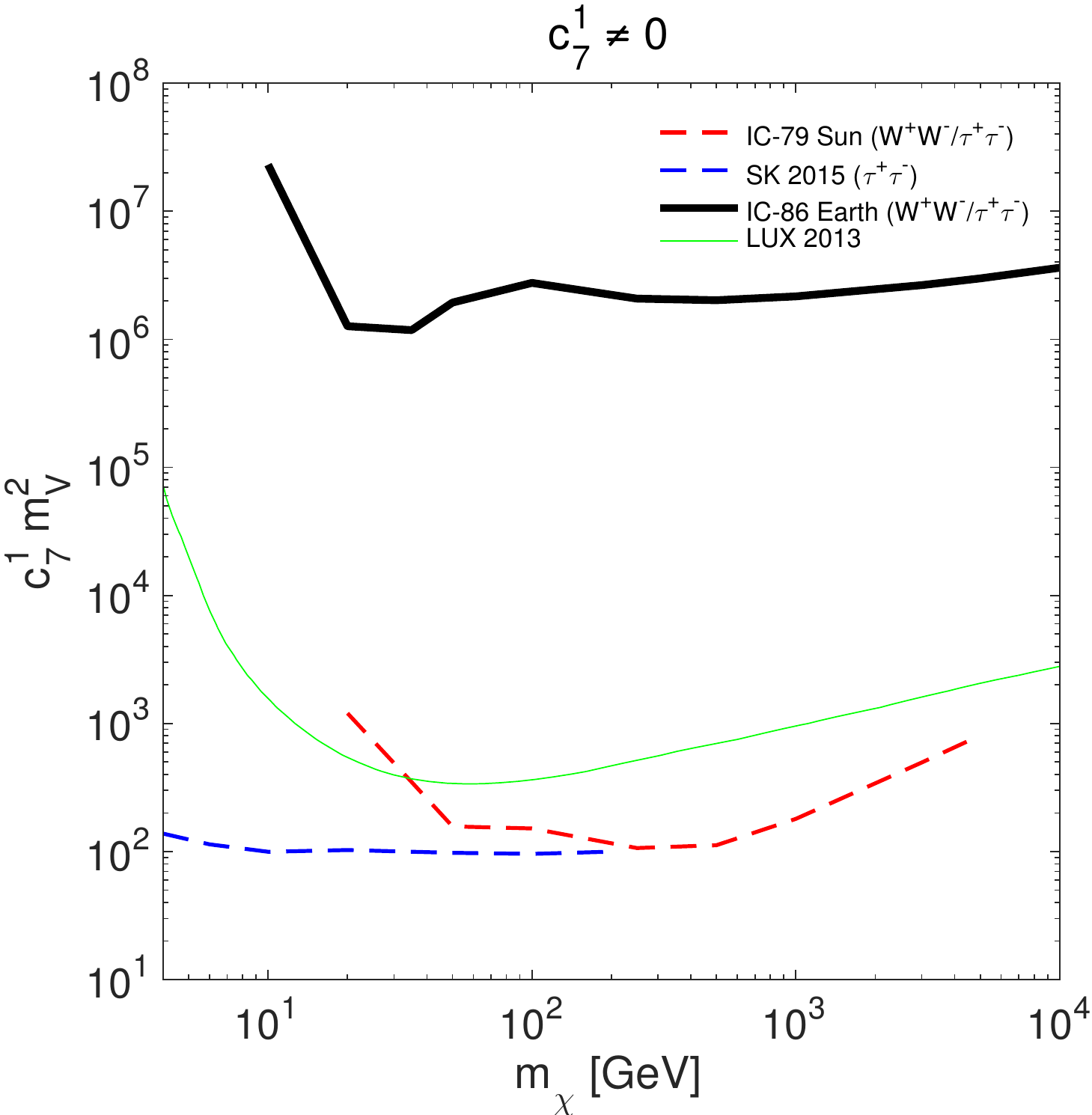}
\end{minipage}
\begin{minipage}[t]{0.45\linewidth}
\centering
\includegraphics[width=\textwidth]{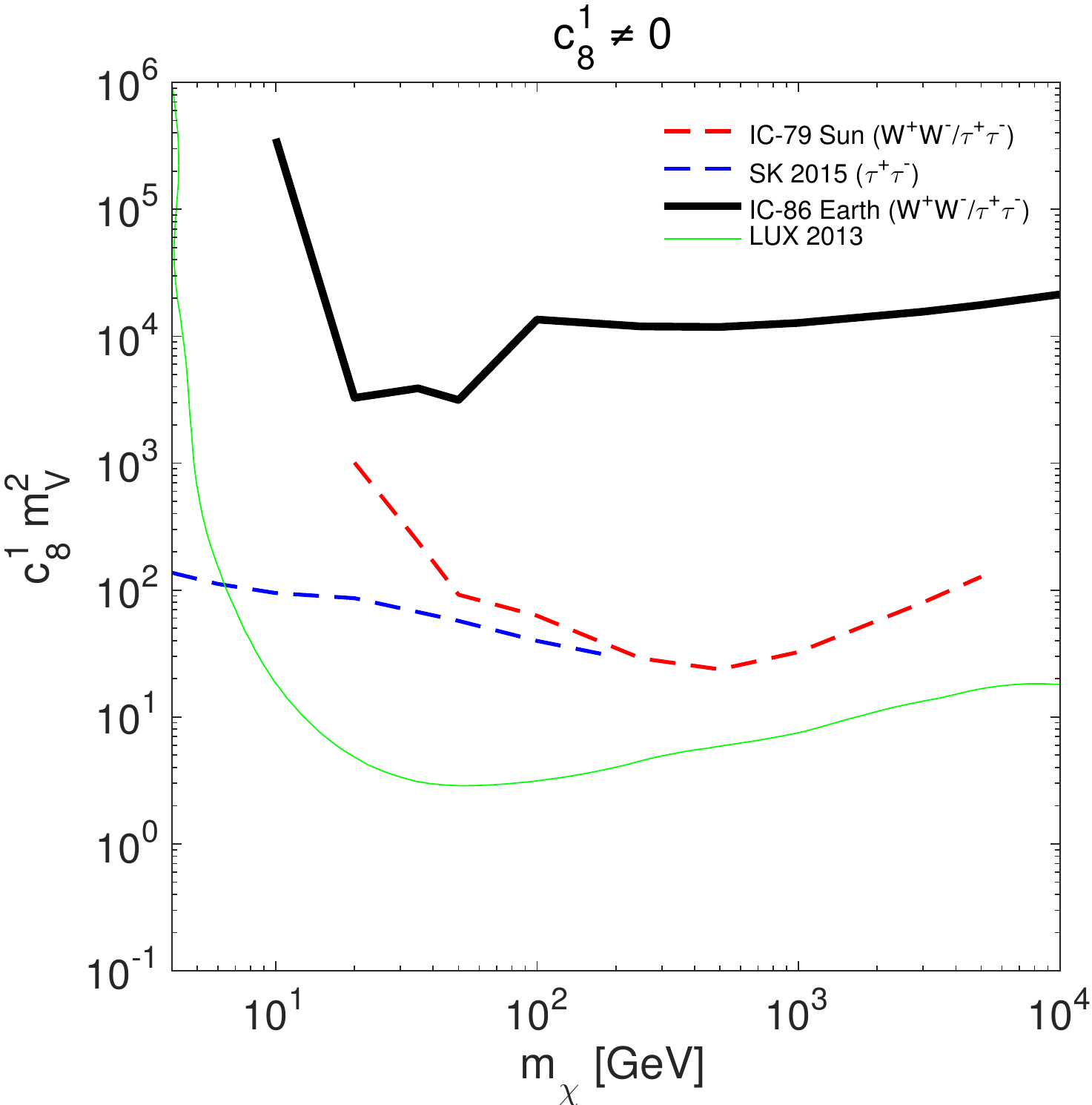}
\end{minipage}
\begin{minipage}[t]{0.45\linewidth}
\centering
\includegraphics[width=\textwidth]{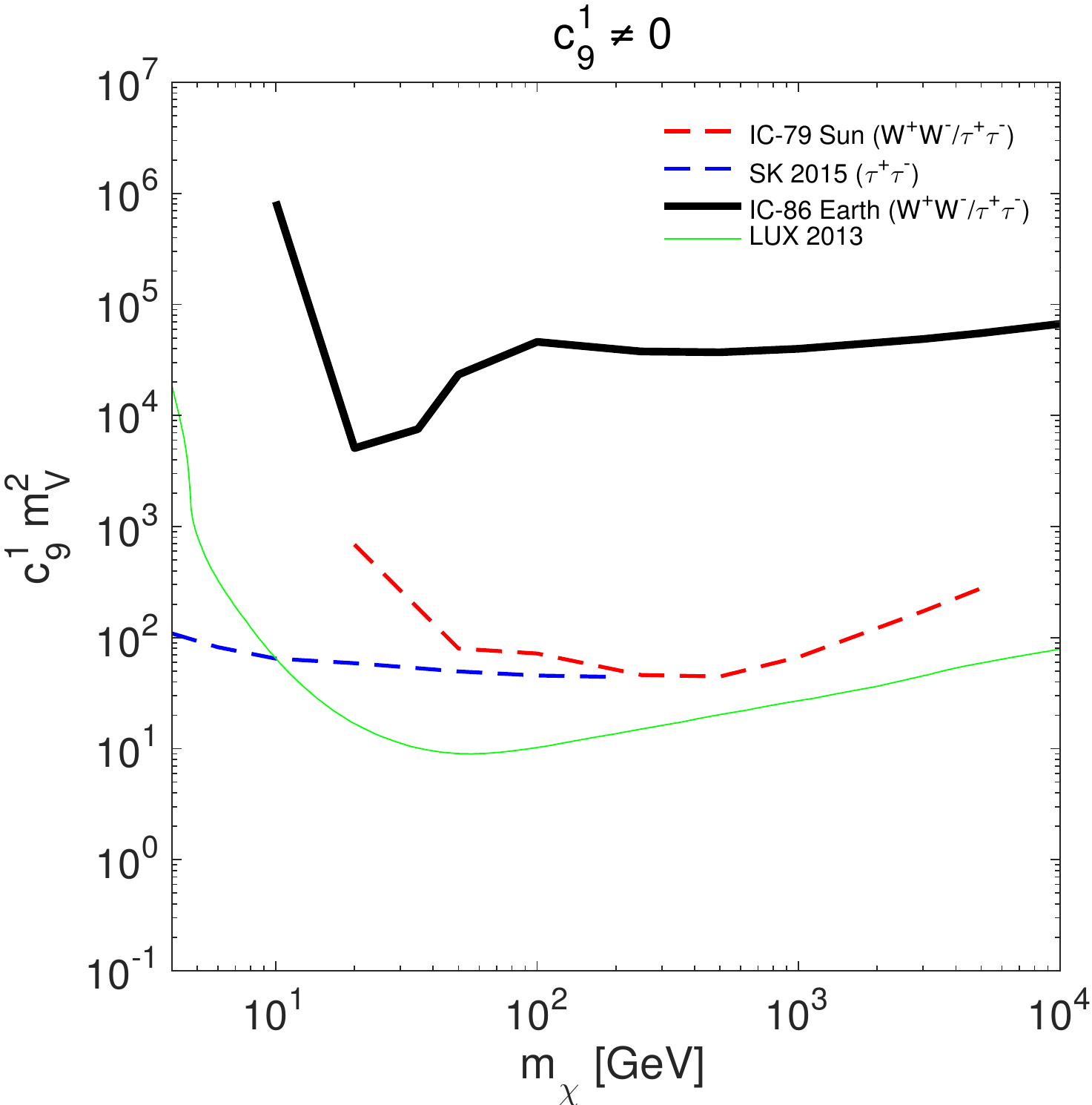}
\end{minipage}
\end{center}
\caption{90\% CL upper limits on the isovector coupling constants of the interaction operators $\hat{\mathcal{O}}_4$, $\hat{\mathcal{O}}_5$, $\hat{\mathcal{O}}_6$, $\hat{\mathcal{O}}_7$, $\hat{\mathcal{O}}_8$, and $\hat{\mathcal{O}}_9$.}
\label{fig:bc3}
\end{figure}

\begin{figure}[t]
\begin{center}
\begin{minipage}[t]{0.45\linewidth}
\centering
\includegraphics[width=\textwidth]{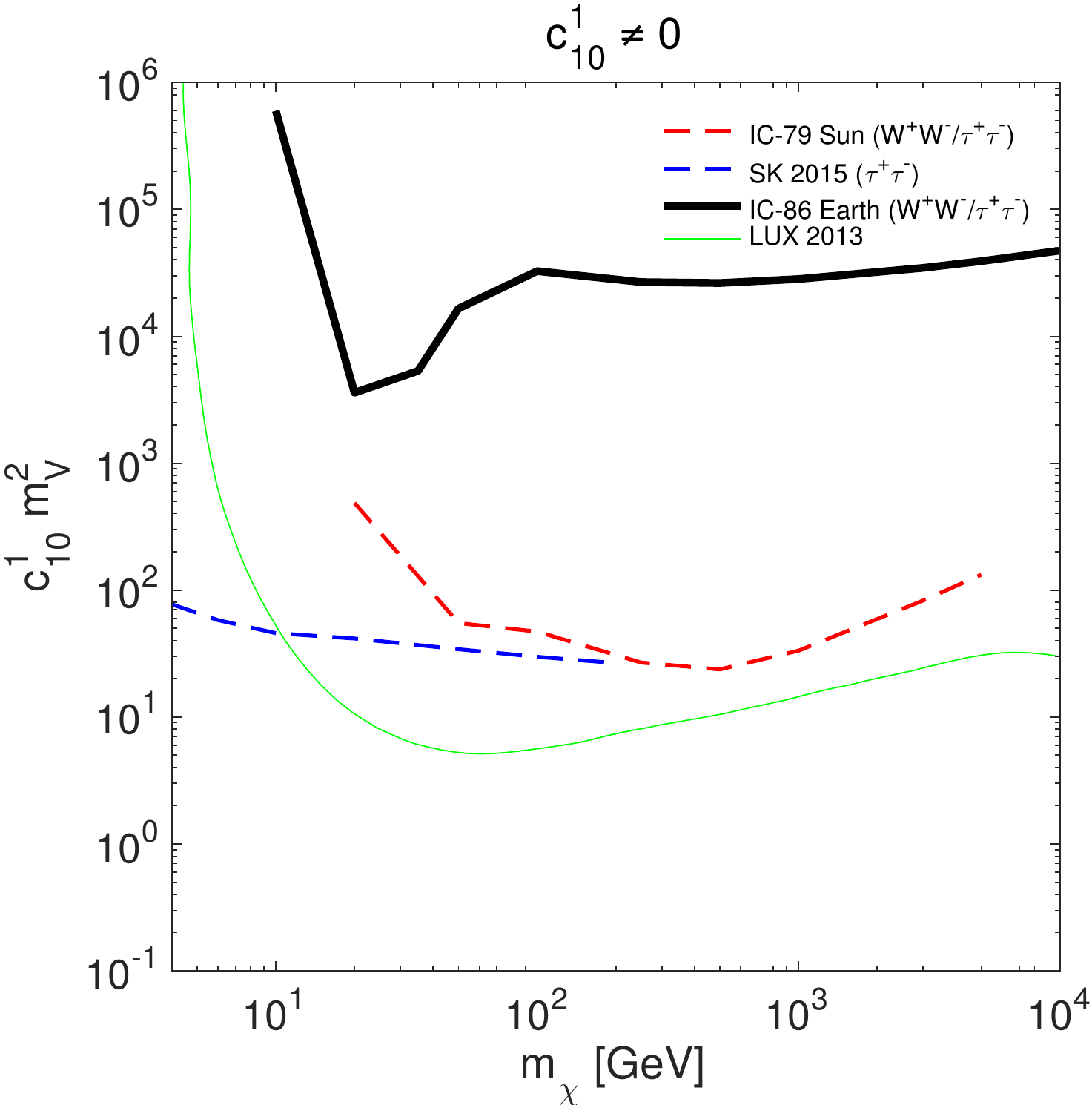}
\end{minipage}
\begin{minipage}[t]{0.45\linewidth}
\centering
\includegraphics[width=\textwidth]{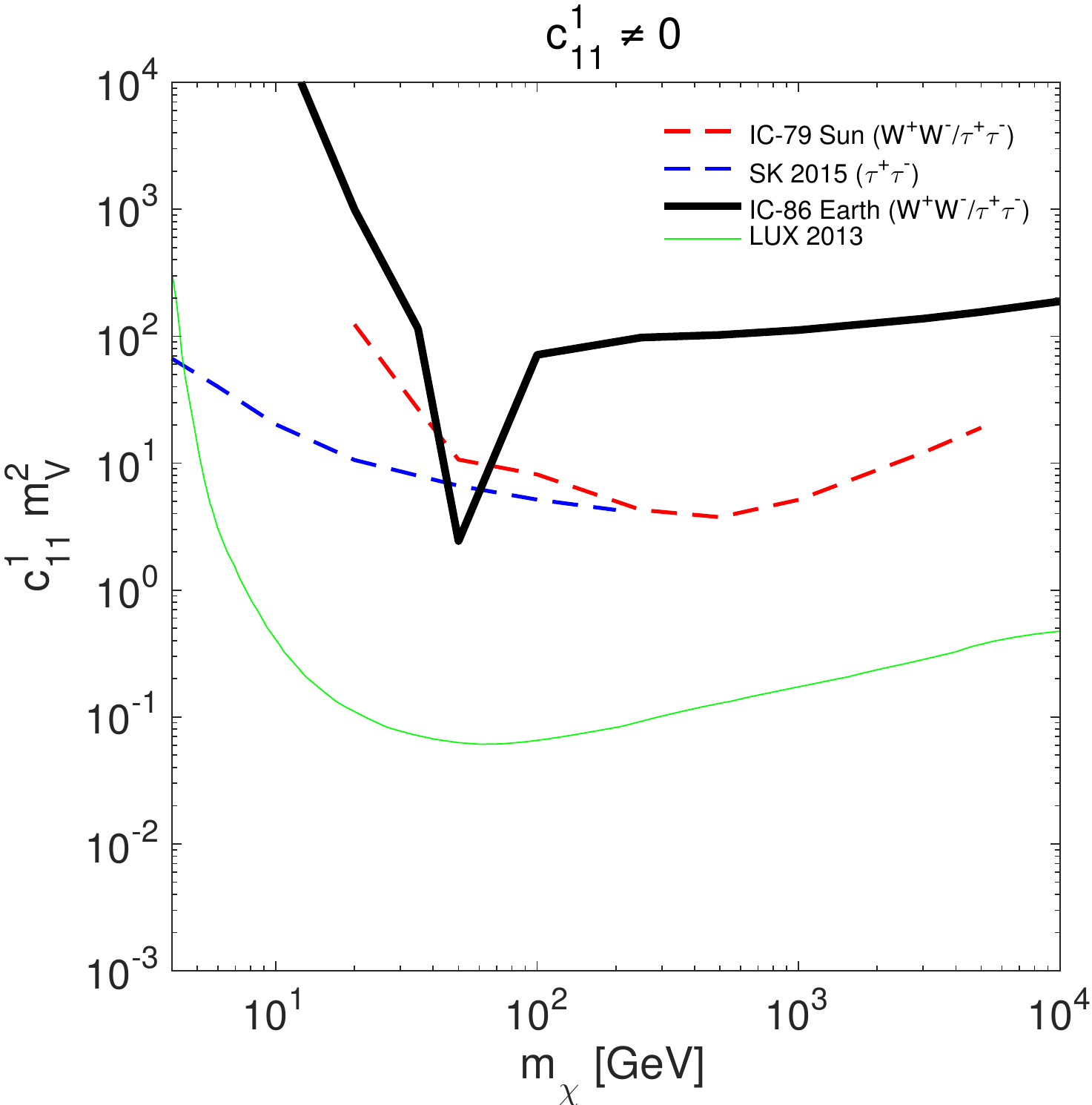}
\end{minipage}
\begin{minipage}[t]{0.45\linewidth}
\centering
\includegraphics[width=\textwidth]{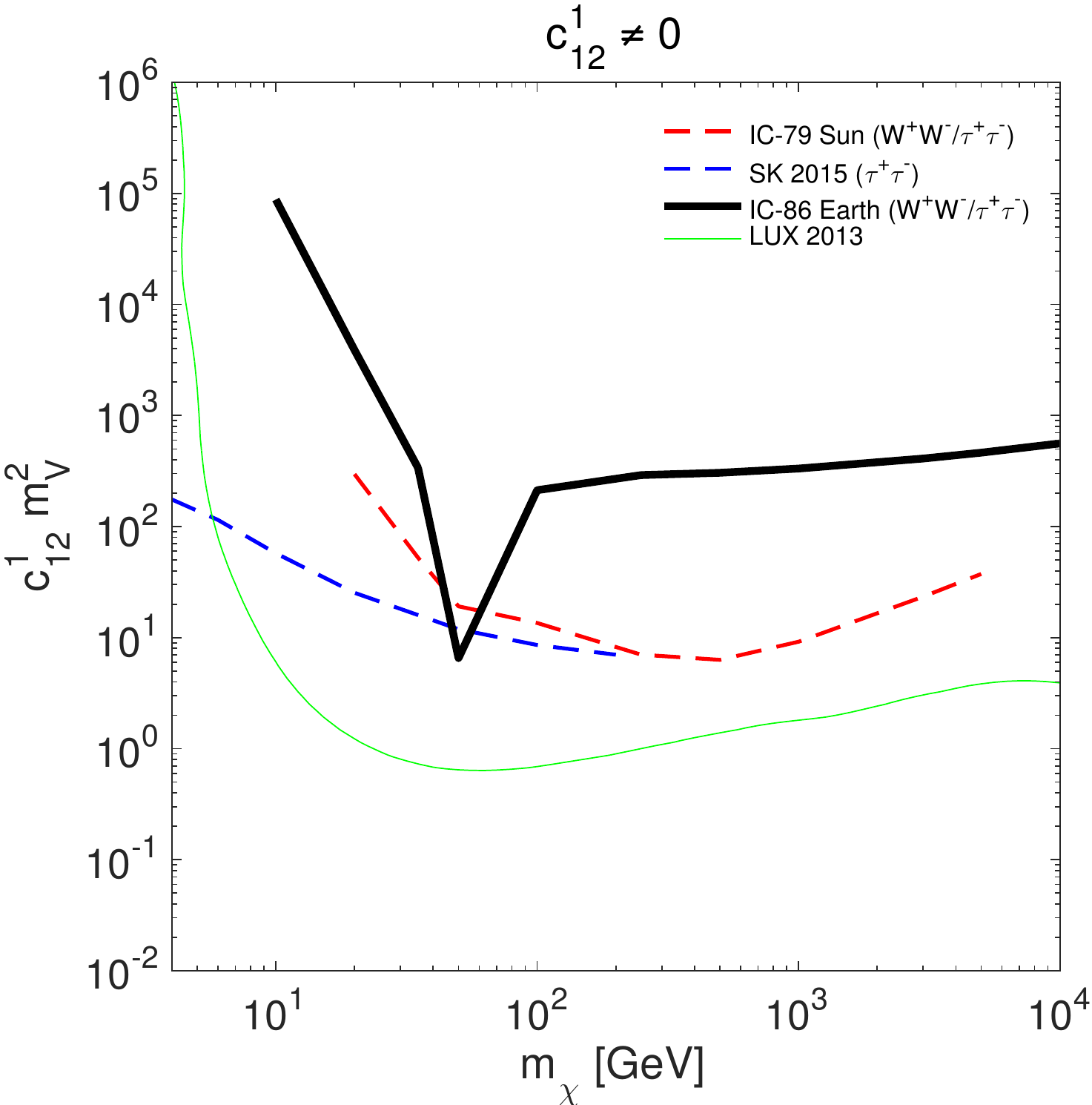}
\end{minipage}
\begin{minipage}[t]{0.45\linewidth}
\centering
\includegraphics[width=\textwidth]{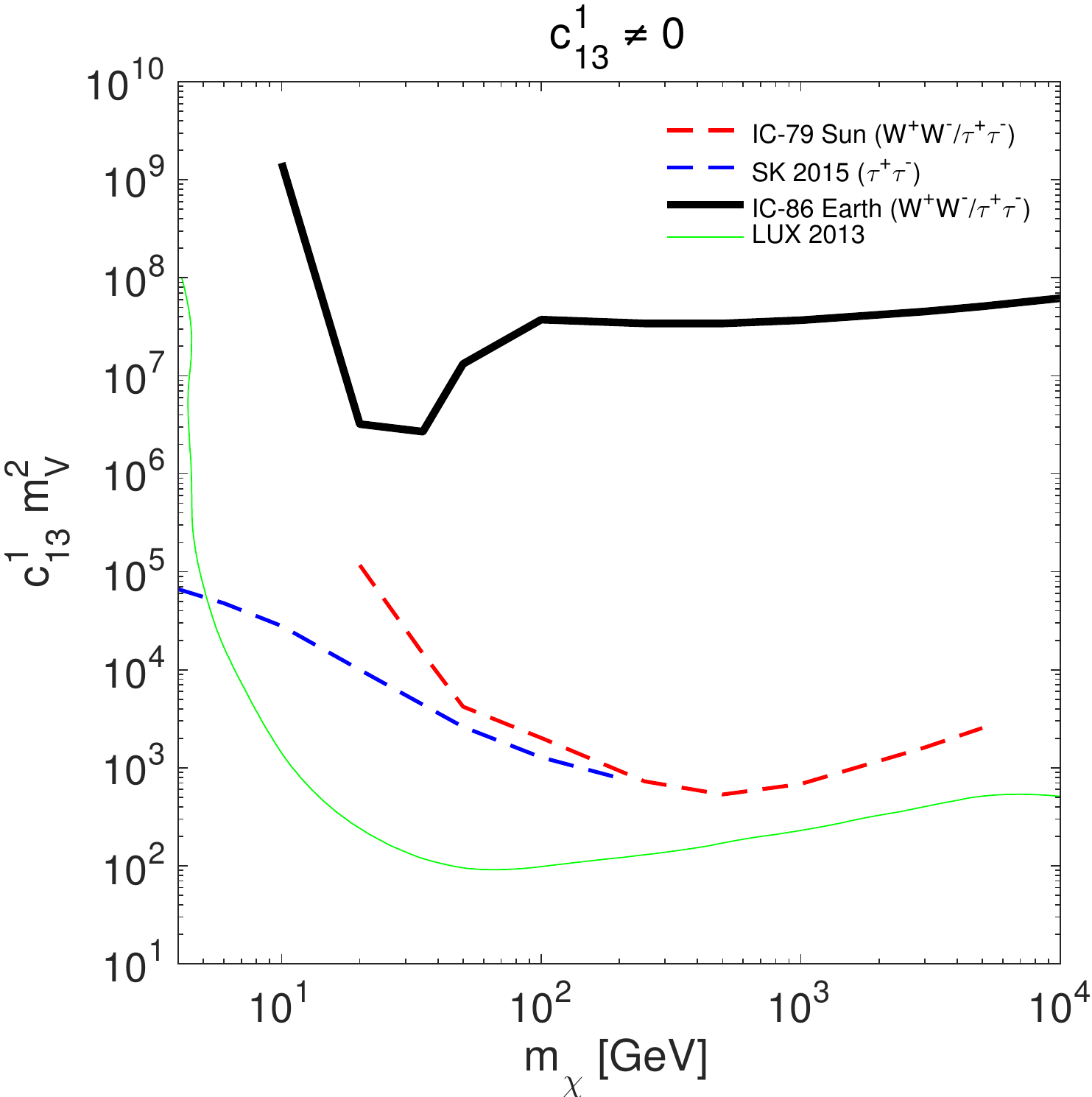}
\end{minipage}
\begin{minipage}[t]{0.45\linewidth}
\centering
\includegraphics[width=\textwidth]{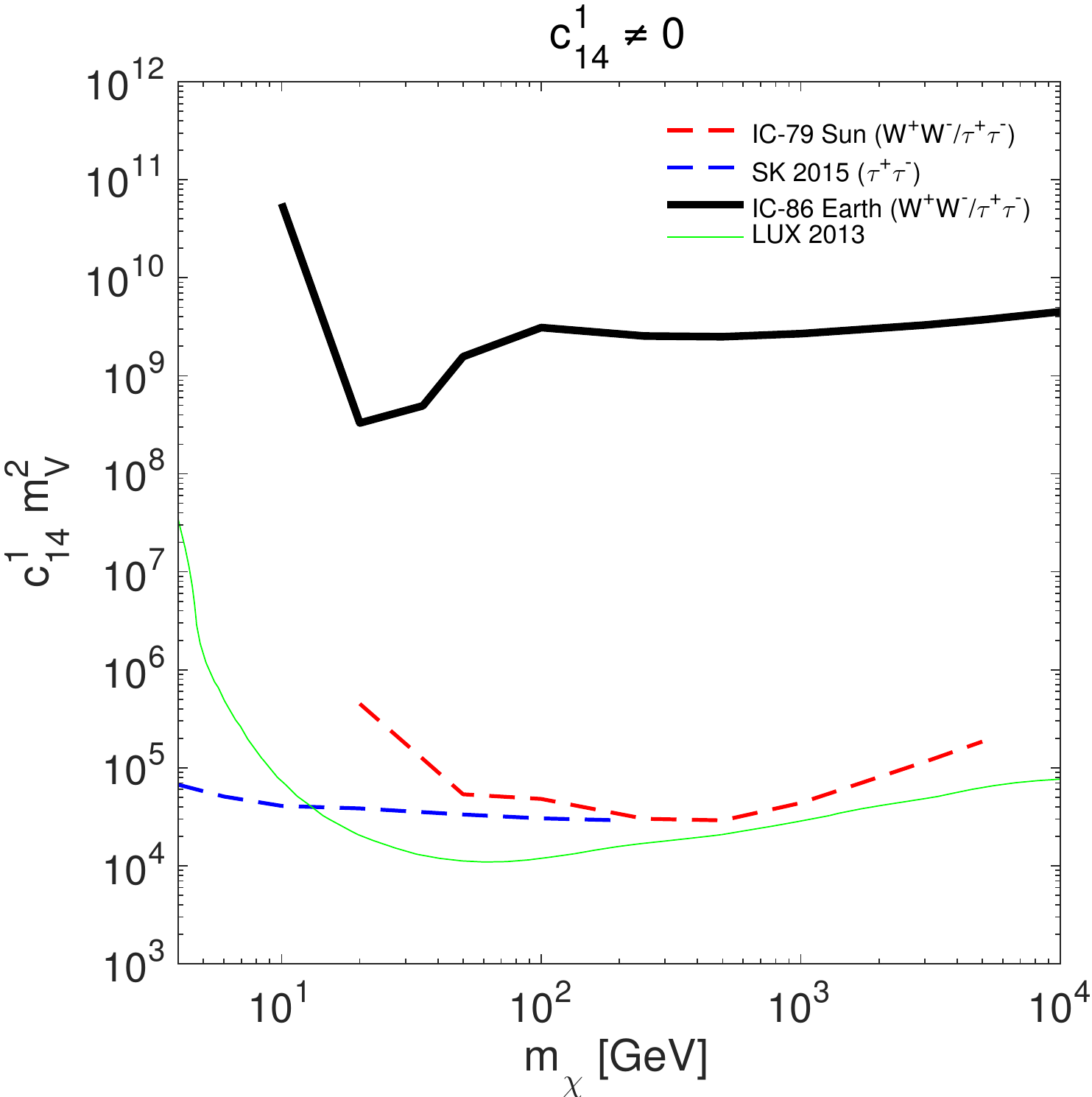}
\end{minipage}
\begin{minipage}[t]{0.45\linewidth}
\centering
\includegraphics[width=\textwidth]{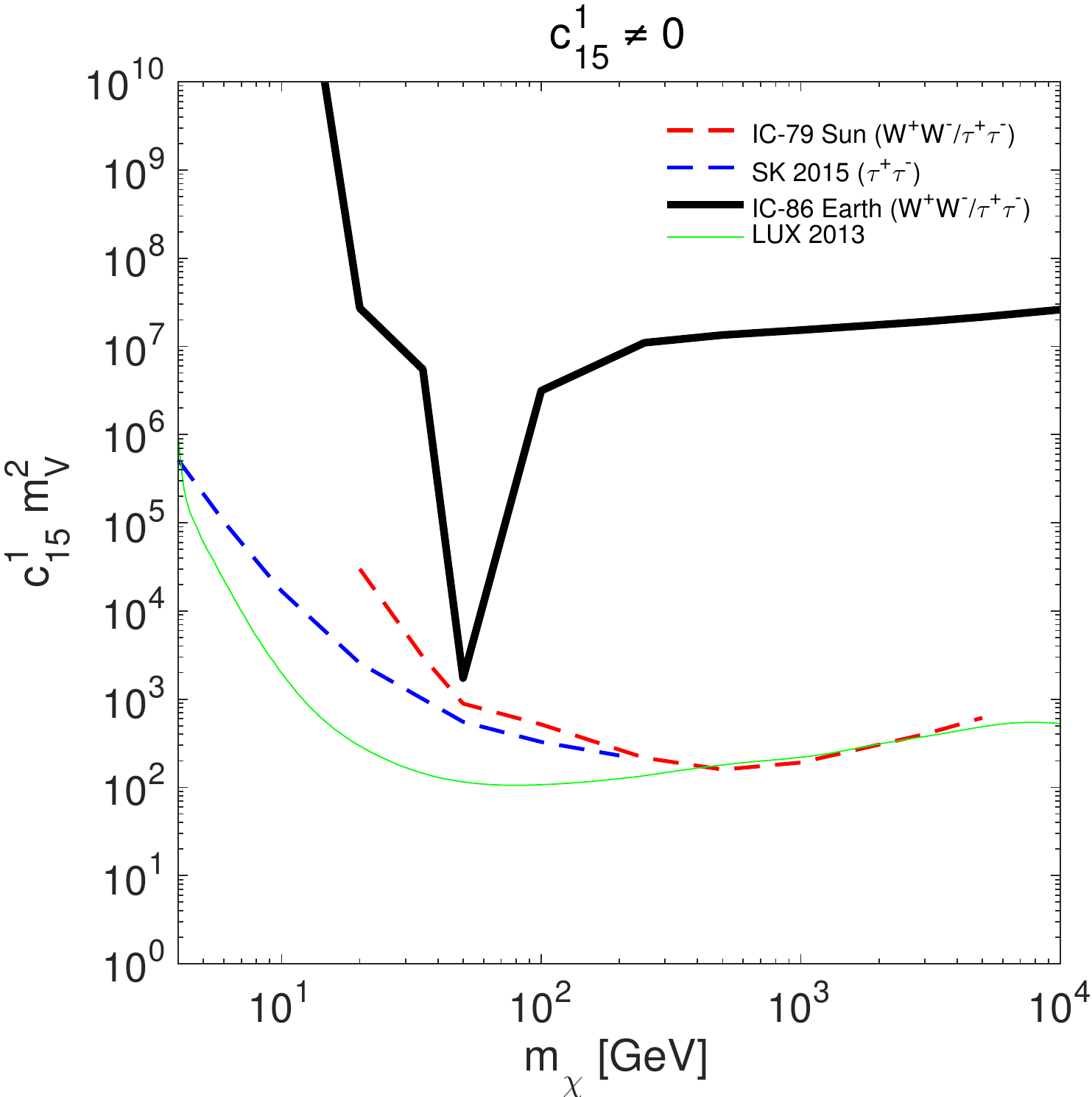}
\end{minipage}
\end{center}
\caption{90\% CL upper limits on the isovector coupling constants of the interaction operators $\hat{\mathcal{O}}_{10}$, $\hat{\mathcal{O}}_{11}$, $\hat{\mathcal{O}}_{12}$, $\hat{\mathcal{O}}_{13}$, $\hat{\mathcal{O}}_{14}$, and $\hat{\mathcal{O}}_{15}$.}
\label{fig:bc4}
\end{figure}

Figs.~\ref{fig:bc1}, ~\ref{fig:bc2}, ~\ref{fig:bc3} and ~\ref{fig:bc4} show the 90\% CL upper limits on the coupling constants of interaction operators not considered in Sec.~\ref{sec:results}.~In the figures the notation is the one used in the body of the paper.

\FloatBarrier


\begin{thebibliography}{10}

\bibitem{Bertone:2010at}
G.~Bertone, \emph{{The moment of truth for WIMP Dark Matter}},
  \href{http://dx.doi.org/10.1038/nature09509}{\emph{Nature} {\bf 468} (2010)
  389--393}, [\href{http://arxiv.org/abs/1011.3532}{{\tt 1011.3532}}].

\bibitem{Bertone:2004pz}
G.~Bertone, D.~Hooper and J.~Silk, \emph{{Particle dark matter: Evidence,
  candidates and constraints}},
  \href{http://dx.doi.org/10.1016/j.physrep.2004.08.031}{\emph{Phys.Rept.} {\bf
  405} (2005) 279--390}, [\href{http://arxiv.org/abs/hep-ph/0404175}{{\tt
  hep-ph/0404175}}].

\bibitem{Jungman:1995df}
G.~Jungman, M.~Kamionkowski and K.~Griest, \emph{{Supersymmetric dark matter}},
  \href{http://dx.doi.org/10.1016/0370-1573(95)00058-5}{\emph{Phys.Rept.} {\bf
  267} (1996) 195--373}, [\href{http://arxiv.org/abs/hep-ph/9506380}{{\tt
  hep-ph/9506380}}].

\bibitem{Bergstrom:2000pn}
L.~Bergstrom, \emph{{Nonbaryonic dark matter: Observational evidence and
  detection methods}},
  \href{http://dx.doi.org/10.1088/0034-4885/63/5/2r3}{\emph{Rept.Prog.Phys.}
  {\bf 63} (2000) 793}, [\href{http://arxiv.org/abs/hep-ph/0002126}{{\tt
  hep-ph/0002126}}].

\bibitem{Catena:2013pka}
R.~Catena and L.~Covi, \emph{{SUSY dark matter(s)}},
  \href{http://dx.doi.org/10.1140/epjc/s10052-013-2703-4}{\emph{Eur.Phys.J.}
  {\bf C74} (2014) 2703}, [\href{http://arxiv.org/abs/1310.4776}{{\tt
  1310.4776}}].

\bibitem{Silk:1985ax}
J.~Silk, K.~A. Olive and M.~Srednicki, \emph{{The Photino, the Sun and
  High-Energy Neutrinos}},
  \href{http://dx.doi.org/10.1103/PhysRevLett.55.257}{\emph{Phys.Rev.Lett.}
  {\bf 55} (1985) 257--259}.

\bibitem{Zentner:2009is}
A.~R. Zentner, \emph{{High-Energy Neutrinos From Dark Matter Particle
  Self-Capture Within the Sun}},
  \href{http://dx.doi.org/10.1103/PhysRevD.80.063501}{\emph{Phys. Rev.} {\bf
  D80} (2009) 063501}, [\href{http://arxiv.org/abs/0907.3448}{{\tt
  0907.3448}}].

\bibitem{Catena:2016ckl}
R.~Catena and A.~Widmark, \emph{{WIMP capture by the Sun in the effective
  theory of dark matter self-interactions}},
  \href{http://arxiv.org/abs/1609.04825}{{\tt 1609.04825}}.

\bibitem{Aartsen:2016exj}
{\scshape IceCube} collaboration, M.~G. Aartsen et~al., \emph{{Improved limits
  on dark matter annihilation in the Sun with the 79-string IceCube detector
  and implications for supersymmetry}},
  \href{http://dx.doi.org/10.1088/1475-7516/2016/04/022}{\emph{JCAP} {\bf 1604}
  (2016) 022}, [\href{http://arxiv.org/abs/1601.00653}{{\tt 1601.00653}}].

\bibitem{Choi:2015ara}
{\scshape Super-Kamiokande} collaboration, K.~Choi et~al., \emph{{Search for
  neutrinos from annihilation of captured low-mass dark matter particles in the
  Sun by Super-Kamiokande}},
  \href{http://dx.doi.org/10.1103/PhysRevLett.114.141301}{\emph{Phys. Rev.
  Lett.} {\bf 114} (2015) 141301}, [\href{http://arxiv.org/abs/1503.04858}{{\tt
  1503.04858}}].

\bibitem{Adrian-Martinez:2016gti}
{\scshape ANTARES} collaboration, S.~Adrian-Martinez et~al., \emph{{Limits on
  Dark Matter Annihilation in the Sun using the ANTARES Neutrino Telescope}},
  \href{http://dx.doi.org/10.1016/j.physletb.2016.05.019}{\emph{Phys. Lett.}
  {\bf B759} (2016) 69--74}, [\href{http://arxiv.org/abs/1603.02228}{{\tt
  1603.02228}}].

\bibitem{Boliev:2013ai}
M.~M. Boliev, S.~V. Demidov, S.~P. Mikheyev and O.~V. Suvorova, \emph{{Search
  for muon signal from dark matter annihilations inthe Sun with the Baksan
  Underground Scintillator Telescope for 24.12 years}},
  \href{http://dx.doi.org/10.1088/1475-7516/2013/09/019}{\emph{JCAP} {\bf 1309}
  (2013) 019}, [\href{http://arxiv.org/abs/1301.1138}{{\tt 1301.1138}}].

\bibitem{Avrorin:2014swy}
{\scshape Baikal} collaboration, A.~D. Avrorin et~al., \emph{{Search for
  neutrino emission from relic dark matter in the Sun with the Baikal NT200
  detector}},
  \href{http://dx.doi.org/10.1016/j.astropartphys.2014.07.006}{\emph{Astropart.
  Phys.} {\bf 62} (2015) 12--20}, [\href{http://arxiv.org/abs/1405.3551}{{\tt
  1405.3551}}].

\bibitem{Catena:2015iea}
R.~Catena, \emph{{Dark matter signals at neutrino telescopes in effective
  theories}},
  \href{http://dx.doi.org/10.1088/1475-7516/2015/04/052}{\emph{JCAP} {\bf 1504}
  (2015) 052}, [\href{http://arxiv.org/abs/1503.04109}{{\tt 1503.04109}}].

\bibitem{Blumenthal:2014cwa}
J.~Blumenthal, P.~Gretskov, M.~Kr{\"a}mer and C.~Wiebusch, \emph{{Effective
  field theory interpretation of searches for dark matter annihilation in the
  Sun with the IceCube Neutrino Observatory}},
  \href{http://dx.doi.org/10.1103/PhysRevD.91.035002}{\emph{Phys.Rev.} {\bf
  D91} (2015) 035002}, [\href{http://arxiv.org/abs/1411.5917}{{\tt
  1411.5917}}].

\bibitem{Liang:2013dsa}
Z.-L. Liang and Y.-L. Wu, \emph{{Direct detection and solar capture of
  spin-dependent dark matter}},
  \href{http://dx.doi.org/10.1103/PhysRevD.89.013010}{\emph{Phys.Rev.} {\bf
  D89} (2014) 013010}, [\href{http://arxiv.org/abs/1308.5897}{{\tt
  1308.5897}}].

\bibitem{Guo:2013ypa}
W.-L. Guo, Z.-L. Liang and Y.-L. Wu, \emph{{Direct detection and solar capture
  of dark matter with momentum and velocity dependent elastic scattering}},
  \href{http://dx.doi.org/10.1016/j.nuclphysb.2013.11.016}{\emph{Nucl.Phys.}
  {\bf B878} (2014) 295--308}, [\href{http://arxiv.org/abs/1305.0912}{{\tt
  1305.0912}}].

\bibitem{Freese:1985qw}
K.~Freese, \emph{{Can Scalar Neutrinos Or Massive Dirac Neutrinos Be the
  Missing Mass?}},
  \href{http://dx.doi.org/10.1016/0370-2693(86)90349-7}{\emph{Phys. Lett.} {\bf
  B167} (1986) 295--300}.

\bibitem{Krauss:1985aaa}
L.~M. Krauss, M.~Srednicki and F.~Wilczek, \emph{{Solar System Constraints and
  Signatures for Dark Matter Candidates}},
  \href{http://dx.doi.org/10.1103/PhysRevD.33.2079}{\emph{Phys. Rev.} {\bf D33}
  (1986) 2079--2083}.

\bibitem{Gaisser:1986ha}
T.~K. Gaisser, G.~Steigman and S.~Tilav, \emph{{Limits on Cold Dark Matter
  Candidates from Deep Underground Detectors}},
  \href{http://dx.doi.org/10.1103/PhysRevD.34.2206}{\emph{Phys. Rev.} {\bf D34}
  (1986) 2206}.

\bibitem{Gould:1987ir}
A.~Gould, \emph{{Resonant Enhancements in WIMP Capture by the Earth}},
  \href{http://dx.doi.org/10.1086/165653}{\emph{Astrophys.J.} {\bf 321} (1987)
  571}.

\bibitem{Gould:1987ww}
A.~Gould, \emph{{Direct and Indirect Capture of Wimps by the Earth}},
  \href{http://dx.doi.org/10.1086/166347}{\emph{Astrophys. J.} {\bf 328} (1988)
  919--939}.

\bibitem{Gould:1991rc}
A.~Gould, \emph{Gravitational diffusion of solar system wimps},
  \href{http://dx.doi.org/10.1086/169726}{\emph{Astrophys. J.} {\bf 368} (feb,
  1991) 610--615}.

\bibitem{Gould:1999je}
A.~Gould and S.~M. Khairul~Alam, \emph{{Can heavy WIMPs be captured by the
  earth?}}, \href{http://dx.doi.org/10.1086/319040}{\emph{Astrophys. J.} {\bf
  549} (2001) 72--75}, [\href{http://arxiv.org/abs/astro-ph/9911288}{{\tt
  astro-ph/9911288}}].

\bibitem{Lundberg:2004dn}
J.~Lundberg and J.~Edsjo, \emph{{WIMP diffusion in the solar system including
  solar depletion and its effect on earth capture rates}},
  \href{http://dx.doi.org/10.1103/PhysRevD.69.123505}{\emph{Phys. Rev.} {\bf
  D69} (2004) 123505}, [\href{http://arxiv.org/abs/astro-ph/0401113}{{\tt
  astro-ph/0401113}}].

\bibitem{Sivertsson:2012qj}
S.~Sivertsson and J.~Edsjo, \emph{{WIMP diffusion in the solar system including
  solar WIMP-nucleon scattering}},
  \href{http://dx.doi.org/10.1103/PhysRevD.85.129905,
  10.1103/PhysRevD.85.123514}{\emph{Phys. Rev.} {\bf D85} (2012) 123514},
  [\href{http://arxiv.org/abs/1201.1895}{{\tt 1201.1895}}].

\bibitem{Chang:2009yt}
S.~Chang, A.~Pierce and N.~Weiner, \emph{{Momentum Dependent Dark Matter
  Scattering}},
  \href{http://dx.doi.org/10.1088/1475-7516/2010/01/006}{\emph{JCAP} {\bf 1001}
  (2010) 006}, [\href{http://arxiv.org/abs/0908.3192}{{\tt 0908.3192}}].

\bibitem{Fan:2010gt}
J.~Fan, M.~Reece and L.-T. Wang, \emph{{Non-relativistic effective theory of
  dark matter direct detection}},
  \href{http://dx.doi.org/10.1088/1475-7516/2010/11/042}{\emph{JCAP} {\bf 1011}
  (2010) 042}, [\href{http://arxiv.org/abs/1008.1591}{{\tt 1008.1591}}].

\bibitem{Fitzpatrick:2012ix}
A.~L. Fitzpatrick, W.~Haxton, E.~Katz, N.~Lubbers and Y.~Xu, \emph{{The
  Effective Field Theory of Dark Matter Direct Detection}},
  \href{http://dx.doi.org/10.1088/1475-7516/2013/02/004}{\emph{JCAP} {\bf 1302}
  (2013) 004}, [\href{http://arxiv.org/abs/1203.3542}{{\tt 1203.3542}}].

\bibitem{Fitzpatrick:2012ib}
A.~L. Fitzpatrick, W.~Haxton, E.~Katz, N.~Lubbers and Y.~Xu, \emph{{Model
  Independent Direct Detection Analyses}},
  \href{http://arxiv.org/abs/1211.2818}{{\tt 1211.2818}}.

\bibitem{Fornengo:2011sz}
N.~Fornengo, P.~Panci and M.~Regis, \emph{{Long-Range Forces in Direct Dark
  Matter Searches}},
  \href{http://dx.doi.org/10.1103/PhysRevD.84.115002}{\emph{Phys.Rev.} {\bf
  D84} (2011) 115002}, [\href{http://arxiv.org/abs/1108.4661}{{\tt
  1108.4661}}].

\bibitem{Menendez:2012tm}
J.~Menendez, D.~Gazit and A.~Schwenk, \emph{{Spin-dependent WIMP scattering off
  nuclei}},
  \href{http://dx.doi.org/10.1103/PhysRevD.86.103511}{\emph{Phys.Rev.} {\bf
  D86} (2012) 103511}, [\href{http://arxiv.org/abs/1208.1094}{{\tt
  1208.1094}}].

\bibitem{Cirigliano:2012pq}
V.~Cirigliano, M.~L. Graesser and G.~Ovanesyan, \emph{{WIMP-nucleus scattering
  in chiral effective theory}},
  \href{http://dx.doi.org/10.1007/JHEP10(2012)025}{\emph{JHEP} {\bf 1210}
  (2012) 025}, [\href{http://arxiv.org/abs/1205.2695}{{\tt 1205.2695}}].

\bibitem{Anand:2013yka}
N.~Anand, A.~L. Fitzpatrick and W.~Haxton, \emph{{Model-independent WIMP
  Scattering Responses and Event Rates: A Mathematica Package for Experimental
  Analysis}},
  \href{http://dx.doi.org/10.1103/PhysRevC.89.065501}{\emph{Phys.Rev.} {\bf
  C89} (2014) 065501}, [\href{http://arxiv.org/abs/1308.6288}{{\tt
  1308.6288}}].

\bibitem{DelNobile:2013sia}
M.~Cirelli, E.~Del~Nobile and P.~Panci, \emph{{Tools for model-independent
  bounds in direct dark matter searches}},
  \href{http://dx.doi.org/10.1088/1475-7516/2013/10/019}{\emph{JCAP} {\bf 1310}
  (2013) 019}, [\href{http://arxiv.org/abs/1307.5955}{{\tt 1307.5955}}].

\bibitem{Klos:2013rwa}
P.~Klos, J.~Menéndez, D.~Gazit and A.~Schwenk, \emph{{Large-scale nuclear
  structure calculations for spin-dependent WIMP scattering with chiral
  effective field theory currents}},
  \href{http://dx.doi.org/10.1103/PhysRevD.89.029901,
  10.1103/PhysRevD.88.083516}{\emph{Phys.Rev.} {\bf D88} (2013) 083516},
  [\href{http://arxiv.org/abs/1304.7684}{{\tt 1304.7684}}].

\bibitem{Peter:2013aha}
A.~H. Peter, V.~Gluscevic, A.~M. Green, B.~J. Kavanagh and S.~K. Lee,
  \emph{{WIMP physics with ensembles of direct-detection experiments}},
  \href{http://dx.doi.org/10.1016/j.dark.2014.10.006}{\emph{Phys.Dark Univ.}
  {\bf 5-6} (2014) 45--74}, [\href{http://arxiv.org/abs/1310.7039}{{\tt
  1310.7039}}].

\bibitem{Hill:2013hoa}
R.~J. Hill and M.~P. Solon, \emph{{WIMP-nucleon scattering with heavy WIMP
  effective theory}},
  \href{http://dx.doi.org/10.1103/PhysRevLett.112.211602}{\emph{Phys.Rev.Lett.}
  {\bf 112} (2014) 211602}, [\href{http://arxiv.org/abs/1309.4092}{{\tt
  1309.4092}}].

\bibitem{Catena:2014uqa}
R.~Catena and P.~Gondolo, \emph{{Global fits of the dark matter-nucleon
  effective interactions}},
  \href{http://dx.doi.org/10.1088/1475-7516/2014/09/045}{\emph{JCAP} {\bf 1409}
  (2014) 045}, [\href{http://arxiv.org/abs/1405.2637}{{\tt 1405.2637}}].

\bibitem{Catena:2014hla}
R.~Catena, \emph{{Analysis of the theoretical bias in dark matter direct
  detection}},
  \href{http://dx.doi.org/10.1088/1475-7516/2014/09/049}{\emph{JCAP} {\bf 1409}
  (2014) 049}, [\href{http://arxiv.org/abs/1407.0127}{{\tt 1407.0127}}].

\bibitem{Catena:2014epa}
R.~Catena, \emph{{Prospects for direct detection of dark matter in an effective
  theory approach}},
  \href{http://dx.doi.org/10.1088/1475-7516/2014/07/055}{\emph{JCAP} {\bf 1407}
  (2014) 055}, [\href{http://arxiv.org/abs/1406.0524}{{\tt 1406.0524}}].

\bibitem{Gluscevic:2014vga}
V.~Gluscevic and A.~H.~G. Peter, \emph{{Understanding WIMP-baryon interactions
  with direct detection: A Roadmap}},
  \href{http://dx.doi.org/10.1088/1475-7516/2014/09/040}{\emph{JCAP} {\bf 1409}
  (2014) 040}, [\href{http://arxiv.org/abs/1406.7008}{{\tt 1406.7008}}].

\bibitem{Panci:2014gga}
P.~Panci, \emph{{New Directions in Direct Dark Matter Searches}},
  \href{http://dx.doi.org/10.1155/2014/681312}{\emph{Adv.High Energy Phys.}
  {\bf 2014} (2014) 681312}, [\href{http://arxiv.org/abs/1402.1507}{{\tt
  1402.1507}}].

\bibitem{Vietze:2014vsa}
L.~Vietze, P.~Klos, J.~Menéndez, W.~Haxton and A.~Schwenk, \emph{{Nuclear
  structure aspects of spin-independent WIMP scattering off xenon}},
  \href{http://arxiv.org/abs/1412.6091}{{\tt 1412.6091}}.

\bibitem{Barello:2014uda}
G.~Barello, S.~Chang and C.~A. Newby, \emph{{A Model Independent Approach to
  Inelastic Dark Matter Scattering}},
  \href{http://dx.doi.org/10.1103/PhysRevD.90.094027}{\emph{Phys.Rev.} {\bf
  D90} (2014) 094027}, [\href{http://arxiv.org/abs/1409.0536}{{\tt
  1409.0536}}].

\bibitem{Catena:2015uua}
R.~Catena and P.~Gondolo, \emph{{Global limits and interference patterns in
  dark matter direct detection}},
  \href{http://dx.doi.org/10.1088/1475-7516/2015/08/022}{\emph{JCAP} {\bf 1508}
  (2015) 022}, [\href{http://arxiv.org/abs/1504.06554}{{\tt 1504.06554}}].

\bibitem{Schneck:2015eqa}
{\scshape SuperCDMS} collaboration, K.~Schneck et~al., \emph{{Dark matter
  effective field theory scattering in direct detection experiments}},
  \href{http://dx.doi.org/10.1103/PhysRevD.91.092004}{\emph{Phys. Rev.} {\bf
  D91} (2015) 092004}, [\href{http://arxiv.org/abs/1503.03379}{{\tt
  1503.03379}}].

\bibitem{Dent:2015zpa}
J.~B. Dent, L.~M. Krauss, J.~L. Newstead and S.~Sabharwal, \emph{{General
  analysis of direct dark matter detection: From microphysics to observational
  signatures}}, \href{http://dx.doi.org/10.1103/PhysRevD.92.063515}{\emph{Phys.
  Rev.} {\bf D92} (2015) 063515}, [\href{http://arxiv.org/abs/1505.03117}{{\tt
  1505.03117}}].

\bibitem{Catena:2015vpa}
R.~Catena, \emph{{Dark matter directional detection in non-relativistic
  effective theories}},
  \href{http://dx.doi.org/10.1088/1475-7516/2015/07/026}{\emph{JCAP} {\bf 1507}
  (2015) 026}, [\href{http://arxiv.org/abs/1505.06441}{{\tt 1505.06441}}].

\bibitem{Kavanagh:2015jma}
B.~J. Kavanagh, \emph{{New directional signatures from the nonrelativistic
  effective field theory of dark matter}},
  \href{http://dx.doi.org/10.1103/PhysRevD.92.023513}{\emph{Phys. Rev.} {\bf
  D92} (2015) 023513}, [\href{http://arxiv.org/abs/1505.07406}{{\tt
  1505.07406}}].

\bibitem{D'Eramo:2016atc}
F.~D'Eramo, B.~J. Kavanagh and P.~Panci, \emph{{You can hide but you have to
  run: direct detection with vector mediators}},
  \href{http://dx.doi.org/10.1007/JHEP08(2016)111}{\emph{JHEP} {\bf 08} (2016)
  111}, [\href{http://arxiv.org/abs/1605.04917}{{\tt 1605.04917}}].

\bibitem{Catena:2016hoj}
R.~Catena, A.~Ibarra and S.~Wild, \emph{{DAMA confronts null searches in the
  effective theory of dark matter-nucleon interactions}},
  \href{http://dx.doi.org/10.1088/1475-7516/2016/05/039}{\emph{JCAP} {\bf 1605}
  (2016) 039}, [\href{http://arxiv.org/abs/1602.04074}{{\tt 1602.04074}}].

\bibitem{Kahlhoefer:2016eds}
F.~Kahlhoefer and S.~Wild, \emph{{Studying generalised dark matter interactions
  with extended halo-independent methods}},
  \href{http://arxiv.org/abs/1607.04418}{{\tt 1607.04418}}.

\bibitem{Catena:2015uha}
R.~Catena and B.~Schwabe, \emph{{Form factors for dark matter capture by the
  Sun in effective theories}},
  \href{http://dx.doi.org/10.1088/1475-7516/2015/04/042}{\emph{JCAP} {\bf 1504}
  (2015) 042}, [\href{http://arxiv.org/abs/1501.03729}{{\tt 1501.03729}}].

\bibitem{Aartsen:2016fep}
{\scshape IceCube} collaboration, M.~G. Aartsen et~al., \emph{{First search for
  dark matter annihilations in the Earth with the IceCube Detector}},
  \href{http://arxiv.org/abs/1609.01492}{{\tt 1609.01492}}.

\bibitem{Aartsen:2012kia}
{\scshape IceCube collaboration} collaboration, M.~Aartsen et~al.,
  \emph{{Search for dark matter annihilations in the Sun with the 79-string
  IceCube detector}},
  \href{http://dx.doi.org/10.1103/PhysRevLett.110.131302}{\emph{Phys.Rev.Lett.}
  {\bf 110} (2013) 131302}, [\href{http://arxiv.org/abs/1212.4097}{{\tt
  1212.4097}}].

\bibitem{Akerib:2013tjd}
{\scshape LUX Collaboration} collaboration, D.~Akerib et~al., \emph{{First
  results from the LUX dark matter experiment at the Sanford Underground
  Research Facility}},  \href{http://arxiv.org/abs/1310.8214}{{\tt 1310.8214}}.

\bibitem{Toivanen:2008zz}
P.~Toivanen, M.~Kortelainen, J.~Suhonen and J.~Toivanen, \emph{{Dark-matter
  detection by elastic and inelastic LSP scattering on Xe-129 and Xe-131}},
  \href{http://dx.doi.org/10.1016/j.physletb.2008.06.057}{\emph{Phys. Lett.}
  {\bf B666} (2008) 1--4}.

\bibitem{Hoferichter:2015ipa}
M.~Hoferichter, P.~Klos and A.~Schwenk, \emph{{Chiral power counting of one-
  and two-body currents in direct detection of dark matter}},
  \href{http://dx.doi.org/10.1016/j.physletb.2015.05.041}{\emph{Phys. Lett.}
  {\bf B746} (2015) 410--416}, [\href{http://arxiv.org/abs/1503.04811}{{\tt
  1503.04811}}].

\bibitem{Catena:2009mf}
R.~Catena and P.~Ullio, \emph{{A novel determination of the local dark matter
  density}}, \href{http://dx.doi.org/10.1088/1475-7516/2010/08/004}{\emph{JCAP}
  {\bf 1008} (2010) 004}, [\href{http://arxiv.org/abs/0907.0018}{{\tt
  0907.0018}}].

\bibitem{Catena:2011kv}
R.~Catena and P.~Ullio, \emph{{The local dark matter phase-space density and
  impact on WIMP direct detection}},
  \href{http://dx.doi.org/10.1088/1475-7516/2012/05/005}{\emph{JCAP} {\bf 1205}
  (2012) 005}, [\href{http://arxiv.org/abs/1111.3556}{{\tt 1111.3556}}].

\bibitem{Bozorgnia:2013pua}
N.~Bozorgnia, R.~Catena and T.~Schwetz, \emph{{Anisotropic dark matter
  distribution functions and impact on WIMP direct detection}},
  \href{http://dx.doi.org/10.1088/1475-7516/2013/12/050}{\emph{JCAP} {\bf 1312}
  (2013) 050}, [\href{http://arxiv.org/abs/1310.0468}{{\tt 1310.0468}}].

\bibitem{Gondolo:2004sc}
P.~Gondolo, J.~Edsjo, P.~Ullio, L.~Bergstrom, M.~Schelke et~al.,
  \emph{{DarkSUSY: Computing supersymmetric dark matter properties
  numerically}},
  \href{http://dx.doi.org/10.1088/1475-7516/2004/07/008}{\emph{JCAP} {\bf 0407}
  (2004) 008}, [\href{http://arxiv.org/abs/astro-ph/0406204}{{\tt
  astro-ph/0406204}}].

\bibitem{raddens}
W.~Mcdonough, \emph{Treatise on Geochemistry, Vol. 2}.
\newblock Elsevier, 2003.

\bibitem{Vincent:2013lua}
A.~C. Vincent and P.~Scott, \emph{{Thermal conduction by dark matter with
  velocity and momentum-dependent cross-sections}},
  \href{http://dx.doi.org/10.1088/1475-7516/2014/04/019}{\emph{JCAP} {\bf 1404}
  (2014) 019}, [\href{http://arxiv.org/abs/1311.2074}{{\tt 1311.2074}}].

\bibitem{Blennow:2007tw}
M.~Blennow, J.~Edsjo and T.~Ohlsson, \emph{{Neutrinos from WIMP annihilations
  using a full three-flavor Monte Carlo}},
  \href{http://dx.doi.org/10.1088/1475-7516/2008/01/021}{\emph{JCAP} {\bf 0801}
  (2008) 021}, [\href{http://arxiv.org/abs/0709.3898}{{\tt 0709.3898}}].

\bibitem{visinelli}
S.~Baum, K.~Freese and L.~Visinelli, \emph{{in preparation}}.

\bibitem{Warburton:1992rh}
E.~Warburton and B.~A. Brown, \emph{{Effective Interactions for the Op1sOd
  nuclear shell model space}},
  \href{http://dx.doi.org/10.1103/PhysRevC.46.923}{\emph{Phys.Rev.} {\bf C46}
  (1992) 923--944}.

\bibitem{Honma:2004xk}
M.~Honma, T.~Otsuka, B.~A. Brown and T.~Mizusaki, \emph{{New effective
  interaction for pf shell nuclei and its implications for the stability of the
  N = Z = 28 closed core}},
  \href{http://dx.doi.org/10.1103/PhysRevC.69.034335}{\emph{Phys.Rev.} {\bf
  C69} (2004) 034335}, [\href{http://arxiv.org/abs/nucl-th/0402079}{{\tt
  nucl-th/0402079}}].

\bibitem{Brown:2001zz}
B.~Brown, \emph{{The nuclear shell model towards the drip lines}},
  \href{http://dx.doi.org/10.1016/S0146-6410(01)00159-4}{\emph{Prog.Part.Nucl.Phys.}
  {\bf 47} (2001) 517--599}.

\bibitem{NuShell}
B.~A. Brown and W.~D.~M. Rae, \emph{Nushell@msu}, {\emph{MSU-NSCL report}
  (2007) }.

\end{thebibliography}

\providecommand{\href}[2]{#2}\begingroup\raggedright\endgroup

\end{document}